\documentclass[12pt]{article}
\usepackage{amsmath, amssymb}

\newtheorem{theorem}{Theorem}[section]

\newtheorem{lemma}[theorem]{Lemma}

\numberwithin{equation}{section}

\newcommand{\R}{\mathbb{R}}

\renewcommand{\Re}{\mathop{\mathrm{Re}}}
\renewcommand{\Im}{\mathop{\mathrm{Im}}}

\newcommand{\norm}[1]{\left\Vert #1 \right\Vert}
\newcommand{\length}[1]{\left| #1 \right|}
\newcommand{\bkA}[1]{\left \langle #1 \right \rangle}
\newcommand{\bke}[1]{\left( #1 \right)}
\newcommand{\bkt}[1]{\left[ #1 \right]}
\newcommand{\bket}[1]{\left\{ #1 \right\}}

\newcommand{\e}{\varepsilon}
\newcommand{\loc}{_{\text{loc}}}

\newcommand{\myproof}{\noindent {\bf Proof.}\quad}
\newcommand{\myendproof}{\hspace*{\fill}{{\bf \small Q.E.D.}} \vspace{10pt}}

\newcommand{\eqlabel}[1]{\label{#1}}
\newcommand{\thlabel}[1]{\label{#1}}

\newcommand{\al}{\alpha}
\newcommand{\ka}{\kappa}
\newcommand{\om}{\omega}
\newcommand{\up}{u_+}
\newcommand{\um}{u_-}

\newcommand{\sA}{A}  %

\renewcommand{\d}{\delta}
\renewcommand{\L}{\mathcal{L}}
\newcommand{\la}{\lambda}
\newcommand{\Pc}{\, \mathbf{P}\! _\mathrm{c} \, \! }
\newcommand{\PcA}{\, \mathbf{P}\! _\mathrm{c} \! ^A \,}
\newcommand{\PcH}{\, \mathbf{P}\! _\mathrm{c} \! ^{H_0} \,}
\newcommand{\PcL}{\, \mathbf{P}\! _\mathrm{c} \! ^\L \,}
\newcommand{\Hc}{\, \mathtt{H} _\mathrm{c} \, \! }
\newcommand{\Eigen}{\mathtt{E}}

\newcommand{\original}{^{(\mathrm{or})}}
\newcommand{\diff}{{\pmb{\delta}}}
\newcommand{\tl}[1]{\bket{#1}}%
\newcommand{\wt}{\widetilde}
\newcommand{\conj}{\mathtt{C}}

\newcommand{\xii}{\xi_\infty}

\newcommand{\pd}{\partial}
\newcommand{\ini}{_{\mathrm{in}}}
\newcommand{\dt} {\Delta T}
\newcommand{\wbar}[1]{\overline{\rule{0pt}{2.4mm} {#1}}}
\newcommand{\ad}{{\,\mathrm{ad}}}

\begin{document}
\noindent February, 2001%\today
\baselineskip 20pt

\vspace{2.5cm} \centerline{\Large{\bf Asymptotic Dynamics of
Nonlinear Schr\"odinger Equations:}} \vspace{.8cm} \centerline{{\bf
\large Resonance Dominated and Radiation Dominated Solutions}}

\vspace{2cm}

\renewcommand{\thefootnote}{\fnsymbol{footnote}}

{\large

\centerline{ Tai-Peng Tsai\footnote[1]{ttsai@cims.nyu.edu},
Horng-Tzer Yau\footnote[2]{Work partially supported by NSF grant
DMS-0072098,  yau@cims.nyu.edu}
\renewcommand{\thefootnote}{\arabic{footnote}}}

\vspace{1.5cm}

\centerline{ Courant Institute }
\centerline{ New York University}
\centerline{ New York, NY, 10012}
}

\bigskip
\bigskip

\begin{abstract}

We consider a linear Schr\"odinger equation with a small nonlinear
perturbation in $\R^3$. Assume that the linear
Hamiltonian has exactly two bound states and its eigenvalues satisfy
some resonance condition. We prove that if the initial data is
near a nonlinear ground state, then the solution approaches to certain
nonlinear ground state as the time tends to infinity. Furthermore,
the difference between the wave function solving the nonlinear Schr\"odinger
equation and its asymptotic profile can have two different types of
decay: 1. The resonance dominated solutions decay as
$t^{-1/2}$. 2.  The radiation dominated solutions decay at least like
$t^{-3/2}$.

\end{abstract}

\newpage

\setcounter{section}{0}
\section{Introduction}
\newcommand{\Li}{{{\cal L}} }
\newcommand{\Lori}{\L \original}

\newcommand{\Lmi}{{L}_-}
\newcommand{\Lpi}{{L}_+}

Let  $H_0$ be the Hamiltonian $H_0 = - \Delta + V- e_0$ with $V$
a smooth  localized potential and $e_0< 0$  the ground state
energy to $- \Delta + V$. Consider the nonlinear  Schr\"odinger
equation
\begin{equation} \eqlabel{Sch} i \partial _t \psi = (-\Delta + V) \psi + \la
|\psi|^2 \psi, \qquad \psi(t=0)= \psi_0
\end{equation}
where $\la$ is a small positive or negative  parameter.
Our goal is to understand its asymptotic evolution as
$t\to \infty$. The nonlinear bound states to the Schr\"odinger
equation \eqref{Sch} are  solutions to the nonlinear equation
\begin{equation}   \eqlabel{Q.eq}
    (-\Delta + V) Q + \la |Q|^2 Q = EQ ,
\end{equation}
For any nonlinear bound state, $\psi_t = Q e^{-i E t } $ is a solution to the nonlinear
Schr\"odinger equation.
We may obtain a family of such bound states from minimizing
the corresponding nonlinear energy functional:
\[
{\cal H}[\phi] =
\int \frac 12 |\nabla \phi|^2 +  \frac 12 V |\phi|^2 + \frac 14 \la |\phi|^4
\, d x ~.
\]
For each $N>0$ and $\lambda$ sufficiently small, subject to the
constraint $\norm{\phi}_{L^2}^2 = N$, there is a unique
positive  minimizer $Q$ of $\cal H$, which solves \eqref{Q.eq} for
some $E=E(N)$ and has exponential decay as  $x \to \infty$. We
call this family nonlinear ground states.

Instead of $N$, we can use $E$ as the parameter. From now on
we will refer to this continuous family as $\bket{Q_E}_{E}$.
Let
\begin{equation}\eqlabel{HE}
  H_E = -\Delta + V -E + \la Q_E^2  ~.
\end{equation}
We have $H_E Q_E = 0$.
Since $\la$ is small, the spectral properties of $H_E$ are similar
to those of $H_0$.

Suppose the initial data of the nonlinear Schr\"odinger equation
$\psi_0$ is near some $Q_E$. If $- \Delta + V$ has only one bound
state, it was proved in \cite{SW1} that the evolution will
eventually settle down to some ground state $Q_{E_\infty}$ with
$E_\infty$ close to $E$. Suppose now that  $- \Delta + V$ has
multiple bound states, say, two bound states: a ground state
$\phi_0$ with eigenvalue $e_0$ and an excited state $\phi_1$ with
eigenvalue $e_1$, i.e., $H_0 \phi_1 = e_{01} \phi_1$ where
$e_{01}=e_1-e_0$. The question is whether the evolution with
initial data $\psi_0$ near some $Q_E$ will eventually settle down
to some ground state $Q_{E_\infty}$ with $E_\infty$ close to $E$?
Furthermore, can we characterize the asymptotic behavior of the
evolution?

The family of  ground states is stable
in the sense that if
\[
   \inf_{\Theta, E} \norm{ \psi_0 -Q_E \, e^{i \Theta} }_{L^2}
\]
is small, it remains so for all $t$. Let $\norm{ \cdot }_{L^2,
loc}$ denote local $L^2$ norm. One expects that this
difference is actually approaching to zero, i.e.,
\begin{equation}\eqlabel{as}
\lim_{t \to \infty}
\inf_{ \Theta, E} \norm{ \psi_t -Q_E \, e^{i \Theta} }_{L^2, loc} = 0
\end{equation}
In this paper, we shall answer this question positively in the case of
two bound states.
Furthermore, we  estimate precisely the rate of relaxation for
certain class of initial data.

We start with a simple question concerning the asymptotic dynamics around a
fixed ground state profile at $t=\infty$. More precisely, we
first fix an $E$ and the corresponding nonlinear ground state
$Q_E$. Our goal is to analyze the detailed asymptotic behavior of
those solutions converging to this nonlinear ground state as
$t \to \infty$. Although this problem seems to be simple, we found
that there are two different types of behaviors for $\psi_t -
Q_{E}e^{i \Theta}$: one is called resonance dominated solutions ;
the other radiation dominated solutions. We first explain the
Resonance condition on $H_0$ and the meanings of these solutions.
Recall that the ground state and the excited state to $H_0$ are denoted
by $\phi_{0}$ and $\phi_1$ respectively. The
following  two conditions are assumed for this paper.

\noindent
{\bf Assumption A1}: Resonance condition. Let $e_{01}= e_1-e_0$ be the
spectral gap of the ground state. We assume that $2e_{01} > |e_0|$
so that $2e_{01}$ is in the continuum spectrum of $H_0$.
Furthermore,
for some constant $\gamma_0>0$ and all real $s$ sufficiently small,
\begin{equation}
 \bke{ \phi_0 \phi_1^2 \, , \, \Im \frac
{1}{H_0 -0i - 2\, e_{01}-s}  \PcH  \phi_0 \phi_1^2       } > \gamma_0>0 ~.
\end{equation}

\noindent
{\bf Assumption A2}:
For $\la$ sufficiently small and $E$ in a small neighborhood of $e_0$,
the bottom of continuum spectrum to $-\Delta+ V+\la Q_E^2$, $0$,
is not a generalized eigenvalue, i.e., not a resonance.
Also, we assume that $V$ satisfies the
assumption in Yajima \cite{Y} so that the $W^{k,p}$ estimates
for $k\le 2$ for the
wave operator $W_H$ holds: for a small $\sigma>0$,
\[
|\nabla^\al V(x)| \le C \bkA{x}^{-5-\sigma}, \qquad
\text{for } |\al|\le 2 ~.
\]
Also, the functions
$(x\cdot \nabla)^k V$, for $k=0,1,2,3$, are $-\Delta$ bounded with
an $-\Delta$-bound $<1$:
\begin{equation*}
  \norm{(x\cdot \nabla)^k V\phi}_2 \le \sigma_0 \norm{-\Delta\phi}_2 +
C\norm{\phi}_2,
  \qquad \sigma_0 < 1 , \quad k=0,1,2,3  ~.
\end{equation*}

\bigskip

Fixed an $E$ and its ground state $Q_{E}$. Let $\Lori$ be the
operator obtained from  linearizing the Schr\"odinger equation
\eqref{Sch} around the trivial evolution $Q e^{-i E t }$, i.e.,
\[
   {\Lori} k = -i \bket{ H_E k + \la Q^2(k + \bar k) } ~,
\]
where $H_E$ is defined in \eqref{HE}.  It is more convenient to
work with operators orthogonal to the ground state $Q$. Let $\Pi$ be
the projection which eliminates the $Q$-direction:
\[
\Pi h= h - ( c_0Q, h) Q ~, \quad c_{0} = (Q,Q)^{-1}
\]
and let $X$ denote its image,
\[
   X=\Pi (L^2) ~.
\]
Define the operator ${\L}$ acting on the space $X=\Pi L^2$ by
\[
        {\L} h= -i \bket{ H h +\la \Pi  Q^2 \Pi (h + \bar h) } ~.
\]
The operator can be written in matrix form
\[
\L= \begin{bmatrix} 0 & {L_-} \\  {- L_+}  & 0 \end{bmatrix},
\qquad {L_-} = H, \qquad {L_+} = H + 2 \la \Pi  Q^2 \Pi.
\]

With respect to $\L$, we define generalized ``eigenspaces'':
\[
   \Eigen_\nu ({\L}) := \bket{ \psi: {\L} ^2 \psi = - \nu ^2 \psi }
        = \bket{ u + iv: u, v \text{ real, }{L_-} {L_+}  u = \nu^2 u,
      {L_+} {L_-} v = \nu^2 v} \ .
\]
Since the original Hamiltonian $H_0$ has only two bound states
with the ground state projected out, there is exactly  one value for
$\nu$, called  $\ka$, and  dim$_\R \Eigen_\kappa = 2$.
From simple
perturbation theory, we know that $\kappa = e_{01}+ O(\lambda)$.
We can normalized $u$ and $v$ such that
$ (u, L_+ u ) = \kappa $ and $v = \kappa^{-1} L_+ u$. Then $(u, v) = 1$ and the space $
\Eigen_\kappa ({\Li})$ is just  $ \text{span}_\R \bket{ u,  iv}$.

Define the inner product
\[
    (( \psi, \phi )) = (\Re \psi, \Re \phi)  +  (\Im \psi , \Im \phi ) \ .
\]
The space of continuum spectrum of $\L$ can be characterized by
\[
   \Hc(\L) := \bket{ \psi: \psi \perp \Eigen_\nu(\L^*) \text{ for all } \nu } \
,
\]
where $\L^*$ is the adjoint w.r.t. the inner product just defined.
It is also clear from the definition that
\[
     X  =  \Eigen_\kappa ({\L}) \oplus \Hc({\L}) \ .
\]
Notice that this is not an orthogonal decomposition.
Explicitly,  the eigenspace $\Eigen_\kappa(\L^*)$ is simply $\text{span}_\R \bket
{ v ,  iu }$. Hence the continuum space is characterized by
\[
    \Hc({\L}) = \bket{w_1 + i w_2:
    w_1 \perp v, w_2 \perp u } \ .
\]
For any initial data $\psi_0$ near $Q$,  we can  solve
$\gamma_0$ and $\ell_0$ uniquely such that
\[
\psi_0 = \bkt{Q + \gamma_0 Q + \ell_0} e^{i\theta_0} , \quad
\ell_0 \perp Q ~.
\]
It is more convenient to express $\psi_0$  in term of $Q$ and
$R=R_E$ defined by
\[
R_E = \partial_E Q_E %
\]
We shall see in next section that $R_E$ is of order
$\lambda^{-1}$. %
We now rewrite the initial data as
\[
\psi_0 = \bkt{Q + a_0 R + h_0} e^{i\theta_0} , \quad h_0 \perp Q
~.
\]
(cf. Lemma 9.1) From the decomposition of $ X $, we can write
\[
   h_0= \al_0  u + \beta_0 iv + \eta_0, \quad \eta_0 \in \Hc(\L) ~.
\]

Denote the $L^2$ norm of $\psi_0 $ by $n$. Since the evolution is
nonlinear, we can always re-scale $\la$ so that $n$, the $L^2$ norm of $\psi_0$, takes
any fixed value.  It is more convenient
to allow  $n$ to be a parameter between, say, $1$ and $10$.
Define ${\cal I}_\lambda$ to be the interval so that the $L^2$ norm
of the ground state $Q_E$ is between $1$ and $10$.
Let $ z_0=\alpha_0 + i\beta_0 $. Define the notations
\[
\langle x \rangle = \sqrt {1 + x^2}, \qquad
 \left\{ t \right \}_\e = \e^{-2} + 2 \Gamma t, \quad
\left\{ t \right \}_\e \sim \max \bket{ \e^{-2}, t }
\]
where $\Gamma$ is a constant to be specified later on.
For the moment, we remark that $\Gamma$ is of order $ 2 \lambda^2$ times the quantity
in (1.5).
When the subscript $\e$ is understood, we shall drop it.

\begin{theorem} \thlabel{mainthm}
$(1) \text{ \bf [Resonance dominated solutions]}$ \quad There exists
small parameters $\lambda_0$ and $\e_0$ such that for any
$E \in {\cal I}_\lambda$ and $|\lambda| \le \lambda_0$
the following hold. Suppose that
$$
0< |z_0|= \e  \le \e _0 ,
$$
$$
\| \eta_0\|_{H^2 \cap W^{2,1}(\R^3)}  \le  C |z_0|^{3/2}  \; .
$$
Then  we can find a small real number $a_0 =a_0 (E,h_0)$
such that the solution $\psi(t)$ to the Schr\"odinger equation
\eqref{Sch} with the initial data $ \psi_0 = Q + a_0 R + h_0$ can
be decomposed as
\begin{equation} \eqlabel{psidec}
   \psi(t) = \bkt{ Q + a(t) R + h(t) } e^{i \Theta(t)}
\end{equation}
with
\[
   a(0) = a_0, \quad  h(0)=h_0, \quad \Theta(0) =0,
\]
and $a(t), h(t) \to 0$ as $t \to \infty$ in the following sense:
Let $h(t) = \zeta(t) + \eta(t)$ with $\zeta(t)$ the component in
$\Eigen_\kappa ({\L})$
and $\eta(t)$ the component in $ \Hc({\L})$. Then we have
\begin{align*}
|a(t)| & \le C \tl{t}^{-1} \\
\zeta(t) & \ldots \text{local, smooth, } \norm{\zeta(t)}_{L^2}
\sim  \tl{t}^{-1/2} ~,
\\
\eta(t) & \ldots \text{dispersive wave, } \norm{\eta(t)}_{L^4} \le
C \tl{t}^{-3/4+ \sigma} , ~ \norm{\eta(t)}_{L^2_{\text{loc}}} \le C
t^{-1} ~,
\end{align*}
where $\tl{t} = \e^{-2} + t$. Also, $\Theta(t)/t \to -E$.

$(2)  \text{ \bf  [Radiation dominated solutions]}$ \quad
For any $\chi_\infty \in H^2 \cap W^{2,1}(\R^3)$ small, there
exist solutions  of  the form
\[
\psi(x,t) =  \left [ Q (x)+ a(t) R(x) \right ]  \, e^{i \Theta(t)} + \chi(x,t)
\]
such that
\[
|a(t)| \le C t^{-2} , \qquad
\norm{\chi(\cdot,t)}_{L^2 \loc} \le C  t^{-3/2}
\]
and
$\chi=\chi_1+\chi_2$, where
$\chi_1 = e^{i \Delta t} \chi_\infty$
and $\norm{\chi_2(\cdot,t)}_{L^2} \le o( t^{-3/2})$.
\end{theorem}

We have used the notation $f\sim g$ for $ C g \le f \le C^{-1} g$
for some constant $C$. Thus the decay of the excited state
for solutions constructed in part (1) is given precisely.
Consequently, the two types of solutions are
completely different.   Return to the general question concerning the
asymptotic mass of the ground state profile at $t=\infty$
\eqref{as}. Instead of minimizing the $L^2$ norm, we prefer to
determine the ground state profile by the condition
\begin{equation}   \label{1.6}
 \bke{\psi_t - Q_{E(t)} \, e ^{i \Theta(t)}\, , \, Q_{E(t)} } = 0 \,
\end{equation}
We shall prove that there is a unique solution to this equation
provided that the initial data is near some $Q\ini=Q_{E\ini}$
(Lemma \ref{th:renormal}). Furthermore, we can determine the
change between  $E(t)$ and $E\ini$.

\begin{theorem}\thlabel{main2}
Let $\lambda_0$ and $\e_0$  be given as in the first theorem.
For any $E\ini \in {\cal I}_\lambda$ with $\lambda\le \lambda_0 $
the following properties hold:

$(1)$ \quad Suppose the initial data
\[
\psi_0 = Q\ini + a\ini R\ini + h\ini
\]
satisfies that
\[
\|\psi_0 - Q\ini\|_{H^2 \cap W^{2,1}(\R^3)}
\sim \| h\ini\|_{H^2 \cap W^{2,1}(\R^3)} + |a\ini| \lambda^{-1}
\le \e_0^{3/2} \; .
\]
Then there are solutions $E(t)$ and $\Theta(t)$ to \eqref{1.6} such
that
\[
\lim_{t \to \infty} E(t)= E_\infty, \qquad  |E(t)-E_\infty|\le C
/ \langle t \rangle \; .
\]
Furthermore, if we decompose $\psi_t$ as in \eqref{psidec} with
$Q=Q_\infty$ being the profile at time $t= \infty$ then we have
\begin{align*}
|a(t)| & \le C \tl{t}^{-1}, \qquad  h(t)
= \zeta(t) + \eta(t) ~, \\
\zeta(t) & \ldots \text{local, smooth, } \norm{\zeta(t)}_{L^2} \le
\tl{t}^{-1/2} ~,
\\
\eta(t) & \ldots \text{dispersive wave, } \norm{\eta(t)}_{L^4} \le
C \tl{t}^{-3/4} , ~ \norm{\eta(t)}_{L^2_{\text{loc}}} \le C
t^{-1} ~.
\end{align*}
where $\tl{t} = \tl{t}_{\e_0}$.

$(2)$ Suppose that
$$
0< \e= |z\ini|\le \e _0 ,
$$
$$
\| \eta\ini\|_{H^2 \cap W^{2,1}(\R^3)}  \le  C \e^{3/2}, \qquad
\la^{-1} |a\ini|\le C \e^{2} \; .
$$
Then $\psi_0$ belongs to the class of resonance dominated
solutions with respect to the final profile $Q_{E_\infty}$.
\end{theorem}

We shall see in next section that $R_E$ is of order
$\lambda^{-1}$. This explains that $a\ini$ has to be order
$\lambda$ in order Theorem \ref{main2} to be correct. Notice that
once we have proved the assertion $(1)$ of Theorem \ref{main2},
the second assertion follows from Theorem ~\ref{mainthm}. The
proof of Theorem \ref{mainthm} is complicated due to
the construction of solutions to the Schr\"odinger equation with the boundary
condition of $h_0$ at time $t=0$ and that of $a$ at the time $t=\infty$.
If we are interested only in Theorem \ref{main2}, we can omit this
construction and the proof will be much simpler. We feel that this
construction may be needed in other contents and so we keep it.

The resonance solutions for nonlinear Klein-Gordon equations were first
proved in an important paper \cite{SW2} by Soffer and Weinstein
(see also \cite{BP}).
They consider real solutions to the nonlinear Klein-Gordon equation
\begin{equation}   \label{KG.eq}
\partial_t^2 u + B^2 u  =  \la u^3, \qquad
B^2 := (-\Delta+V + m^2),
\end{equation}
with $\la$ a small nonzero number. Assume that $B^2$ has only {\it one}
 eigenvector (the ground state) $\phi$, $B^2 \phi = \Omega^2 \phi$,
with the  resonance condition $\Omega < m < 3 \Omega$
(and some positivity assumption similar to that appears in assumption A1).
Rewrite real solutions to equation \eqref{KG.eq}
as $u =a \phi + \eta$ with
\[
   a(t) = \Re A(t) e^{i \Omega t} , \quad
    \Re A^\prime (t) e^{i \Omega t} =0 .
\]
Then $A$ and $\eta$ satisfy the equations
\begin{align}\eqlabel{A}
\dot A &= \frac 1 {2 i \Omega} \, e^{-i \Omega t} \bke{ \phi, \la
(a\phi+\eta)^3 }
\\
(\partial_t^2  + B^2) \eta &= P_c \la (a\phi+\eta)^3
\end{align}
Theorem 1.1 in \cite{SW2} states that {\it all} solutions
decay as
\[
   A(t) \sim \bkA{t}^{-1/4} , \quad
   \norm{\eta(t)}_{L^\infty} \sim  \bkA{t}^{-3/4} .
\]
In particular, the  ground state is
unstable and will decay as a resonance with rate $t^{-1/4}$.

We first remark that the proof in \cite{SW2} has only established the
upper bound $t^{-1/4}$. Furthermore,
an universal lower bound of the form $t^{-1/4}$ is in fact
incorrect. From the previous work of \cite{B, FTY},
it is clear that radiation dominated solutions decaying much faster than $t^{-1/4}$ exist.
Similar to Theorem \ref{mainthm}, we have two cases:
\begin{align*}
& 1. \quad \eta(0) \ll A(0) : \;
\text {the dominant term on  the right side of \eqref{A} is } \la a^3 \phi^3.  \\
& 2. \quad \eta(0) \gg A(0): \;
\text {the dominant term    is } \la \eta ^3 \; .
\end{align*}
In case 2,  another type of solutions arises, namely, those with
decay rate
\[
   A(t) \sim \bkA{t}^{-2} , \quad
   \norm{\eta}_{L^\infty} \sim  \bkA{t}^{-3/2} .
\]
We shall sketch a construction of such solutions at the end of section 10.

Notice that all solutions in \cite{SW2} decay as a function of $t$. Therefore, we can view
\cite{SW2} as a study of asymptotic dynamics  around vacuum. Although most works concerning
asymptotic dynamics of nonlinear evolution equations have been concentrated
on cases  with vacuum as the unique profile at $t= \infty$, more interesting and relevant
cases are
asymptotic dynamics around solitons (such as the Hartree equations \cite{FTY},
see also \cite{BP}).
The soliton dynamics  have extra complications
involving translational invariance. The current setting
of nonlinear Schr\"odinger equations
with local potentials eliminates the translational invariance
and constitutes an useful intermediate step. This simplifies greatly the analysis
but  preserves
a key difficulty %
which we now explain.

Recall that we need to approximate the wave function $\psi_t$ by
nonlinear ground states for all $t$. Since we aim to show that the
error between them decay like $t^{-1/2}$, we have to track the
nonlinear ground states with accuracy at least like $t^{-1/2}$.
While the nonlinear ground states approximating the wave function
$\psi_t$ can in principle be defined, say, via equation \eqref{as}
or (1.6), neither characterizations are useful unless we know the
wave functions $\psi_t$ precisely. Furthermore, even assuming we
can track the approximate  ground states reasonably well, the
linearized evolution around these approximate  ground states will
be based on {\it time-dependent, non-self adjoint} operators
$\L_t$. At this point we would like to mention the approach of
\cite{SW1} based on perturbation around the unitary evolution
$e^{i t H_0}$, where $H_0$ is the original self-adjoint
Hamiltonian. While we do not know whether this approach can be
extended to the current setting by combining with ideas of
\cite{SW2} (It was announced in \cite{SW2} that its method can be
extended to (1.1) as well.), such an
approach can be difficult to extend to the Hartree or other
equations with non-vanishing solitons. The main reason is that
these dynamics are not perturbation of linear dynamics.
We believe that perturbation around the profile %
at $t= \infty$ is a more nature setup.
In this approach, at least we do not have to worry about the time
dependence of approximate  ground states in the beginning. But the
linearized operator $\L=\L_\infty$ is still non-self adjoint and
it does not commute with the multiplication by $i$. So
calculations and estimates based on $\L$ are  rather complicated.
Our first idea is to map this operator to a self-adjoint operator
by a bounded transformation in Sobolev spaces. This map simplifies
many calculations and in a sense brings the problem back to the
self-adjoint case at least as far as estimates are concerned. The
price to pay is that the new operator involves square root of
operators like $H$. From the standard formula for the square root
of an operator, we can represent it as an integral of resolvents
of $H$. Thus the linear analysis of the non-self-adjoint $\L$ is
reduced to the analysis of resolvents of $H$ where standard
methods applied,  see section 3.

The next step is to identify and calculate the leading oscillatory
terms of the nonlinear systems involving the bound states
components and the continuum spectrum components.
%Since we have map
%$\L$ to a self-adjoint one, the integration by parts is now standard.
The leading order terms however depend on the relative sizes of
these components and thus we have two different asymptotic
behaviors: the resonance dominated solutions and the radiation
dominated solutions.  Finally we represent the continuum spectrum
component in terms of the bound states components and this leads to
a system of
ordinary differential-integral equations for the
bound states components. This system
can be put into a normal form and the size of the
excited state component can be seen to decay as $t^{-1/2}$. Notice
that the phase and the size of the excited state
component decay differently.  It is thus important to isolate
the contribution of the phase in the system. Finally we estimate the
error terms using estimates of the linearized operators.

The estimates obtained from integrating the equation from
$t= \infty$ can be viewed as uniform
bounds for all $T$,  provided that  the approximate
nonlinear ground states for all $T$ are known. On the other hand, to pin down
the approximate
nonlinear ground states, we need to have precise estimates on the wave functions.
It is thus nature to consider continuity method by assuming that
the approximate
nonlinear ground states  and various estimates on the wave function are known
up to time $T$. %
We then show that these
estimates continue to hold up to time $T + \delta T$, with $\delta T$ small but fixed,
provided that  all estimates are re-adjusted
w.r.t. to the new nonlinear ground state at $T + \delta T$.

From this outline,  it seems that  the resonance
dominated solutions and the radiation dominated solutions occur on
equal footing. On the other hand,
we believe that the radiation dominated solutions
are in fact of lower dimension in the space of solutions.
A proof is still lacking.
Although we restrict to  the small coupling constant
case, it can be replaced by spectral
assumptions on the operators $\L$ with some modifications.
The details will appear in a future publication.
This paper is divided into $10$
sections:

\begin{verse}
Section 2. {The set-up of the problem}

Section 3. {Properties of the linearized operator}

Section 4. {Main oscillation terms}

Section 5. {Estimates on dispersive wave}

Section 6. {Excited state equation and normal form}

Section 7. {Change of the mass of the ground state}

Section 8. {Contraction mapping}

Section 9.  Dynamical renormalization of mass

Section 10.  {Radiation dominated solution}

\end{verse}

It is a great pleasure to thanks M. Weinstein for explaining to us the beautiful idea
in the  work \cite{SW2} and, in particular, to call our attention
to the toy model in \cite{SW2} which contains the basic idea of the resonance decaying
in a very illuminating way.

\section{The set-up of the problem}

\subsection{Ground state family}

We first review the construction of  the ground state family
$\{Q_E \}_E$ mentioned in Section 1. The results we reviewed in
this subsection follow from simple perturbation theory and we
shall not give details.  Denote the standard $L_2$ inner product
by
\[
  (f,g) \equiv \int \bar f g \, d^3x ~.
\]
Let $ Q=Q_{E}$  satisfy
\[
(-\Delta + V-e_0) Q + \la Q^3 = E' Q ~.
\]
where $E'= E- e_0$.
Let $Q_E = w \phi_0 + h$ with  $h$ real and orthogonal to $\phi_0$.
Then $h$ satisfies
\[
   (H_0 - E') h + \la (w \phi_0 + h)^3 = E' w \phi_0,
\]
We can solve $w$ and $h$ so that
\begin{equation}
w^2 = O(\lambda^{-1} E'), \qquad \|h \|_2 = O(\lambda w^3)
\end{equation}
where we have used that the spectral gap of the operator
$-\Delta + V -e_0$ is of order one.

Let $R=\partial_E Q_E$.
Recall
\[
   L_+\original = -\Delta + V -E + 3\la Q^2 .
\]
Then by differentiating the
equation of $Q_E$ w.r.t $E$, we have
\[
L_+\original R_E = Q_E
\]
Denote by  $|0\rangle $  the ground state to  $L_+\original$
of norm $1$.
We also have
\[
Q_{E}= w |0\rangle  + O(\lambda  w)
\]
Hence
\begin{equation}\eqlabel{2.2}
R =  (L_+\original)^{-1} \left [ w |0\rangle + O(\lambda  w) \right ]
  = O(\la^{-1}w^{-1} ) \, |0\rangle + O(\la w)
  = O(\la^{-1} w^{-2} ) \, Q + O(w)~.
\end{equation}

\subsection{The set-up of the problem}

We consider solutions $\psi(t)$ to the equation \eqref{Sch}
The picture is that the solution $\psi(t)$ can be decomposed to
two parts: one represents a soliton, the other one the radiation.
The radiation part will disperse to infinity. The soliton part
will converge to a soliton $Q_{N_\infty}$, while its $L^2$-norm is
changing in time. Hence it is natural to consider solutions of the
form
\begin{equation} \eqlabel{Eq.3.2}
   \psi = (Q_{E(t)} + h(t))e^{i \Theta(t)} ~,
\end{equation}
and study its evolution. If we consider a minimization problem
\[
   \inf_{ \Theta, E} \norm{ \psi -Q_E \, e^{i \Theta} }_{L^2} ~.
\]
Then $\psi = (Q_E + h) \, e^{i \Theta}$ with $\Im h \perp Q_E$,
$\Re h \perp R_E$. Hence we almost have $h(t) \perp Q_{E(t)}$ in
our problem, with some small correction.

Since this setup would introduce a time-dependent linear operator,
we try to find a good approximation with a fixed $Q=Q_{E}$. For
this fixed $Q$, let
\[
   H = -\Delta + V - E + \la Q^2 ~,
\]
(we have $HQ=0$), and let $\Pi$ be the projection which eliminates
the $Q$-direction:
\[
\Pi h= h - ( c_0Q, h) Q ~, \quad c_0 = (Q,Q)^{-1}
\]
Let $X$ denote its image,
\[
   X=\Pi (L^2) ~.
\]
If we assume
\[
   \psi = (Q  + k(t))e^{i [-Et + \theta(t)]} ~.
\]
Then the equation for $k(t)$ is
\begin{equation} \eqlabel{Eq.3.3}
    \partial _t k = \L\original k -i F(k) - i \dot \theta (Q+k) ~,
\end{equation}
where
\[
   \L\original k = -i \bket{ H k + \la Q^2(k + \bar k) } ~,
\]
and
\[
F(k) = \la Q(2 |k|^2 + k^2) + \la |k|^2 k ~.
\]

In view of \eqref{Eq.3.2}, it is natural to assume
\[
   k = a R + h, \quad R = \partial _E   Q_E ~.
\]
Here $a(t)$ is small and compensates the change of $L^2$ norm of
the soliton. Since we would like $h(t) \perp Q$ for all $t$. Hence
we look for solutions of the form
\[
   \psi = (Q + a R + h) \, e^{i \Theta} ,\quad
(\Theta = -Et + \theta(t)) ~.
\]
We note that $\psi$ can be written in this form if $\psi$ is
sufficiently close to $Q$. Specifically, we let $e^{i \Theta}$ be
the phase of $P_Q \psi$ and choose $a$ so that $\norm{ P_Q
\psi}_{L^2} = \norm{ Q + a P_Q R}_{L^2}$. Then $h$ is obtained by
$h=\psi \, e^{-i\Theta} - Q - aR \in X$.

We also note, by differentiating the equation of $Q$ w.r.t.~$E$,
\[
   \L \original R  = - i Q ~.
\]
We substitute $k=aR+h$ into \eqref{Eq.3.3} and obtain
\[
     \partial _t h = -\dot a R -a i Q + \L\original h -i F(k) - i \dot \theta (Q+aR+h) ~.
\]
Since we would like $h \perp Q$, we choose $\dot a$ and $\dot
\theta$ so that the right side of the above equation is
perpendicular to $Q$. Using  $h \perp Q$ and  $Hh \perp Q$ we
obtain the equations for $a$ and $\theta$:
\begin{align*}
   \dot a &=(c_1 Q, \Im  F(k) ) ~, \qquad c_1 = (Q, R)^{-1} ~,
\\
   \dot \theta &=  - (1 + c_0c_1^{-1}a) ^{-1} \,
\bkt{ a + (c_0Q, \la Q^2(h + \bar h))  +(c_0Q, \Re F(k) )} ~,
\end{align*}
where $c_0=(Q, Q)^{-1}$. Then the equation of $h$ is
\begin{align*}
\partial _t h &= \L h + \Pi F_{all} ~,
\\
\L h &= -i \bket{ H h + \Pi \la Q^2  \Pi (h+ \bar h) } ~,
\\
F_{all} &=- i \dot \theta h
  -i F(k)  - \bkt{ (c_1 Q, \Im F) + i a \dot \theta } R_\Pi ~,
\quad R_\Pi := \Pi R ~.
\end{align*}
Here we have used the equation for $\dot a$.

Next we compute $F(k)=F(h+aR)$:
\[
 F(k) = \la Q(2|h|^2 + h^2) + 2 \la aQR (2h+ \bar h) + 3\la a^2 QR^2 +
\la (aR+ h)^2(aR + \bar h) ~.
\]
With respect to $\L$, we can decompose $X$ as
\[
   X = \Eigen_\ka \oplus \PcL(X) ~,
\]
where the direct sum is not orthogonal, and
\[
   \Eigen_\ka = \bket{ \al u + \beta iv: \al, \beta \in \R }, \quad
   u, v \text{ real}, L_+u=\ka v, L_-v=\ka u, (u,v)=1.
\]
We write
\[
   h(t)  = \al (t) u + \beta (t) iv + \eta (t) ~.
\]
Their equations are
\begin{align*}
\dot \al &= \ka \beta + (v, \Re F_{all}) ~,
\\
\dot \beta &= - \ka \al + (u,  \Im F_{all}) ~,
\\
\partial _t \eta &= \L \eta + \Pc^\L \Pi (F_{all}) ~.
\end{align*}
(Note $(v,\Pi f) = (v, f)$, and similarly for $u$.) The linear
parts of the equations for $\al(t)$ and $\beta(t)$ form a
rotation. To single out the rotation, we study
\[
z(t)=\alpha(t) + i\beta (t) = p \mu^{-1}, \quad \mu(t)=e^{i\ka t}.
\]
Here $\mu^{-1}$ captures the rotation part and we expect a slower
oscillation in $p$. Note that $z(t)$ can be obtained from $h(t)$
by
\[
   z(t) = (v , \Re h(t)) + i \, (u, \Im h(t)) ~.
\]

The function $p(t)$ satisfies
\begin{align*}
\mu^{-1} \dot p &= (v,\Re F_{all}) + i(u,\Im  F_{all})
\\
&= (v,\Im (F + \dot \theta h) - (c_1 Q, \Im F) R_\Pi)
        + i(u, -\Re (F + \dot \theta h)  -  a \dot \theta R_\Pi)
\\
&= (v,\Im F) - (v,R_\Pi)\, (c_1 Q, \Im F)+ i(u, -\Re  F) + [(v,
\Im h) + i (u,-\Re h)] \dot \theta
 - i c_3 a \dot \theta
\\
&= (\tilde v,\Im F ) + i(u, -\Re  F)
  + [(v, \Im h) + i (u,-\Re h) -ic_3 a] \dot \theta
\\
&= (\tilde v,  (F-\bar F)/2i) + i(u,  -(F+\bar F)/2)
 + [(v, (h-\bar h)/2i) + i (u, -(h+\bar h)/2) -ic_3 a] \dot \theta
\\
&= -i  \bkt{ (\tilde \up, F) + (\tilde \um, \bar F) + \bket{(\up,
h) + (\um, \bar h) +c_3 a} \dot \theta \, } ~,
\end{align*}
where $c_3=(u,R_\Pi)$, $\tilde v = v - (v, R_\Pi) c_1 Q \not \in
X$, and
\begin{align*}
 \up &= \tfrac 12 (u+  v)=O(1), \qquad
 \um = \tfrac 12 (u- v)=O(\la),
\\
\tilde \up &= \tfrac 12 (u+ \tilde v)=O(1), \qquad \tilde \um =
\tfrac 12 (u-\tilde v)=O(\la).
\end{align*}
Their orders are  derive from the simple facts: $(v,R_\Pi)=O(1)$,
$c_2=O(\la)$ and $c_3=(u,R_\Pi)=O(1)$.

Summarizing, we have
\begin{align}
\psi & = (Q + a R + h) \, e^{i \Theta}, \nonumber
\\
 h &= \zeta + \eta =(z \up +
\bar z \um) + \eta   ~,
\nonumber \\
  \dot a &=(c_1 Q, \Im  F(k) ) ~, \qquad c_1 = (Q, R)^{-1} ~,
\nonumber\\
   \dot \theta &=  - (1 + c_0c_1^{-1}a) ^{-1} \,
\bkt{ a + (c_0Q, \la Q^2(h + \bar h))  +(c_0Q, \Re F(k) )} ~,
\eqlabel{eqsum} \\
\partial _t \eta &= \L \eta + \Pc^\L \Pi (F_{all})
\nonumber \\
z(t)& =\alpha(t) + i\beta (t) = p \mu^{-1}, \quad \mu(t)=e^{i\ka
t},
\nonumber \\
\mu^{-1} \dot p &= -i  \bkt{ (\tilde \up, F) + (\tilde \um, \bar
F) + \bket{(\up, h) + (\um, \bar h) +c_3 a} \dot \theta \, } ~,
\nonumber
\end{align} where
\begin{align}
F_{all} &=- i \dot \theta h
  -i F(h+a R)  - \bkt{ (c_1 Q, \Im F) + i a \dot \theta } R_\Pi ~,
\quad R_\Pi := \Pi R ~, \eqlabel{Fall}
\\
F(h+a R) & = \la Q(2|h|^2 + h^2) + 2 \la aQR (2h+ \bar h) + 3\la
a^2 QR^2 + \la (aR+ h)^2(aR + \bar h) ~. \eqlabel{F}
\end{align}
This is the equation we shall solve for the rest of this paper.

Convention: for a double-index $\al=(\al_0, \al_1)$, we denote
\[
   z^\al = z^{\al_0}\, \bar z^{\al_1} , \quad
   |\al| = \al_0+\al_1 , \quad
   [\al] = -\al_0+\al_1 ~.
\]
For example, $z^{(32)}=z^3 \bar z^2$.
Hence $z^\al= \mu^{[\al]} \, p^\al$. In what follows $|\al|=2$, $|\beta|=3$,
$|\gamma|=4$.
\section{Properties of the linearized operator}

\subsection{Spectral decomposition}

Recall that, for given $E$,
\[
        H_E = -\Delta + V - E + \la Q_E^2
\]
and $H_EQ_E=0$. Recall  $\Pi$ be the projection which eliminates
the $Q$-direction:
\[
\Pi h= h - (h, Q) Q ~.
\]
and the  operator $\L$ acting on the space $X=\Pi L^2$ by
\[
        \L h= -i \bket{ H h + \Pi \la Q^2 \Pi (h + \bar h) } ~.
\]
We also have $L_- = H$, $L_+ = H + 2 \Pi \la Q^2 \Pi$. With
respect to $\L$, we can define generalized ``eigenspaces'':
\[
   \Eigen_\nu (\L) := \bket{ \psi: \L ^2 \psi = - \nu ^2 \psi }
        = \bket{ u + iv: u, v \text{ real, } L_-L_+ u = \nu^2 u,
      L_+L_- v = \nu^2 v} \ .
\]
In our case, the only value for $\nu$ is $\ka$.  It is natural to
define
\[
   \Hc(\L) := \bket{ \psi: \psi \perp \Eigen_\nu(\L^*) \text{ for all } \nu } \
,
\]
since it is invariant under $\L$.
The orthogonality is defined with respect to the inner product
\[
    (( \psi, \phi )) = (\Re \psi, \Re \phi)  +  (\Im \psi , \Im \phi ) \ ,
\]
and we know
\[
    \Eigen_\nu (\L^*) = \bket{ v + iu: u, v \text{ real, }  u+iv \in
\Eigen_\nu(\L) } \ .
\]
It is also clear from the definition that
\[
    X = \mathop{\oplus} _\nu \Eigen_\nu(\L) \oplus \Hc(\L) \ .
\]
Hence,
\[
    \Hc(\L) = \bket{\phi_1 + i \phi_2:
    \phi_1 \perp v, \phi_2 \perp u,
 \text{ for all $v$, $u$ such that } L_+ L_- v = \nu^2 v,  L_- L_+ u =
\nu^2 u } \ .
\]
In our situation, since  we only have two simple eigenvalues for
$H_0$ and the ground state is factored out,  we can prove by
perturbation argument that there are  real $u$ and $v$ such that
\[
   \Eigen_\kappa = \bket{ au + biv: a, b \in \R}, \quad
   L_+ u = \kappa v, \quad
   L_- v = \kappa u, \quad
   (u,v)=1 \ .
\]

Since $L_+ L_-$ and $L_-L_+$ are not self-adjoint, it is not
convenient to use them to characterize $\Hc(\L)$. We define two
operators
\[
        B = \Pi(L_-)^{1/2}\Pi, \quad \sA = \sqrt{BL_+ B \, } .
\]
Note that both $\sA$ and $B$ are self-adjoint and
\textbf{positive} on $X$. With these operators, if we define  $w
=\kappa^{-1/2} B v$, we have $w$ is the normalized eigenvector for
the operator $A$ and
\[
    u = \nu^{-1/2} B w, \quad v = \nu^{1/2} B^{-1} w , \quad
    \sA w= \kappa w , \quad (w,w)=1.
\]
Hence the continuum spectrum of $\Hc(\L)$ can be characterized by
\begin{align*}
\Hc(\L)
&= \bket{ \psi = \psi_1 + i \psi_2: \psi_1, \psi_2 \text{ real },
        \psi_1 \perp B^{-1} w_\nu , \psi_2 \perp B w_\nu \text{ for all }
\nu }
\\
&= \bket{ \psi = \psi_1 + i \psi_2: \psi_1, \psi_2 \text{ real },
        B^{-1}\psi_1, B \psi_2 \in \Hc(\sA) }
\\
&= \bket{ \psi = B \phi_1 + i B^{-1} \phi_2: \phi_1, \phi_2 \text{ real },
        \phi_1, \phi_2 \in \Hc(\sA) } \ .
\end{align*}
The maps $B$ and $B^{-1}$ change the differentiability of the
functions. A better way is to use $\sA^{-1/2}B$ and
$\sA^{1/2}B^{-1}$ instead.
Let $U$ be the operator
\begin{equation}        \eqlabel{U.def}
   U(f+ig) = \sA^{1/2} B^{-1}f + i \sA^{-1/2} Bg, \quad
   \text{for real } f, g \in X.
\end{equation}
We will prove in next subsection that
these operators are bounded in Sobolev spaces
and weighted $L^2$-spaces.
We have established a one-to-one
correspondence between the spectral decomposition of $X$ with
respect to $\L$ and that with respect to $\sA$:
\begin{align*}
\psi \in \Eigen_\kappa (\L) & \longleftrightarrow U \psi \in
\Eigen_\kappa (\sA),
\\
\psi \in \Hc(\L) & \longleftrightarrow U \psi \in \Hc(\sA).
\end{align*}
%
%Moreover, the operators $U$ and $U^{-1}$ are bounded in weighted
%Sobolev spaces $H^{s,p} \cap X$, see Lemma \ref{th:3-3}.

We have several ways to decompose $\L$ as products:
\begin{alignat*}{2}
\L &=
V_1^{-1} \begin{bmatrix} 0  & 1 \\ -\sA ^2 & 0 \end{bmatrix} V_1
\qquad \qquad
&V_1 &= \begin{bmatrix} B^{-1} & 0 \\  0 & B \end{bmatrix}
\\
& = V_2^{-1} \begin{bmatrix} j\sA  & 0 \\ 0 & - j\sA \end{bmatrix} V_2
&j&=\sqrt{-1}, \quad
V_2 = \frac 1{\sqrt{2}}
\begin{bmatrix}
 1 & \frac 1 { j\sA} \\ -j\sA  & 1 \end{bmatrix} V_1
\\
& = U^{-1} \begin{bmatrix} 0 & \sA  \\ -\sA & 0\end{bmatrix} U
&U &= \begin{bmatrix} \sA^{1/2}B^{-1} & 0 \\  0 & \sA^{-1/2}B \end{bmatrix}
\end{alignat*}
We prefer using the last one since it has a simpler form:
\[
   \L = U^{-1} (-i \sA)\, U ~,
\]
with $U$ given in \eqref{U.def}.

This  decomposition will be especially useful when we integrate
integrals of the form
\[
   \int_0^t e^{(t-s) \L} \PcL e^{2 i \ka s} \phi \, ds ~.
\]
For this purpose, it is convenient to decompose
\begin{equation}  \eqlabel{U.dec}
 U = U_+ + U_- \conj , \qquad
 U_\pm = \sA^{1/2} B^{-1} \pm \sA^{-1/2} B ~.
\end{equation}
Here $\conj$ is the conjugation operator, and both $U_+$ and $U_-$
are self adjoint. Hence $[U,i] \not = 0$ but $[U_\pm ,i]=0$.
More detailed properties of $A$ and $U$ are collected in next
subsection.

\subsection{Lemmas}

Here we collect four lemmas. Recall
\[
H=-\Delta -E + V+\la Q^2, \qquad A=(H^2 + \Pi H^{1/2} \la Q^2
H^{1/2} \Pi )^{1/2}.
\]
Also, $X$ is the subspace of all functions in $L^2(\R^3)$ that are
orthogonal to $Q$, and $\Pi$ is the orthogonal projection from
$L^2(\R^3)$ onto $X$.

\begin{lemma}[decay estimates for $e^{iAt}$] \label{3.1}
For $q=4, 8$,
\begin{equation}  \label{eq:32-1}
\norm{  e^{-it\sA} \, \PcA \Pi \phi }_{L^q} \le C \,|t|^ {-3
\bke{\frac 12 - \frac 1q}} \norm{\phi}_{L^{q'}} ~.
\end{equation}
For smooth local functions $\phi$ and sufficiently large $\beta$, we have
\begin{equation} \label{eq:32-2}
\norm{ \bkA{x}^{-\beta} \, e^{-it\sA} \, \frac 1{\sA - 0i -2\ka}
\PcA \Pi \bkA{x}^{-\beta} \phi }_{L^2} \le C \bkA{t}^{-6/5}
\end{equation}
\end{lemma}
The estimate \eqref{eq:32-2} for $A=\sqrt H$ was proved by
Soffer-Weinstein \cite{SW2}.

\begin{lemma} \label{th:3-2}
\begin{equation}
\bke{ Q \up^2 \, , \, \Im \frac {1}{\sA -0i - 2\ka} \PcA \Pi Q
\up^2   } =
\bke{ \phi_0\phi_1^2 \, , \, \Im \frac {1}{H_0 -0i - 2\ka} \PcH
\phi_0\phi_1^2 } + O(\la) > 0 ~.
\end{equation}
\end{lemma}
This Lemma is a perturbation result. Notice that if $\lambda=0$
then the statement of this lemma follows the assumption A.1. Since
$\lambda$ is small, by continuity it holds for small $\lambda$. We
define
\begin{equation}
\Gamma \equiv 2 \la^2 \bke{ Q \up^2 \, , \, \Im \frac {1}{\sA -0i
- 2\ka} \PcA \Pi Q \up^2  } > 0 ~.
\end{equation}

\begin{lemma} [operator $U$]   \label{th:3-3}
$(a)$
The operators  $U$ and $U^{-1}$ are bounded operators in
$W^{k,p}\cap X$ for $k\le 2$, $1\le p < \infty$,
and in $H^{0,r} \cap X$ for $r\le 3$.
($H^{0,r}$ is the weighted $L^2$ space with
$\norm{f}_{H^{0,r}} = \norm{\bkA{x}^r \!f}_{L^2}$.)

\noindent $(b)$ The commutator $[U,i]$ is a local operator in the sense
\begin{equation}
   \norm{ [U,i] \, \Pi \phi }_{L^{8/7}} \le O(\la) \norm{\phi}_{L^4} ~.
\end{equation}
\end{lemma}

We denote the  wave operators for $\L$ (resp.~$A$ and $H$) by
$W_\L$, (resp.~$W_A$ and $W_H$).

\begin{lemma}[wave operators]   \label{th:3-4}
The wave operators $W_\L$  and $W_A$ exist and satisfy $W^{k,p}$
estimates for $k\le 2$, $1\le p < \infty$:
(Similar estimates hold for their adjoints.)
\[
\norm{W_\L \PcL}_{(W^{k,p}, W^{k,p})} \le C , \qquad
\norm{W_A \PcA \Pi}_{(W^{k,p}, W^{k,p})} \le C .
\]
\end{lemma}
The statement on $W_\L$ was proved in \cite{C}, following
the proof of \cite{Y}. Hence we only need to prove the statement
on $W_A$.

\bigskip

\noindent {\bf Proof of these lemmas}

\bigskip

We now proceed to prove these lemmas. To simplify the
presentation, we will assume $\la >0$. The proof for the case $\la
<0$ is exactly the same.  Recall
\[
H=-\Delta -E + V+\la Q^2, \qquad A=(H^2 + \Pi H^{1/2} \la Q^2
H^{1/2} \Pi )^{1/2}.
\]
%In this subsection, and this subsection only, we denote $H_* =
%-\Delta -E$. ($H_*$ has a different meaning in the previous
%sections.)
We also denote
\[
H_* =-\Delta -E ~.
\]
 Note that, if $W_A$ exists, we have the intertwining property that
$f(A)P_c(A)=W_A f(H_*) W_A^*$ for suitable functions $f$.
We also have similar property for $\L$.

Recall $X$ is the space of all functions in $L^2(\R^3)$ that are
orthogonal to $Q$, and $\Pi$ is the orthogonal projection from
$L^2(\R^3)$ onto $X$. In what follows we will only consider the
restrictions of $H$ and $A$ on $X$. Hence we often omit the
projection $\Pi$ in the definition of $A$. (It should be noted,
however, $H_*$ acts on $L^2(\R^3)$.)
We denote by $H^{0,r}$ the weighted $L^2$ space with norm
$\norm{f}_{H^{0,r}} = \norm{\bkA{x}^r \!f}_{L^2}$.
In the remaining
of the subsection, when we write $L^2$, $W^{k,p}$, or $H^{0,r}$,
we often mean their intersection with $X$: $L^2 \cap X$,
$W^{k,p}\cap X$, or $H^{0,r}\cap X$.

\bigskip

{\bf Assumption on $V$}: We recall our Assumption A2.
We assume that $0$ is not an eigenvalue nor a resonance
for $-\Delta+V$. We also assume that $V$ satisfies the
assumption in Yajima \cite{Y} so that the $W^{k,p}$ estimates for $k\le 2$
for the wave operator $W_H$ holds:
for a small $\sigma>0$,
\[
|\nabla^\al V(x)| \le C \bkA{x}^{-5-\sigma}, \qquad
\text{for } |\al|\le 2 ~.
\]
Also, the functions
$(x\cdot \nabla)^k V$, for $k=0,1,2,3$, are $-\Delta$ bounded with
an $-\Delta$-bound $<1$:
\begin{equation} \label{eq:33-01}
  \norm{(x\cdot \nabla)^k V\phi}_2 \le \sigma_0 \norm{-\Delta\phi}_2 +
C\norm{\phi}_2,
  \qquad \sigma_0 < 1 , \quad k=0,1,2,3  ~.
\end{equation}
By the assumption,
the following operators
\begin{equation} \label{eq:33-02}
H_*^{-1/2}(x\cdot \nabla)^k V H_*^{-1/2}, \quad (x\cdot \nabla)^k
V H_*^{-1}, \quad H_*^{-1}(x\cdot \nabla)^k V
\end{equation}
 for $k=0,1,2,3$, are bounded in $L^2$.

Since $Q$ is the ground state of $H$ with $V$ satisfying the previous
assumptions,
$Q$ is a smooth function with exponential decay at infinity. Hence the
above statements on $V$ hold also for $Q$ and $Q^2$. Since $V+\la
Q^2$ and $V$ have same properties, in what follows we will replace
$V+\la Q^2$ in $H$ by $V$ and write $H=H_* + {V}$ to make the
presentation simpler. So it should be kept in mind that the
potential $V$ in this subsection is in fact $V+\la Q^2$.

\bigskip

For two operators $S$ and $T$, $S$ is said to be {\it
$T$-bounded}\, if $ST^{-1}$ is a bounded operator. If both $S$ and
$T$ are self-adjoint, this implies $T^{-1}S$ is also bounded. A
deeper result says $S^{1/2}$ is $T^{1/2}$-bounded, see [RS2]. We
say $S$ and $T$ are {\it mutually bounded}\, if both $ST^{-1}$ and
$TS^{-1}$ are bounded operators. This is the case if
$\norm{(S-T)T^{-1}}_{(L^2,L^2)}=\theta < 1$ for some $\theta$. (It
implies immediately $\norm{ST^{-1}} < 2$. Since $\norm{T\phi}\le
\norm{S\phi}+\norm{(T-S)\phi}\le \norm{S\phi}+\theta\norm{T\phi}$,
we have $\norm{T\phi}\le C \norm{S\phi}$, which implies $T$ is
$S$-bounded.)

\begin{lemma} \label{th:3-5}
For each $k=\tfrac 12, 1, \tfrac 32, 2, 3$, the operators
$H_*^k$, $H^k$ and $A^k$ are mutually bounded.
\end{lemma}
\myproof
That $H_*^k$ and $H^k$ are mutually bounded follows from our assumption
on $V$ by standard argument.
To show  $H^k$ and $A^k$ are mutually bounded, it suffices to prove
the cases $k=2$ and $k=3$ by the previous remark.
We first show
$\norm{(A^2-H^2)H^{-2}}<1$, which implies the case $k=2$.
\[
\norm{(A^2-H^2)H^{-2}}= \norm{H^{1/2} \la Q^2 H^{1/2} H^{-2}} \le
\norm{H^{1/2} \la Q^2 H^{-1}} \le \norm{H_*^{1/2} \la Q^2 H_*^{-1}} \le
1/2.
\]
The last inequality can be obtained by writing $H_*^{1/2}  Q^2
H_*^{-1}=H_*^{-1/2} Q^2+H_*^{1/2}[ Q^2,H_*^{-1}]=H_*^{-1/2}
Q^2+H_*^{-1/2}[ Q^2,H_*]H_*^{-1}$, and noting $[ Q^2,H_*]=\Delta
Q^2 + 2 \nabla Q^2 \cdot \nabla$.

To prove the case $k=3$, it suffices to prove $A^6\le CH^6$ and
$H^6 \le C A^6$. Note
\[
(f A^6 f)=(f A^2 A^2 A^2 f) \le (f A^2 H^2 A^2 f) ~.
\]
Since $A^2 = H^2 + H^{1/2}\la Q^2 H^{1/2}$, we have
\[
(f A^2 H^2 A^2 f)\le C (f H^2 H^2 H^2 f) +
C (f (H^{1/2} \la Q^2 H^{1/2}) H^2 (H^{1/2} \la Q^2 H^{1/2}) f)
\]
where the cross terms are estimated by Schwarz inequality. To show
that the last term is bounded by $C (f H^6 f)$, we shall show that
$H^{3/2}Q^2 H^{-5/2}$ is bounded in $X$. Rewrite
\begin{align*}
H^{3/2}Q^2 H^{-5/2}&=H^{3/2}H^{-2}Q^2 H^{-1/2}+H^{3/2}[Q^2,H^{-2}]
H^{-1/2}\\&=H^{-1/2}Q^2 H^{-1/2}+H^{-1/2}[Q^2,H^2] H^{-5/2}
\end{align*}
Since $[Q^2,H^2]$ is of the form $\sum_{|\al|\le 3} G_\al(x) \nabla ^\al$,
the operators on the right side of the equation are bounded in $X$.
This shows  $A^6\le CH^6$. That $H^6 \le C A^6$ is
proved similarly. \myendproof

%%%%%%%%%%%%%%%%%%%%%%%%%%%%%%%%%%%%%%%%%%%%%%%%%%%%%%%%%%%%

Recall the standard formula:
\begin{equation} \label{eq:sq-root}
T^{-\sigma} = \int_0^\infty \frac 1{s +T}
 \, \frac {ds }{s ^{\sigma}} , \quad \qquad 0<\sigma <1 ~.
\end{equation}
The operator $T$ in the above formula will be $A^2$ or $H$.
Hence we also need to estimate operators of the form
$\frac {H^m}{s +H^2}$.
Clearly, for $s\ge 0$,
\begin{equation} \label{eq:33-1}
\norm{\frac {H^2}{s +H^2} } _{\bke{W^{k,p},W^{k,p}}}\le 1 , \qquad
\norm{H^{-1/2}}_{\bke{W^{k,p},W^{k,p}}} \le C ~.
\end{equation}

\begin{lemma}  \label{th:3-6}
Let $s\ge 0$.
The operator $\displaystyle \frac H{s +H^2}$ is bounded in $W^{k,p}\cap X$
with
\begin{equation} \label{eq:33-2}
\norm{\frac H{s +H^2} } _{\bke{W^{k,p},W^{k,p}}} \le C\bkA{s }^{-1/2}
\end{equation}
Also, for $k=\pm 1,\pm 2,\pm 3$,
\begin{equation} \label{eq:33-3}
\norm{\bkA{x}^k\frac 1{s +H} \, \bkA{x}^{-k} } _{\bke{L^2, L^2}
} \le C \bkA{s }^{-1} , \quad
\norm{\bkA{x}^k\frac H{s +H^2} \, \bkA{x}^{-k} } _{\bke{L^2, L^2}
} \le C \bkA{s }^{-1/2} ~.
\end{equation}
\end{lemma}

\myproof
We can rewrite
\[
\frac H{s +H^2} = \frac 1{H+\sqrt s i} +\frac 1{H-\sqrt s i} ~.
\]
Therefore, to prove statements for $\frac H{s +H^2}$, it suffices to
prove the corresponding statements for $\frac 1{H \pm \sqrt s i}$.
We first prove \eqref{eq:33-2} for $k=0$. Let $\ka_1$ denote the
eigenvalue of the excited state of $H$ and
$P_1$ denote the projection onto the corresponding eigenspace. We can write
\[
\frac 1{H+\sqrt s i}\Bigg|_X= \frac 1{\ka_1+\sqrt s i} P_1
+W_H\, \frac {1}{ p^2-E+\sqrt s i} \, W_H^{*} \Pc^H
\]
where $p=-i\nabla$ and $W_H$ is the wave operator of $H$. Note $E < 0$.
Since $W_H$ and $W^*_H$ are bounded in  $W^{k,p}$ for sufficiently nice $V$,
(\cite{Y}), it is sufficient to prove that
$\frac {1}{ p^2-E \pm \sqrt s i}$ are bounded in $W^{k,p}$. However,
$\frac {1}{ p^2-E \pm \sqrt s i}$ are
convolution operators with explicit Green functions:
\[
  \frac C{|x|} \, e^{-|x|(-E \pm \sqrt si)^{1/2} } ~.
\]
Since $| e^{-|x|(-E \pm \sqrt si)^{1/2} }| \le
e^{-c|x|\bkA{s }^{1/4}}$, the $L^1$-norms of the Green functions are
bounded by $\bkA{s }^{-1/2}$. By Young's inequality we have
\[
\norm{\frac {1}{p^2-E \pm \sqrt s i}}_{(L^p,L^p)} \le C \bkA{s }^{-1/2} ~,
\]
which proves \eqref{eq:33-2} for $k=0$.
For $k\ge 1$ and for $\phi \in W^{k,p}$, we have
\begin{align*}
\norm{ H^{k/2}\frac H{s +H^2} \phi}_{W^{k,p}}
&\sim \norm{ H^{k/2}\frac H{s +H^2} \phi}_{L^{p}}
=  \norm{ \frac H{s +H^2} H^{k/2}  \phi}_{L^{p}}
\\
&\le C\bkA{s }^{-1/2} \norm{ H^{k/2} \phi}_{L^{p}}
\sim \bkA{s }^{-1/2}\norm{ \phi}_{W^{k,p}} ~.
\end{align*}
This proves \eqref{eq:33-2} for $k \ge 1$.

For \eqref{eq:33-3}, we prove the second part.
The proof for the first part is similar.  For $k>0$, since
$\bkt{\bkA{x}^k, \frac 1{H+\sqrt si}}=\frac 1{H+\sqrt s
i} \, [\bkA{x}^k, H +\sqrt si]\, \frac 1{H+\sqrt si}$ and
\[
 [\bkA{x}^k, H +\sqrt si] = 2 \nabla^* (\nabla \bkA{x}^k) -
 (\Delta \bkA{x}^k)
\]
we have
\[
\norm{\bkt{\bkA{x}^k, \frac 1{H+\sqrt si}} \bkA{x}^{-k} } \le
C\norm{\frac 1{H+\sqrt si}(\nabla^*+1)}\cdot
\norm{\bkA{x}^{k-1}\frac 1{H+\sqrt si} \, \bkA{x}^{-k+1}} \le C
\bkA{s }^{-1/2}
\]
by induction in $k$. We have the same estimate for
$\bkt{\bkA{x}^k, \frac 1{H-\sqrt si}}$ and hence \eqref{eq:33-3}
holds for positive $k$. The proof for the case $k<0$ is similar.
 \myendproof

%%%%%%%%%%%%%%%%%%%%%%%%%%%%%%%%%%%%%%%%%%%%%%%%%%%%%%%%%%%%%%%%%%%%%%%%%

 Recall $H^{0,r}$ is the weighted $L^2$ space with norm
$\norm{f}_{H^{0,r}} = \norm{\bkA{x}^r \!f}_{L^2}$.

\begin{lemma} \label{th:3-7}
The operators $H^{1/2}A^{-1/2}$, $A^{-1/2}H^{1/2}$,
$H^{-1/2}A^{1/2}$ and $A^{1/2}H^{-1/2}$ are bounded operators in
$W^{k,p}\cap X$ and $H^{0,r} \cap X$.
\end{lemma}

\myproof
By \eqref{eq:sq-root} we can write
\begin{align*}
H^{1/2}A^{-1/2} &= H^{1/2}\int_0^\infty
\frac 1{s+ H^2 + H^{1/2} \la Q^2 H^{1/2}} \, \frac{ ds }{s ^{1/4}}
\\
&=H^{1/2}\int_0^\infty \bkt{
\frac {1}{s+ H^2 } +\frac {1}{s+ H^2
}H^{1/2}\la Q \sum_{n=0}^\infty\bke{\la Q\frac {H}{s+ H^2 } Q}^n
QH^{1/2}\frac {1}{s+ H^2 } } \frac{ ds }{s ^{1/4}}
\\
&=1+ \int_0^\infty \bkt{ \frac {H}{s+ H^2 }\la Q
\sum_{n=0}^\infty\bke{\la Q\frac {H}{s+ H^2 } Q}^n Q\frac {H}{s+
H^2 }H^{-1/2} } \frac{ ds }{s ^{1/4}}
\end{align*}
Since $\norm{\frac {H}{s+ H^2 }}\le \bkA{s }^{-1/2}$ by Lemma
\ref{th:3-6}, we have
\[
\norm{H^{1/2}A^{-1/2} }_{\bke{W^{k,p},W^{k,p}}}
 \le 1+C \int_0^\infty \bkA{s }^{-1/2}\la\sum_{n=0}^\infty
 (\la\bkA{s }^{-1/2})^n
 \bkA{s }^{-1/2}\frac{ ds }{s ^{1/4}}
\le 1+C  \la ~.
\]
Similarly
\[
\norm{A^{-1/2}H^{1/2} }_{\bke{W^{k,p},W^{k,p}}}\le 1+C  \la ~.
\]
Also, using \eqref{eq:33-3}, for $r\le 3$ we have
\[
\norm{H^{1/2}A^{-1/2} }_{\bke{H^{0,r},H^{0,r}}} +
 \norm{A^{-1/2}H^{1/2}}_{\bke{H^{0,r},H^{0,r}}}
\le 1+C \la ~.
\]
The above proves that $H^{1/2}A^{-1/2}$ and $A^{-1/2}H^{1/2}$ are bounded in
$W^{k,p}$ and $H^{0,r}$. Indeed, we have proved
\begin{equation} \label{eq:33-5}
\norm{\bkA{x}^3(H^{1/2}A^{-1/2}-1)\bkA{x}^3 }_{\bke{L^2,L^2}} +
 \norm{\bkA{x}^3(A^{-1/2}H^{1/2}-1)\bkA{x}^3}_{\bke{L^2,L^2}}
\le C \la ~.
\end{equation}

We now consider $H^{-1/2}A^{1/2}$ and $A^{1/2}H^{-1/2}$.
Since $A^{1/2}=A^2 A^{-3/2}=A^2\int_0^\infty \frac 1{s+
A^2}\frac{ ds }{s ^{3/4}}$, we have
\begin{align*}
H^{-1/2}A^{1/2} &= H^{-1/2}(H^2 + H^{1/2} \la Q^2
H^{1/2})\int_0^\infty \frac {1} {s+ H^2 + H^{1/2} \la Q^2 H^{1/2}}
\, \frac{ ds }{s ^{3/4}}
\\
&= (H^{3/2}+ \la Q^2 H^{1/2})\int_0^\infty  \frac {1}{s+ H^2 +
H^{1/2} \la Q^2 H^{1/2}} \, \frac{ ds }{s ^{3/4}}=I_1+\la Q^2 I_2
\end{align*}
The main term is $I_1$. The term $I_2$ is similar to
$H^{1/2}A^{-1/2}$, and its integrand has a better decay in $s$ for
large $s $. Hence
\[
\norm{\la Q^2I_2}\le C\la \norm{I_2} \le C\la ~.
\]
For the main term $I_1$,
\begin{align*}
I_1=1+\int_0^\infty \frac {H^2}{s+ H^2 }Q
\sum_{n=0}^\infty\bke{Q\frac {H}{s+ H^2 } Q}^n Q\frac {H}{s+
H^2 }H^{-1/2} \frac{ ds }{s ^{3/4}} ~.
\end{align*}
Hence
\[
\norm{I_1}  \le 1+C \int_0^\infty \la \sum_{n=0}^\infty
(\la \bkA{s }^{-1/2})^n  \bkA{s }^{-1/2}\frac{ ds }{s ^{3/4}}
\le 1+C \int_0^\infty \la \bkA{s }^{-1/2}\frac{ ds }{s ^{3/4}}
\le 1+C\la ~.
\]
Here the norms are taken in $(W^{k,p},W^{k,p})$ and $(H^{0,r},H^{0,r})$.
Hence we have proved Lemma \ref{th:3-7}.
\myendproof

In fact, the last part of the above proof also shows
\begin{equation} \label{eq:33-6}
\norm{\bkA{x}^3(H^{-1/2}A^{1/2}-1)\bkA{x}^3 }_{\bke{L^2,L^2}} +
 \norm{\bkA{x}^3(A^{1/2}H^{-1/2}-1)\bkA{x}^3}_{\bke{L^2,L^2}}
\le C \la ~.
\end{equation}

The above lemma proves part (a) of Lemma \ref{th:3-3} stating that
$U$ and $U^{-1}$ are bounded in $W^{k,p}$ and $H^{0,r}$.
Moreover, \eqref{eq:33-5} and \eqref{eq:33-6} mean that
$U-1$ and $U^{-1}-1$ are ``local'' operators. In particular, they
imply part (b) of Lemma \ref{th:3-3} and that, for any $\phi\in L^2$,
\begin{equation} \label{eq:33-7}
(U^{\pm 1} -1) e^{-it H_*} \phi \to 0 \qquad \text{in $L^2$, \quad as }
t \to \infty ~.
\end{equation}

%%%%%%%%%%%%%%%%%%%%%%%%%%%%%%%%%%%%%

We now prove Lemma \ref{th:3-4}. We only need to prove the statement on
$W_A$. Notice
\[
W_A=\lim_{t\to \infty} e^{itA} e^{-itH_*}
=\lim_{t\to \infty} Ue^{t\L}U^{-1} e^{-itH_*}
= \lim_{t\to \infty} Ue^{t\L} e^{-itH_*}
+ \lim_{t\to \infty} Ue^{t\L}(U^{-1}-1) e^{-itH_*} ~.
\]
By \eqref{eq:33-7} we have
\[
W_A=\lim_{t\to \infty} Ue^{t\L} e^{-itH_*} = U W_\L ~.
\]
The bounededness of $W_A$ follows from that of $U$ and $W_\L$.
This proves Lemma \ref{th:3-4}.

\bigskip

We now prove Lemma \ref{3.1}. Since
\[
e^{-itA} \PcA \phi = W_A e^{-itH_*} W_A^* \PcA \phi ~,
\]
the estimate \eqref{eq:32-1} follows from the usual $(L^p,L^q)$ estimate
for $e^{-itH_*}$ and the boundedness of $W_A$ and $\PcA$ in
$L^p$-spaces. To prove \eqref{eq:32-2}, either we prove the
bounededness of $W_A$ in weighted spaces $H^{0,r}$,
or we use the Mourre estimate. We will follow the second approach
and the argument in \cite{SW2}.

Let $a=2\ka$. We consider intervals $\Delta=(a-r,a+r)$. Let
$g_\Delta(t)=g_0((t-a)/r)$, where $g_0$ is a fixed smooth function
with support in $(-2,2)$ and $g_0(t)=1$ for $|t|<1$. We will
consider $g_\Delta(A)$ with $r$ small enough. Let $D=xp+px$, $(p=-i\nabla)$,
and the commutators
\[ %
\ad_D^0(A)=A, \qquad \ad_D^{k+1}(A)=[\ad_D^{k}(A),D] ~.
\]
We need to prove the following lemma.

\begin{lemma} \label{th:3-8}
For $\Delta$ small enough, the Mourre estimate
\[
g_\Delta(A)[i A,D] g_\Delta(A) \ge \theta g_\Delta(A)^2
\]
holds for some $\theta>0$. Also,
$g_\Delta(A) \ad_D^k (A) g_\Delta(A)$ are bounded operators in
$L^2$ for $k=0,1,2,3$.
\end{lemma}

We will use the following lemma.
\begin{lemma} \label{th:3-9}
The operators
\[
H^{-3}D^k H^{m/2} \bkA{x}^{-3} \quad \text{ and } \quad
\bkA{x}^{-3}H^{m/2}D^k H^{-3}
\]
are bounded in $L^2$, for $k,m=0,1,2,3$.
\end{lemma}
\myproof
This is standard and we only sketch the proof.
If $m$ is even, we can compute the commutator $[D^k, H^{m/2}]$ explicitly
and estimate
\[
H^{-3} D^k H^{m/2}\bkA{x}^{-3} = H^{-3}H^{m/2} D^k \bkA{x}^{-3}
+ H^{-3} [D^k, H^{m/2}]\bkA{x}^{-3} ~.
\]
If $m$ is odd, we write
\[
H^{-3}D^k H^{m/2} \bkA{x}^{-3} = \int_0^\infty
H^{-3}D^k H^{(m+1)/2} \frac 1{s +H}
\bkA{x}^{-3} \, \frac {ds }{\sqrt s}
\]
and proceed as in the case $m$ is even, by using \eqref{eq:33-3}.
Here we have used the formula \eqref{eq:sq-root}.
\myendproof

\medskip

\noindent{\bf Proof of Lemma \ref{th:3-8}}

\medskip

Let $G=A-H$ and we write $A=H+G$. Since
\[
[i A,D]=[H_*+V+G,i D]=-\Delta + [V+G,i D]=A-V-G + [V+G,i D]
\]
and $g_\Delta(A)A g_\Delta(A) \ge 2\theta g_\Delta(A)^2$ for some
$2\theta>0$, it suffices to show that, for $M=-V+[V, iD]$,$-G$
and $[G,D]$, the operators
\[
g_\Delta(A)M g_\Delta(A)= (g_\Delta(A)H_*^2)\, (H_*^{-2}MH_*^{-2}) \,
(H_*^2 g_\Delta(A))
\]
are bounded by $g_\Delta(A)^2$ and the bound goes to zero when the
interval $\Delta$ shrinks to zero. Since both
$g_\Delta(A)H_*^2=(g_\Delta(A)A^2) (A^{-2}H_*^2)$ and $H_*^2
g_\Delta(A)=(H_*^2A^{-2})(A^{2}g_\Delta(A))$ are bounded and converges
to zero weakly when $\Delta$ shrinks to zero, this
will be true if one can show that $H_*^{-2}MH_*^{-2}$ is compact.
 The case $M=-V+ [V,iD]$ is standard and follows from our assumption, so we
only consider $H_*^{-2}G H_*^{-2}$ and $H_*^{-2}[G, D] H_*^{-2}$.

We proceed to find an explicit form of $G$. By
\eqref{eq:sq-root} with $T=A^2$, $\sigma=1/2$, we write
\begin{align*}
A^{-1} &=\int_0^\infty \frac 1{s+ H^2 + H^{1/2} \la Q^2 H^{1/2}} \,
\frac{ ds }{\sqrt s}
\\
&= \int_0^\infty \frac {1}{s+ H^2 } +\frac {1}{s+ H^2
}H^{1/2} \la Q \sum_{n=0}^\infty\bke{\la Q\frac {H}{s+ H^2 } Q}^n
QH^{1/2}\frac {1}{s+ H^2 } \frac{ ds }{\sqrt s}
\\
&=H^{-1}+ {H^{-1/2}} \bkA{x}^{-3} J_0 \bkA{x}^{-3} {H^{-1/2}}
\end{align*}
where
\[
 J_0 =\int_0^\infty\bkA{x}^{3}\frac {H}{s+ H^2 }\la Q
 \sum_{n=0}^\infty\bke{\la Q\frac {H}{s+ H^2 } Q}^n
Q  \frac {H}{s + H^2 }\bkA{x}^{3} \frac{ ds }{\sqrt s} ~.
\]
By Lemma \ref{th:3-6}, $\norm{J_0}_{(L^2,L^2)}\le \int_0^\infty
\bkA{s }^{-1/2} \cdot \la \cdot \bkA{s }^{-1/2} s ^{-1/2}\, ds
\le C \la$.  Hence
\begin{align}
A&= A^2 A^{-1}=(H^2+H^{1/2}\la Q^2H^{1/2}) \, \bke{H^{-1}+ {H^{-1/2}}
\bkA{x}^{-3} J_0 \bkA{x}^{-3} {H^{-1/2}}}=H + G \nonumber
\\
G&=H^{1/2}\la Q^2H^{-1/2} + {H^{3/2}} \bkA{x}^{-3} J_0 \bkA{x}^{-3}
{H^{-1/2}} + H^{1/2}\la Q^2 \bkA{x}^{-3} J_0 \bkA{x}^{-3} {H^{-1/2}}
\label{eq:33-31}
\end{align}

Since $H^{-1/2}\bkA{x}^{-1}$ and $\bkA{x}^{-1} H^{-1/2}$
are compact, from \eqref{eq:33-31} $H_*^{-2} G H_*^{-2}$ is
compact.  We can also write
\[
 H_*^{-2} G D H_*^{-2}
 =\bket{H_*^{-2}GH^{1/2}\bkA{x}} \cdot \bket{\bkA{x}^{-1}H^{-1/2}DH_*^{-2}}
\]
The second operator is bounded by Lemma \ref{th:3-9}. The first
is compact since its terms are of the form: $H^{-m} \cdot  \bkA{x}^{-k}
\cdot $ (bounded operator).  Similarly $H_*^{-2} D G
H_*^{-2}$ is also compact. Hence we conclude the Mourre estimate.

 To show that $g_\Delta(A) \ad_D^k (A) g_\Delta(A)$ are bounded for
$k=0,1,2,3$, we rewrite
\[
g_\Delta(A) \ad_D^k (A) g_\Delta(A)=
(g_\Delta(A) A^3)\,( A^{-3}  H^3) \, (H^{-3} \ad_D^k (A)  H^{-3}) \,
(H^{3} A^{-3})\, (A^{3} g_\Delta(A)) ~.
\]
We only need to show that $H^{-3} \ad_D^k (A) H^{-3}$ are bounded
since the other terms are bounded by  Lemma \ref{th:3-5}.
Recall $A=H+G$. It is standard to prove that $H^{-3} \ad_D^k (H)  H^{-3}$
are bounded. For
$H^{-3} \ad_D^k (G)  H^{-3}$,  since it is a sum of terms of the form
\[
H^{-3} D^k G D^{m} H^{-3} ,\qquad k+m\le 3~,
\]
it suffices to show that these terms are bounded.
By the explicit form \eqref{eq:33-31} of $G$ and Lemma \ref{th:3-9},
they are indeed bounded. For example,
\begin{align*}
&H^{-3} D^2  \bket{{H^{3/2}}\bkA{x}^{-3} J_0 \bkA{x}^{-3}
{H^{-1/2}}} D^{1} H^{-3} \\
&\qquad = \bket{ H^{-3} D^2 {H^{3/2}}\bkA{x}^{-3}} J_0
 \bket{\bkA{x}^{-3} H^{-1/2} D^{1} H^{-3}},
\end{align*}
a product of three bounded operators. We conclude that
$g_\Delta(A) \ad_D^k (A) g_\Delta(A)$ are bounded for $k=0,1,2,3$.
\myendproof

With  Lemma \ref{th:3-8},  (cf.~the remark in \cite{SW2}, p.27),
 the minimal velocity estimate in \cite{HSS} and Theorem 2.4
of \cite{Sk} implies
\[
\norm{ F(D \le \theta t/2) \, e^{-iAt} g_\Delta(A) \bkA{D}^{-3/2} }
_{(L^2,L^2)} \le C \bkA{t}^{-5/4} .
\]
The same argument in \cite{SW2} then gives
the desired decay estimate \eqref{eq:32-2} in Lemma \ref{3.1}.

\bigskip

Finally we prove Lemma \ref{th:3-2}. Let $\psi_\la=\PcA \Pi Q
\up^2$ and $\psi_0= \PcH\phi_0\phi_1^2$.
Recall $H_0 = - \Delta + V -e_0$.
We have $\psi_\la=\psi_0 + O(\la)$.
We write $\psi_\la=\psi_0 + b \phi_1 + \eta$, where $\eta\in \Hc(H_0)$
and $b,\eta=O(\la)$.
Rewrite
\begin{align} \nonumber
&\bke{ \psi_\la \, , \,  \Im \frac {1}{\sA -0i - 2\ka}  \psi_\la}
= \Im i \int_0^\infty
\bke{ \psi_\la \, , \, e^{-it(\sA -0i - 2\ka)}  \psi_\la} \, dt
\\
&\qquad =\Im i \int_0^\infty
\bke{ \psi_\la \, , \, e^{-it(H_0 -0i - 2\ka)}  \psi_\la} \, dt
\label{eq:3-16}
\\
&\quad \qquad +\Im i \int_0^\infty \int_0^t
\bke{ \psi_\la \, , \, e^{-i(t-s)(\sA -0i - 2\ka)}
\,(\la Q^2+G)\, e^{-is(H_0 -0i - 2\ka)}  \psi_\la} \, ds \, dt ~.
\label{eq:3-17}
\end{align}
The main term lies in \eqref{eq:3-16}. It is
\[
\Im i \int_0^\infty
\bke{ \psi_0 \, , \, e^{-it(H_0 -0i - 2\ka)}  \psi_0} \, dt
= \bke{ \psi_0 \, , \,  \Im \frac {1}{H_0 -0i - 2\ka}  \psi_0}
\]
which is the desired main term in Lemma \ref{th:3-2}.
We want to show that the rest of \eqref{eq:3-16} and \eqref{eq:3-17}
are integrable and of order $O(\la)$.
Recall we write $\psi_\la=\psi_0 + b \phi_1 + \eta$.
For the term $\eta$ in $\psi_\la$, by decay estimate we have
\begin{equation} \label{eq:3-20}
|\bke{ \psi_\la \, , \, e^{-it(H_0 -0i - 2\ka)}  \eta}|
\le C \bkA{t}^{-3/2} \norm{\psi_\la}_{L^1\cap L^2} \,
\norm{\eta}_{L^1\cap L^2} \le C \bkA{t}^{-3/2} \la ~,
\end{equation}
hence this term is integrable.
Also, since $H_0 \phi_1 = e_{01} \phi_1$,
\[
\bke{ \psi_\la \, , \, e^{-it(H_0 -0i - 2\ka)}  b\phi_1}
= \bke{ \psi_\la \, , \, e^{-it(e_{01}-2\ka-0i)}  b\phi_1} ~,
\]
so we can integrate this oscillation term explicitly. (The
boundary term at $t=\infty$ vanishes due to the
decay of $e^{-it(-0i)}$.)  We conclude that
the rest of \eqref{eq:3-16} are integrable and of order $O(\la)$.

For  \eqref{eq:3-17}, it suffices
to show its integrability since $\la Q^2 + G$ gives the order $O(\la)$.
Rewrite the last $\psi_\la$ in \eqref{eq:3-17}
as $b\phi_1 + \PcH \psi_\la$.
For the part containing $b\phi_1$, we have
\begin{align*}
&\bke{ \psi_\la \, , \, e^{-i(t-s)(\sA -0i - 2\ka)}
\,(\la Q^2+G)\, e^{-is(H_0 -0i - 2\ka)}  b\phi_1}
\\
&\quad =  \bke{ \psi_\la \, , \, e^{-it(\sA -0i - 2\ka)} \,e^{is(A-e_{01})}
\,(\la Q^2+G)\, b\phi_1} ~.
\end{align*}
Integration in $s$ gives
\[
\bke{ (A-e_{01})^{-1} \psi_\la \, , \, e^{-it(\sA -0i - 2\ka)}
\,(\la Q^2+G)\, b\phi_1} ~.
\]
Since $e_{01}$ lies outside the continuous spectrum of $A$, the last
expression is integrable in $t$ following the same argument as
\eqref{eq:3-20}.
For the part containing $\PcH \psi_\la$, since $(\la Q^2+G)$ is
a ``local'' operator
in the sense that it sends $L^\infty$ functions to $L^1$, we have
\[
|\bke{ \psi_\la \, , \, e^{-i(t-s)(A -0i - 2\ka)}
\,(\la Q^2+G)\, e^{-is( H_0 -0i - 2\ka)}   \PcH \psi_\la}|
\le C \la \bkA{t-s}^{-3/2} \bkA{s}^{-3/2}\norm{\psi_\la}_{L^1\cap L^2}^2
\]
which can be integrated in $s$ and $t$.
Hence we have proved Lemma \ref{th:3-2}.

%%%

%%%

%
%

%
%
%
\section{Main oscillation terms}

We now identify the main oscillation terms in equation
\eqref{eqsum}. We shall use the complex amplitude of the excited
state, $z$, as the reference. Recall $z(t)= e^{-i \kappa t} p(t)$.
Eventually we will have $p(t) \sim t^{-1/2}$ and its oscillation
is much smaller than $\kappa$. The change of mass on the direction
of the nonlinear ground state is given by $a$. We will also show
that  $a=O(z^2)$ and the order of the dispersive wave $\eta(t)$ is
also of order $ O(z^2)$.  Assuming these orders, the second order
term in $F$, $F^{(2)}$, is given explicitly by:
\begin{equation*}
F^{(2)} = \la Q(2|\zeta|^2 + \zeta^2) = z^2  \phi_{(20)} + z\bar z
\phi_{(11)}
 + \bar z ^2  \phi_{(02)}
\end{equation*}
where  $\zeta = z \up + \bar z \um$ and
\begin{align}
\phi_{(20)}&=\la Q(\up^2 + 2 \up \um) =\la Q\up^2 + O(\la^2),
\nonumber \\
\phi_{(11)}&=2\la Q(\up^2 + \um^2 + \up \um)=O(\la), \eqlabel{phi}
\\
 \phi_{(02)}&=\la Q(\um^2 + 2 \up \um) =O(\la^2).
\nonumber
\end{align}

We shall write $F^{(2)}= z^\al \phi_\al$, where $\alpha$ is a
double indices $(ij)$ with $i+j=2$ and $i, j \ge 0$. The repeated
indices mean summation. We shall use $\beta$ later on to denote
double indices summing to $3$ and $\gamma$ for $4$.

\subsection{Main oscillation term in  $a$ and $F$}

We start with identifying  the main oscillation terms of $a(t)$.
We have fixed the boundary condition of $a$ at $t=\infty$ and set
$a(\infty) = 0$.  Thus  we can rewrite the equation for $a(t)$
into the following equivalent integral equation:
\[
a(t) =  \int_\infty^ t (c_1Q, \Im F(k)) \, ds ~.
\]
As the oscillation term  of order $z^2$ in $F$ comes from
$F^{(2)}$, we have up to second order in $z^2$
\[
(c_1Q, \Im F) \sim A^{(2)}
\]
where
\begin{equation*}
A^{(2)} = \bke{c_1 Q, \, \la Q \Im \zeta^2 }
\end{equation*}
Since $\zeta = z \up + \bar z \um$, we have
\begin{equation*}
\Im \zeta = \Im z \, ( \up - \um), \quad \Im \zeta^2= (\Im z^2)\,
(\up^2 - \um^2) ~.
\end{equation*}
Therefore, we have
\begin{equation*}
A^{(2)} = C_1 \,\Im z^2, \qquad C_1 = \bke{ c_1Q, \, \la Q (\up^2
- \um^2)} = O(\la^2) ~,
\end{equation*}
We can integrate $A^{(2)}$ by parts to have:
\begin{align*}
\int_\infty^t A^{(2)} \, ds &=  C_1 \Im \int_\infty^t  \,  z^2 \,
ds = C_1 \Im \int_\infty^t  \,  \mu^{-2} p^2 \, ds
\\
&=  C_1 \Im \frac 1{-2i\ka} \bket{ { \mu^{-2} p^2 } -
\int_\infty^t  \,  \mu^{-2} 2p \dot p \, ds }
\\
\end{align*}
As we shall prove later on, the last integral is higher order
term. The first term on the right hand side can be written
explicitly as
\begin{equation*}
C_1 \Im \frac 1{-2i\ka}  \mu^{-2} p^2 =  a_{20} (z^2 + \bar z^2)
~,
\end{equation*}
where
\begin{equation} \eqlabel{a20}
a_{20} = \frac {C_1}{4\ka} = \frac \la {4 \ka } \bket { c_1Q, \, Q
(\up^2 - \um^2)} = O(\la^2) ~,
\end{equation}

We shall prove later on that the last term $a_{20}(z^2 + \bar
z^2)$ is the main oscillatory term in $a$. We denote the rest by
$b$, i.e.,
\begin{equation} %
   a = a_{20} (z^2 + \bar z^2) + b
\end{equation}
As shall prove $a, b, \dot a=O(z^2)$, but $\dot b=O(z^3)$. In
other words, $b$ is the part of $a$ that has slower oscillation.
From the equation of $a$, we have the following equation for $b$:
\begin{equation}  %
\dot b = \left (c_1 Q, \Im(F-F^{(2)} ) \right ) -4  \Re a_{20}
z \mu^{-1} \dot p ~.
\end{equation}

Assuming that $b$ and $\eta$ are of order $z^2$, we can decompose
$F$ into
\[
F = F^{(2)} + F^{(3)}  + \widetilde F^{(3)}
 + F^{(4)}
\]
where $F^{(2)}$ and $F^{(3)}$ denote terms of order $z^2$ and
$z^3$, respectively, and $F^{(4)}$ denotes higher order terms:
\begin{align}
F^{(2)} &= \la Q(2|\zeta|^2 + \zeta^2) = z^2  \phi_{(20)} + z\bar
z \phi_{(11)}
 + \bar z ^2  \phi_{(02)}
\nonumber \\
F^{(3)} &= 2\la Q[(\zeta +\bar \zeta) \eta^{(2)} +  \zeta \bar
\eta^{(2)}] + \la |\zeta|^2 \zeta + 2 \la a_{20}(z^2 + \bar z^2)
QR(2\zeta+\bar \zeta)
\nonumber \\
\widetilde F^{(3)} &=2 \la b QR(2\zeta+\bar \zeta)
\eqlabel{Fcom}  \\
F^{(4)} &= 2\la Q[(\zeta +\bar \zeta) \eta^{(3)} +  \zeta \bar
\eta^{(3)}] + \la Q \bkt{ 2 |\eta|^2 + \eta^2} + 2 \la a QR (
2\eta + \bar \eta)
\nonumber \\
&\quad + 3 \la a^2 QR^2 + \la \bkt{ |k|^2k- |\zeta|^2 \zeta } ~,
\nonumber
\end{align}
We can also rewrite the equation of $\theta$ into
\begin{equation} %
\dot \theta = c_2(z + \bar z) + F_\theta ~,
\end{equation} where
$c_2= -(c_0 Q, \la Q^2 u)=O(\la )$ and
\[
F_\theta=\frac {-1}{1+c_0 c_1^{-1}a}
\bket{ c_0 c_1^{-1} c_2 a (z+\bar z) +
 \bkt{ \rule{0pt}{4.5mm}
a + (c_0Q, \la Q^2(\eta + \bar \eta)) +(c_0Q, \Re F(k) )} }
\]
Hence, since $z= \mu^{-1}p$,
\begin{align}
\theta(t) &= \int_0^t 2c_2 \Re (z) + F_\theta \, ds
= 2 c_2 \Re z/(-i\ka) + \int_0^t -2c_2 \Re (\mu^{-1}\dot p) /(-i\ka)
+ F_\theta \, ds   \nonumber
\\
&= \frac {2 c_2}{\ka}\Im z +
\int_0^t -\frac {2 c_2}{\ka} \Im (\mu^{-1}\dot p)
+ F_\theta \, ds   \eqlabel{eq:theta}
\end{align}
\subsection{Main oscillation term in  $\eta$}

We now identify the main oscillation term in $\eta$. We first
rewrite the equation of $\eta$ using the operator $A$ as
\begin{align}
   \partial _t \eta &= \L \eta - \PcL i \dot \theta \eta
+\PcL \Pi  F^ \sharp ~, \nonumber
\\
F ^\sharp &= - i \dot \theta \zeta - i F(k) - [(c_1Q,\Im F) + ia
\dot \theta ] R_\Pi ~. \eqlabel{Fsharp.def}
\end{align}
Notice that $F^\sharp $ and $F^{\text {all}}$ differs only by the
term $i \dot \theta \eta$. Observe also that $- i \dot \theta
\zeta$ is not killed by $\PcL$ since
 $[\L , i] \not = 0$.
Let
\[
   \eta ^\diamond = U \eta ~.
\]
Since $\L = U^{-1} (-i \sA) U$ and $U\PcL=\PcA U$, we have
\begin{align*}
\partial _t \eta^\diamond &=  -i \sA\eta^\diamond -  \PcA U i \dot \theta U^{-1}\eta^\diamond
+ \PcA U \Pi  F^ \sharp
\\
&=-i \sA\eta^\diamond -  i \dot \theta \eta^\diamond -   \PcA
[U,i]\dot \theta U^{-1} \eta^\diamond + \PcA U \Pi  F^ \sharp ~.
\end{align*}

Let $\tilde \eta = e^{i \theta}\eta^\diamond$ and use $U^{-1} \eta
^\diamond = \eta$, we get
\begin{equation}        \eqlabel{Eq.eta0}
\partial _t \tilde \eta  = -i\sA \tilde \eta
+e^{i \theta} \PcA U \Pi  F^\sharp - e^{i \theta} \PcA [U,i]\dot
\theta  \eta ~.
\end{equation}
Hence
\begin{equation}        \eqlabel{Eq.eta}
\tilde \eta(t) = e^{-i\sA t} \tilde \eta_0 + \int_0^t e^{-i\sA
(t-s)} \PcA \bket{e^{i \theta}  U \Pi  F^\sharp  -e^{i \theta}
[U,i] \dot \theta  \eta } \, ds ~.
\end{equation}
Since $\eta = U^{-1} e^{-i \theta} \tilde \eta $ and $U$ is
bounded in Sobolev spaces, for the purpose of estimation we can
treat $\eta$ and $\tilde \eta$ the same.

From the the definition of $\tilde \eta$ \eqref{Eq.eta}, the
integrand is $e^{i \theta} U \Pi F^\sharp-e^{i \theta} [U,i]\dot
\theta \eta$. We first identify the main term in $F^\sharp$:
\begin{align} \nonumber
F^\sharp &= - i \dot \theta \zeta - i F(k) - [(c_1Q,\Im F) + ia
\dot \theta ] R_\Pi
\\
&=  i \, z^\al \phi ^\sharp_\al + F^{\sharp \sharp}.  \label{eq:Fsharp}
\end{align}
We have already decomposed $F$ into orders in $z$. To identify the
main term of $F^\sharp$, it remains to decompose $- i \dot \theta
\zeta$ and $(c_1Q,\Im F)R_\Pi$. From the equation of $\theta$
\eqref{eqsum}, we have
$$
 - i \dot \theta \zeta= -i c_2 (z+\bar z)(zu_++ \bar z u_-)
 - i F_\theta \zeta
$$
Also
$$
(c_1Q,\Im F)R_\Pi=(c_1Q, \phi_{20} - \phi_{02} ) R_\Pi \Im z^2 +
O(z^3) .
$$
Recall the decomposition of $F$ in \eqref{Fcom} and  $F^{(2)}=
z^\al \phi_\al$.  Now we have \eqref{eq:Fsharp} with
\[
F^{\sharp\sharp} =
 - i F_ \theta \zeta - i (F-F^{(2)} ) - [(c_1Q,\Im (F-F^{(2)}) )
+ ia \dot \theta ] R_\Pi
\]
and $\phi_\al ^\sharp$ are defined as follows:
\begin{alignat}{2}
\phi ^\sharp _{20} &= - \phi_{20} - c_2 \up + \tfrac 12 (c_1Q,
\phi_{20} - \phi_{02}  ) R_\Pi &&=- \phi_{20} - c_2 \up +
O(\la^2)~,                \nonumber
\\
\phi ^\sharp _{11} &= - \phi_{11}- c_2 (\up+\um) &&= O(\la) ~,
\\
\phi ^\sharp _{02} &= - \phi_{02} - c_2 \um + \tfrac 12 (c_1Q,
\phi_{20} - \phi_{02}  ) R_\Pi &&= O(\la^2) ~.
       \nonumber
\end{alignat}
Here we have used $\phi_{20} = O(\la)=\phi_{11}$, $\phi_{02} =
O(\la^2)$, $c_2=O(\la)$ and $R_\Pi=O(1)$. Also note that they are
all real.

Recall \eqref{U.dec}. Since $Ui = (U_+ + U_- C) i = i(U_+ - U_-
C)$, We have
\[
\PcA e^{i \theta} U i z^\al \Pi \phi ^\sharp _\al = \PcA e^{i
\theta} i(U_+ - U_- C)  z^\al \Pi \phi ^\sharp _\al = e^{i \theta}
i z^\al \Phi_\al
\]
where
\begin{align}
\Phi_{20} &= \PcA \bket{ U_+  \Pi \phi ^\sharp_{20} - U_-  \Pi
\phi ^\sharp_{02} },\quad \Phi_{11} = \PcA \bket{ (U_+ - U_-)  \Pi
\phi ^\sharp_{11} },    \label{eq:4-13}
\\
\Phi_{02} &= \PcA \bket{ -U_-  \Pi \phi ^\sharp_{20} + U_+  \Pi
\phi ^\sharp_{02} } ~.  \nonumber
\end{align}

Hence we can rewrite the integrand in \eqref{Eq.eta} as
\[
   e^{i \theta} i z^\al \Phi_\al +
\PcA \bket{ e^{i \theta} U\Pi F^{\sharp \sharp}
        - e^{i \theta} [U,i] \dot \theta \eta } ~.
\]
 The leading term in $\tilde \eta$ is
\begin{align}
(I) &\equiv \int_0^t  e^{-i\sA(t-s)} e^{i \theta} i z^\al \Phi_\al
\, ds \nonumber
\\
&=\int_0^t  e^{-it\sA} e^{is (\sA - 0i)} \bke{ \mu^{[\al]} (e^{i
\theta} p^2)(s) i\Phi_\al } \, ds \qquad  (\mu = e^{i \ka s})
\nonumber
\\
&= e^{-it\sA}\bkt{ \frac { e^{is (\sA - 0i)} [e^{i \theta}
z^\al](s)}{i(\sA -0i +[\al]\ka)} i\Phi_\al }       ^t_{s=0} + (II)
\nonumber
\\
&= \tilde \eta^{(2)}  - e^{-it\sA} (e^{i \theta} z^\al)(0) \wt
\eta_\al
 +(II) ~,                                               \eqlabel{Eq.(I)}
\end{align}
where
\begin{equation} \label{eq:4-15}
\tilde \eta^{(2)}= e^{i \theta} z^\al \wt \eta_\alpha,  \qquad \wt
\eta _\al= \frac {1}{\sA -0i +[\al]\ka} \Phi_\al
\end{equation}
and $(II)$ is the error from integration by parts,
\[
(II) = -\int_0^t e^{-i(t-s)\sA} \bket{ \mu^{\al} \frac d{ds} (
e^{i \theta} p^\al)  \wt \eta_\al} \, ds ~.
\]
Also note the sign of $0i$ is so that the last two terms in
\eqref{Eq.(I)} decay as $t \to \infty$.

We have identify the main oscillation term in $\tilde \eta$ and we
denote the remaining term in by $\eta^{(3)}$:
\begin{equation}\eqlabel{etad}
 \tilde \eta= \tilde \eta^{(2)} + \tilde\eta^{(3)}.
\end{equation}
Notice that $\tilde \eta^{(2)}$ is not in $L^2$ and
this is not an $L^2$ decomposition. It is, however, very useful
for local behavior as it identifies the local oscillation.

From \eqref{Eq.eta} we obtain the equation for $\tilde\eta^{(3)}$:
\begin{align}
\tilde \eta^{(3)}(t) &= e^{-i\sA t} \tilde \eta_0 - e^{-i\sA
t}(e^{i \theta}z^\al) (0) \wt \eta_\al
-\int_0^t e^{-i\sA(t-s)} \bket{ \mu^{[\al]} \frac d{ds} ( e^{i
\theta} p^\al)  \wt \eta_\al} \, ds
\nonumber \\
&\quad + \int_0^t  e^{-i\sA(t-s)} \PcA \bket{ e^{i \theta} U\Pi
\bkt{ F^{\sharp \sharp} +i\eta^2\bar \eta}
        - e^{i \theta} [U,i] \dot \theta \eta }\, ds
\eqlabel{eta3} \\
&\quad + \int_0^t  e^{-i\sA(t-s)} \PcA
 e^{i \theta} U\Pi (-i\eta^2\bar \eta)  \, ds
\nonumber \\
&= \tilde \eta^{(3)}_1 + \tilde \eta^{(3)}_2 + \tilde \eta^{(3)}_3
  + \tilde \eta^{(3)}_4 + \tilde \eta^{(3)}_5 ~.
\nonumber
\end{align}
We treat $\tilde \eta^{(3)}_5$ separately because $\eta^2 \bar
\eta$ is a non-local term. Notice that $\tilde \eta^{(2)} \not \in
L^2$, but is still ``orthogonal'' to the eigenvector of $\sA$.

From the decomposition of $\tilde \eta$ and the relation $\eta=
U^{-1} e^{-i \theta} \tilde \eta$, we have the corresponding
decomposition for $\eta$:
\[
   \eta(t) = \eta^{(2)}(t) + \eta^{(3)}(t) ~.
\]
where
\[
\eta^{(2)}= U^{-1} e^{-i \theta} \tilde \eta ^{(2)} = U^{-1} z^\al
\wt \eta_\al , \quad
\eta^{(3)}= U^{-1} e^{-i \theta} \tilde \eta ^{(3)} \ .
\]

Summarizing, we have decompose $a$,  $F$  and $\eta$ into terms in
order of $z$. The main oscillatory terms of order $z^2$ in $a$ is
$a_{20}(z^2 +\bar z^2)$ and
\begin{equation} \eqlabel{b.def}
   a = a_{20} (z^2 + \bar z^2) + b
\end{equation}
with
\begin{equation}\eqlabel{b.eq0}
\dot b = \left (c_1 Q, \Im(F-F^{(2)} ) \right ) -4  \Re a_{20} z
\mu^{-1} \dot p ~.
\end{equation}
The nonlinear term $F$ is decomposed into orders in \eqref{Fcom}
with the second order term $F^{(2)}$ explicitly given. We also
rewrite the equation of $\theta$ into
\begin{equation}\eqlabel{thetadot}
\dot \theta = c_2(z + \bar z) + F_\theta ~,
\end{equation} where
$c_2=(c_0 Q, \la Q^2 u)=O(\la )$ and
\begin{equation}\eqlabel{Ft}
F_\theta=\frac {-1}{1+c_0 c_1^{-1}a}
\bket{ c_0 c_1^{-1} c_2 a (z+\bar z) +
 \bkt{ \rule{0pt}{4.5mm}
a + (c_0Q, \la Q^2(\eta + \bar \eta)) +(c_0Q, \Re F(k) )} }
\end{equation}
The dispersive wave $\eta$ is related to $\tilde \eta$ by the
relation $\eta= U^{-1} e^{-i \theta} \tilde \eta$ with $\tilde
\eta$ satisfying
\begin{equation} \eqlabel{Eq.eta2}
\tilde \eta(t) = e^{-i\sA t} \tilde \eta_0 +
\int_0^t e^{-i\sA (t-s)} \PcA \bket{e^{i \theta}  U \Pi  F^\sharp
-e^{i \theta} [U,i] \dot \theta  \eta } \, ds ~.
\end{equation}
Furthermore, it can be decomposed into $\tilde \eta^{(2)} +
\tilde\eta^{(3)}$ with
\[
\tilde \eta^{(2)}= e^{i \theta} z^\al \wt \eta_\alpha,  \qquad \wt
\eta _\al= \frac {1}{\sA -0i +[\al]\ka} \Phi_\al
\]
and   $\tilde\eta^{(3)}$ satisfies the equation \eqref{eta3}. For Theorem \ref{mainthm}
the boundary conditions are
\begin{align}
a(\infty) & = 0 =b(\infty)
\nonumber \\
 0 & < |z_0|\le \e _0
\eqlabel{ab} \\
\| \eta_0\|_{H^2 \cap W^{2,1}(\R^3)} &  \le   |z_0|^2 \nonumber
\end{align}

We now collect a few properties of the operator $U$. We can expand
$U_\pm$ in order of $\la$ as
\[
   U_+ = 1 + O(\la), \quad U_- =  O(\la).
\]
Notice that $\PcA U c_2\up=O(\la^2)$ since $U \up$ is almost
orthogonal to $\Hc(A)$. Hence we have
\[
\Phi_{20} = -\PcA \Pi \phi_{20} + O(\la^2),\quad \Phi_{11} =
O(\la) ,\quad \Phi_{02} =O(\la^2) ~.
\]
We may decompose
\begin{equation}   \eqlabel{U.dec2}
   U^{-1} = U_+^\diamond + U_-^\diamond \conj , \qquad
   U_\pm^\diamond = \tfrac 12 (B\sA^{-1/2} \pm B^{-1}\sA^{1/2}) ~.
\end{equation}
then
\[
\eta^{(2)} =\bke{U_+^\diamond + U_-^\diamond \conj} z^\al \wt
\eta_\al = z^2 \eta_{20} + z \bar z  \eta_{11} + \bar z^2
\eta_{02} ~,
\]
where
\[
\eta_{20}=U_+^\diamond  \wt \eta_{20} + U_-^\diamond  \wt
\eta_{02} ,\quad \eta_{11}=U^{-1} \wt \eta_{11} , \quad
\eta_{02}=U_+^\diamond  \wt \eta_{02} + U_-^\diamond  \wt
\eta_{20} ~.
\]
If we expand
\[
   U_+ = 1 + O(\la), \quad U_- =  O(\la), \quad
   U_+^\diamond = 1 + O(\la), \quad U_-^\diamond =  O(\la),
\]
then we get, by \eqref{eq:4-15} and   \eqref{eq:4-13},
\[
\eta_{20}=  \wt \eta_{20} + O(\la^2) = \frac {-1}{\sA -0i -
2\ka}\PcA \Pi \phi_{20} +O(\la^2) ~,\quad \eta_{11}=O(\la) ~,
\quad \eta_{02}=  O(\la^2) ~.
\]

\section{Estimates of dispersive wave}

We now estimate solutions to the equations \eqref{eqsum} with the
decompositions into main oscillatory  and higher order terms in
section 4. We first need to choose a suitable norm. Define
\[
 \tl{t} = \e^{-2} + 2 \Gamma t, \quad
 \tl{t} \sim \max \bket{ \e^{-2}, t }.
\]
and, for $\eta^{(3)}_{2-5} \equiv \sum_{j=2}^5 \eta^{(3)}_j=\eta^{(3)}-
\eta^{(3)}_1$,
\begin{equation}
M(T) := \sup _{0\le t \le T} \bket{\tl{t}^{1/2} |z(t)| +
\tl{t}^{3/4-\sigma} \norm{\eta(t)}_{L^4} + \tl{t}^{1+ \sigma/4}
\norm{\eta^{(3)}_{2-5}(t)}_{L^2 \loc}  }
\end{equation}
Recall that $b$, the slow oscillation part of $a$, is defined via
\eqref{b.def} and satisfies the equation \eqref{b.eq0}. We assume
that the function $b$ is in the following space ${\cal B}_T$:
\begin{equation} \eqlabel{bb}
 {\cal B}_T = \bket{ b(t): |b(t)| \le D \tl{t}^{-1}, \, 0
\le t \le T } ~,
\end{equation}
where $D= 2 B_{22}/\Gamma$ is some constant. The precise form of
the constant is not important, it merely has to be an order one
constant bigger than the true behavior of $b$ which we shall
derive. The class ${\cal B}$ is nonempty since it contains the
constant function $0$.

Our setting is thus given at the end of section~4 except we now
assuming the estimate \eqref{bb} on $b$ (and thus on $a$ too) up
to time $T$. From now on, we fix $T$.

\begin{theorem}\thlabel{M}
Suppose $\eta, h, a, \theta, z$ are solutions to the equations
\eqref{eqsum}.  %
Assuming the estimate  \eqref{bb} on $b$,  we have
\[
        M(t) \le 2 \qquad \text{for all }\quad t\le T ~.
\]
Moreover, if we further assume $|z_0|=\e>0$ and
$\norm{\eta_0} \le \e ^{3/2}$, then $|z(t)| \ge c \tl{t}^{-1/2}$.
\end{theorem}

Our strategy is to show that $M(0) \le 3/2$ and that $M(t) \le
3/2$ if $M(t) \le 2$. By continuity of $M(t)$, this would imply
that $M(t) \le 3/2$ for all $t\le T$. So for the rest of this  and next
sections, we shall use freely that $M(t)\le 2$ to prove that $M(T)
\le 3/2$. The proof of this theorem will be completed after Lemma
\ref{th:p1}. We now start the proof.

\bigskip

Due to the presence of the non-local term $h^3$ in $F$,  we first
need a global norm estimate on $\eta$, which we choose to be
$\norm{\eta(t)}_{L^4}$. Our goal is to prove that
\[
        \norm{\eta(t)}_{L^4} \le C \tl{t}^{-3/4} \, \log \tl{t}
\]
which agrees  with that of free evolution.

We first recall some basic facts concerning the Schr\"odinger equation
which  provide some useful feeling on the size of various quantities. Since we
shall proceed with iteration scheme, our $a$, $z$ and $\eta$ do
not solve the Schr\"odinger equation and  we will not use these
facts.

The $H^1$ norm of $\psi_t$ is uniformly bounded if $\la$ is
sufficiently small. (It can be proved by using the conservation of
the Hamiltonian and the Gagliado-Nirenberg inequality.) Hence
$\norm{h(t)}_{H^1}$ is uniformly bounded since $h=\psi e^{i
\theta}-Q$. Assuming $z(t)$ is bounded, then $\zeta(t)$ and
$\eta(t)$ are both bounded in $H^1$, uniformly in $t$,
 since $\zeta=z \up + \bar z \um$ and $\eta = h - \zeta$.

Return to a global estimate for $\eta$. Since $\eta = U^{-1} e^{-i
\theta} \tilde \eta $ and $U$ is bounded in Sobolev spaces, for
the purpose of estimation we can treat $\eta$ and $\tilde \eta$
the same.
Recall  $[U,i]$ is a local operator satisfying the estimate at
Lemma 3.3.

We will need the following calculus lemma:

\begin{lemma}   \thlabel{th:eta1}
Let $0<d < 1 < m$. $\tl{t} \equiv \e^{-2} + 2\Gamma t $.
\[
\int _0^t |t-s|^{-d}   \tl{s} ^{-m}  \, ds
\le C \e ^{2m-2}  \tl{t}^{-d} ~.
\]
Also,
\[
\int _0^t |t-s|^{-d}   \tl{s} ^{-1}  \, ds
\le C  \tl{t}^{-d} \log (\e^2\tl{t}) ~.
\]

If, instead, $d \ge m >1$,
\[
\int _0^t \bkA{t-s}^{-d}   \tl{s} ^{-m}  \, ds
\le C \tl{t}^{-m} ~.
\]
\end{lemma}

\myproof
Denote the first integral by (I).
If $t\le \e^{-2}$, then $\tl{t} \sim \e^{-2}$ and
\[
(I) \sim \int_0^t |t-s|^{-d}   \e^{2m}  \, ds  \lesssim \e^{2m} \e^{-2(1-d)}
= \e^{2m-2} \e^{2d} \sim \e^{2m-2} \tl{t}^{-d} ~.
\]
If $t \ge \e^{-2}$, then $\tl{t} \sim 2\Gamma t$ and
\begin{align*}
(I) &\le \int_0^{t/2} C t^{-d} \tl{s}^{-m} \, ds +
\int_{t/2}^t C |t-s|^{-d} C \tl{t}^{-m} \, ds
\\
&\le C t^{-d} \e^{2(m-1)} +  Ct^{1-d} \tl{t}^{-m}
\sim C \tl{t}^{-d} \e^{2(m-1)} +  C \tl{t}^{1-d-m}
\\
&\le C\e^{2m-2} \tl{t}^{-d} ~.
\end{align*}

For the second case, denote the second integral by (II).
If $t\le \e^{-2}$, then $\tl{t} \sim \e^{-2} $ and
\[
(II) \sim \int_0^t \bkA{t-s}^{-d}   \e^{2m}  \, ds
\le \e^{2m} \cdot C \sim C\tl{t}^{-m} ~.
\]
If $t \ge \e^{-2}$, then $\tl{t} \sim \Gamma \bkA{t}$ and
\[
(II) \sim \int_0^t \bkA{t-s}^{-d} \Gamma^{-m} \bkA{s}^{-m}    \, ds  \le
C \Gamma^{-m} \bkA{t}^{-m} \sim C\tl{t}^{-m} ~.
\]
We conclude the lemma.
\myendproof

\subsection{Estimates in $L^4$ and $L^2$}

\begin{lemma} \thlabel{etaglobal}
 Suppose that  $\tilde \eta$ is given by equation
\eqref{Eq.eta} and recall $\eta = U^{-1} e^{-i \theta} \tilde \eta
$. Assuming  the estimate  \eqref{bb} on $b$ and  $M(T)\le 2$, we
have
\[
\norm{\eta(t)}_{L^4}  \le C  \tl{t}^{-3/4} \log \tl{t} ~.
\]
Moreover, we have
\[
\norm{ \eta(t)}_{L^2} \le \e^{1/2} ~.
\]
\end{lemma}

\myproof From the defining equation of $\tilde \eta$, we have
\[
\norm{\wt \eta(t)}_{L^4}
\le C \norm{\wt \eta_0}_{H^1 \cap L^{4/3}} \bkA{t}^{-3/4}
+ \int_0^t C |t-s|^{-3/4}
\bket{  \norm{F^\sharp(s)}_{L^{4/3}}
+ |\dot \theta| \norm{\eta(t)}_{4}  } \, ds ~.
\]
From the definitions of $F^ \sharp$ and $\zeta$, we have
\[
\norm{F^\sharp(s)}_{L^{4/3}}
+ |\dot \theta| \norm{\eta(t)}_{4}
\le  C |\dot \theta| \norm{\zeta(t)}_{4/3}
+ \norm{F(s)}_{L^{4/3}} + C |a \dot \theta|  ~,
\]
From the Holder inequality and the definition of
$\norm{F(s)}_{L^{4/3}}$ we can bound $\norm{F(s)}_{L^{4/3}}$ by
\[
\norm{F(s)}_{L^{4/3}}
\le C \bke{
     \norm{h}_{L^{4}} ^2
  + |a|\norm{h}_{L^{4}}   + |a|^2
+ \norm{h^3}_{L^{4/3}}   + |a|^3 }~.
\]
Since $\norm{h^3}_{L^{4/3}} = \norm{h}_{L^{4}}^3$ and
\[
\norm{h(s)}_{L^4} \le \norm{\zeta(s)}_{L^4} +
\norm{\eta(s)}_{L^4},
\]
we have from the assumption $M\le 2$ that
\[
\norm{h(s)}_{L^4}
 \le C \tl{s}^{-1/2} + C\tl{s}^{-3/4} \log
\tl{s} \le C \tl{s}^{-1/2} ~,
\]
Therefore we have
$\norm{F(s)}_{L^{4/3}} \le C \tl{s}^{-1}$
and thus
\[
 \norm{F^\sharp(s)}_{L^{4/3}}  \le C \tl{s}^{-1} ~.
\]
Since $\bkA{t}^{-1} \le \e^{-2} \tl{t}^{-1}$ and
$\norm{\eta(t)}_{L^4} \sim \norm{\wt \eta(t)}_{L^4}$,
we conclude
\[
\norm{\eta(t)}_{L^4}
\le C \e^{-3/2}\norm{\eta_0}_{H^1 \cap L^{4/3}} \tl{t}^{-3/4}
+ \int_0^t |t-s|^{-3/4} C\tl{s}^{-1} \,ds
\le C  \tl{t}^{-3/4} \log \tl{t} ~.
\]
Here we have used Lemma \ref{th:eta1} in the last integration.

We now bound $\norm{\eta(t)}_{L^2}$. Since
\[
\frac d{2dt} (\tilde \eta , \tilde \eta )
= \Re (\tilde \eta , \partial_t \tilde \eta )
= \Re (\tilde \eta ,
e^{i \theta} \PcA U \Pi  F^\sharp
- e^{i \theta} \PcA [U,i]\dot \theta  \eta )~,
\]
we have
\begin{align*}
\frac d{dt} \norm{\tilde \eta}_{L^2}^2
&\le C \norm{\tilde \eta}_{L^4} \cdot
\bket{ \norm{F^\sharp(s)}_{L^{4/3}}
+ |\dot \theta| \norm{\eta(t)}_{4}  }
\\
&\le C \tl{t}^{-3/4} \log \tl{t} C \tl {t}^{-1}
= C \tl{t}^{-7/4} \log \tl{t} ~.
\end{align*}
In the second inequality we use our previous estimates.
Hence
\[
\norm{\tilde \eta}_{L^2}^2
\le C \e^3
 + \int_0^\infty C \tl{t}^{-7/4} \log \tl{t} \le \e ~,
\]
and we conclude $\norm{\tilde \eta(t)}_{L^2} \le \e^{1/2} $, and
so is $\norm{\eta(t)}_{L^2}$.
\myendproof

\subsection{Local decay of $\eta^{(3)}$}

Recall $\eta^{(3)}= U^{-1} e^{-i \theta}\tilde\eta^{(3)}$ and
$\tilde\eta^{(3)}$ satisfies the equation \eqref{eta3}. We want to
show that $\tilde \eta^{(3)}$ is smaller than $\tilde \eta^{(2)}$
locally. Define the local $L^2$ norm by
\[
\norm{f}_{L^2\loc} = \norm{\bkA{x}^{-\beta_0} f}_{L^2}
\]
for a fixed sufficiently large $\beta_0 > 0$, which will become
clear later on.

\begin{lemma} \thlabel{eta3b}
Assuming  the estimate  \eqref{bb} on $b$ and  $M(T)\le 2$ for all
$T$, we have
\begin{align} \nonumber %
\norm{\tilde \eta^{(3)}_{2-5} }_{L^2\loc}  &\le C \tl{t}^{-1-\sigma/2}
\le C \e^{\sigma/2} \tl{t}^{-1-\sigma/4} ~,
\\
\norm{ \eta^{(3)}_{2-5} }_{L^2\loc}  &\le C \e^{\sigma}
\tl{t}^{-1-\sigma/4}
\eqlabel{eta3.decay}
\end{align}
In particular, for a local function $\phi$ we have
\begin{align}
\big| (\phi, \eta^{(3)}) \big| &\le C  \tl{t}^{-1-\sigma/4} ~,
\\
\big| (\phi, |\eta|^2+|\eta|^3) \big| &\le C \norm{\eta}_{L^2\loc}
^2 + C\norm{\eta}_{L^2\loc} \norm{\eta}_{L^4}^2 \le C  \tl{t}^{-2}
~.
\end{align}

\end{lemma}

\myproof We first estimate $\tilde \eta^{(3)}_j$ appearing on the
right side of the equation for $\tilde\eta^{(3)}$ \eqref{eta3}.
Note we did not include $\tilde \eta^{(3)}_1$ in the above Lemma:
\[
\norm{\tilde \eta^{(3)}_1 }_{L^2\loc}
 \le C \norm{\eta_0}_{H^1 \cap L^{8/7}} \, \bkA{t}^{-9/8} ~.
\]
This term would be bounded by $  \e^{\sigma/4} \tl{t}^{-9/8}$ if
we assumed $\norm{\eta_0} \le \e^{2+\sigma}$. However, since we only
assume $\norm{\eta_0} \le \e^{3/2}$, this term needs to be treated
separately.

For  $\tilde \eta^{(3)}_2$ and  $\tilde \eta^{(3)}_3$,
the two terms involving $\eta_\al$, ($\eta_\al \not \in L^2$),
from the definition and the estimates Lemma \ref{3.1} on $A$, we
have
\[
\norm{\tilde \eta^{(3)}_2 }_{L^2\loc}
   \le C \tl{t}^{-9/8}  ~.
\]
\[
\norm{\tilde \eta^{(3)}_3 }_{L^2\loc}
   \le C \int_0^t \bkA{t-s}^{-9/8}  \tl{s}^{-3/2} \, ds
\le C \tl{t}^{-9/8}~.
\]

To estimate $\tilde \eta^{(3)}_4$, we recall the definition of
$F^{\sharp \sharp}$ and rewrite
\begin{align*}
F^{\sharp \sharp} +i\eta^2\bar \eta
&=\bke{F^{\sharp } - iz ^\al \phi^\sharp_\al } +i\eta^2\bar \eta
\\
&= - i[\dot \theta - c_2(z+\bar z)] \zeta
 - i \bket{ F - F^{(2)} - \eta^2\bar \eta}
\\
& \quad
- \bkt{ (c_1 Q, \Im (F- F^{(2)}) ) + i a \dot \theta } R_\Pi ~.
\end{align*}
Since $M\le 2$, we have
$|\dot \theta - c_2(z+\bar z)| \le C \tl{s}^{-1}$,
$|a| \le C \tl{s}^{-1}$,
$\norm{\bkA{x}^{\beta_0} (F - F^{(2)}-\eta^2\bar\eta) } _{L^2} \le
C\tl{s}^{-3/2} $, and
$|(c_1 Q, \Im (F- F^{(2)}) ) | \le \tl{s}^{-3/2}$.
 Hence, by Lemma \ref{3.1} we have
\begin{align*}
\norm{\tilde \eta^{(3)}_4 }_{L^2\loc}
&\le \int_0^t \bkA{t-s}^{-3/2}
\norm{ \bkA{x}^{\beta_0} \bke{F^{\sharp \sharp} +i\eta^2\bar
\eta}(s)}_{L^2} \, ds
\\
&\le \int_0^t \bkA{t-s}^{-3/2} C\tl{s}^{-3/2} \, ds
\\
&\le C\tl{t}^{-3/2} ~.
\end{align*}

We now consider $\tilde \eta^{(3)}_5$. Denote the integrand $
e^{-i\sA(t-s)} \PcA e^{i \theta} U\Pi (-i\eta^2\bar \eta)$ by $
X(t,s)$ and decompose the time interval into $(0,{t-\sigma})$ and
$({t-\sigma}, t)$. From the triangle inequality, we have
\begin{equation*}
\norm{\tilde \eta^{(3)}_5 }_{L^2\loc} \le  \int_0^{t-\sigma}
\norm{ X(t,s)  \, ds }_{L^2\loc} +  \int_{t-\sigma}^t \norm{
X(t,s) \, ds}_{L^2\loc}
\end{equation*}
We can bound the ${L^2\loc}$ norm by either $L^8$ or $L^4$ norm.
Thus
\begin{equation*}
\norm{\tilde \eta^{(3)}_5 }_{L^2\loc} \le C \int_0^{t-\sigma}
\norm{ X(t,s) }_{L^8} \, ds + C \int_{t-\sigma}^t \norm{ X(t,s)
}_{L^4} \, ds
\end{equation*}
From Lemma \ref{3.1}, we have
\[
\norm{ X(t,s) }_{L^8} = \norm{e^{-i\sA(t-s)} \PcA e^{i \theta}
U\Pi (-i\eta^2\bar \eta) }_{L^8} \le \frac 1{|t-s|^{9/8}}
\norm{\eta^3}_{L^{8/7}}
\]
and
\[
\norm{ X(t,s) }_{L^4} = \norm{e^{-i\sA(t-s)} \PcA e^{i \theta}
U\Pi (-i\eta^2\bar \eta) }_{L^8} \le \frac 1{|t-s|^{9/8}}
\norm{\eta^3}_{L^{4/3}}
\]
From the Holder inequality and the global estimate of $\eta$ in
Lemma \ref{etaglobal} we have
\begin{align*}
 \norm{\eta^3(s)}_{L^{8/7}}
&\le \norm{\eta}_{2}^{1/2}
 \norm{\eta}_{4}^{5/2}
\le C \e^{1/4} \bke{\tl{s}^{-3/4} \ln \tl{s} }^{5/2}
\le C \e^{1/4} \tl{s}^{-7/4} ~,
\\
\norm{ \eta^3(s)}_{L^{4/3}}
&\le \norm{ \eta}_{L^{4}}^3
\le \bke{\tl{s}^{-3/4} \ln (2+s)}^{3}
\le C \tl{s}^{-2} ~.
\end{align*}

If $t \ge 2 \e^{-2}$, then $\tl{t} \sim t$.  We choose $\sigma =
t/2$ and thus
\begin{align*}
\norm{\tilde \eta^{(3)}_5 }_{L^2\loc}
&\le C \int_0^{t/2} \frac 1{|t-s|^{9/8}} C \tl{s}^{-7/4} \, ds
+  C \int_{t/2}^t \frac 1{|t-s|^{3/4}} C \tl{s}^{-2}  \, ds
\\
&\le C \int_0^{t/2} C\tl{t}^{-9/8} \tl{s}^{-7/4} \, ds
+  C \int_{t/2}^t \frac 1{|t-s|^{3/4}} C \tl{t}^{-2}  \, ds
\\
&\le  C\tl{t}^{-9/8}  (\e^{-2})^{-3/4} +  C \tl{t}^{-2} t^{1/4}
\le C \e^{5/4} \tl{t}^{-9/8} ~.
\end{align*}
On the other hand, if $t < 2 \e^{-2}$, then $\tl{t} \sim \e^{-2}$ and
we choose $\sigma = t$ to get
\begin{align*}
\norm{\tilde \eta^{(3)}_5 }_{L^2\loc}
&\le C \int_{0}^t \frac 1{|t-s|^{3/4}} C (\e^{-2})^{-2}  \, ds
\\
&\le C \e^4 t^{1/4} \le C \e ^{7/2}
\sim C \e^{5/4} \tl{t}^{-9/8} ~.
\end{align*}
Combining two cases, we conclude
\[
\norm{\tilde \eta^{(3)}_5 }_{L^2\loc} \le C \e^{5/4} \tl{t}^{-9/8} ~.
\]
The lemma follows from all the estimates on $\eta^{(3)}_j$, $j=1,
\cdots, 5$.
\myendproof

\setcounter{section}{5}
\section{Excited state equation and normal form}

\subsection{Excited state equation}

Return to the basic equation \eqref{eqsum} and recall the equation
for $\dot p$:
\begin{equation} \eqlabel{p.eq1}
\dot p = -i \mu \bkt{ (\tilde \up, F) + (\tilde \um, \bar F) +
\bket{(\up, h) + (\um, \bar h) +c_3 a} \dot \theta \, } ~,
\end{equation}
where $c_3=(u,R_\Pi)$, $c_1=(Q,R)^{-1}$. Recall that
\[
   z,p=O(z), \quad a, \eta=O(z^2), \quad \eta^{(3)}=O(z^3) ~.
\]
We now expand the right hand side of the equation for $\dot p$
into terms in order of $z$:
\begin{equation} \eqlabel{p.eq2}
   \dot p = \mu
\bket{ c_\al z^\al + d_\beta z^\beta +d_1 b z + d_2 b \bar z +
P^{(4)} }
\end{equation}
The coefficients will be computed later on and their properties are
summarized in the following lemma. The proof of this lemma is just
straightforward computation and the reader can check it rather
easily.

\begin{lemma}  \label{th:p1}
We can rewrite  the equation of $p$ into the form \eqref{p.eq2} such
that the coefficients $d_1$, $d_2$ and all $c_\al$ are purely
imaginary. Moreover, $\Re d_{21} = - \Gamma + O(\la^3)$, with
\begin{equation} \label{Gamma.def}
\Gamma \equiv 2 \la^2  \bke{ Q \up^2 , \Im \frac {1}{\sA -0i - 2\ka}
\PcA \Pi Q \up^2 } \ge 0 ~.
\end{equation}
\end{lemma}

\myproof There are two parts in $p$ equation:
 $(\tilde \up, F) + (\tilde \um, \bar F)$ and
 $\bket{(\up, h) + (\um, \bar h) +c_3 a} \dot \theta $.
We first consider the second part:
\begin{align*}
&\bket{(\up, h) + (\um, \bar h) +c_3 a } \, \dot \theta
\\
&= \bket{ (\up, \zeta+ \eta^{(2)}) + (\um, \bar \zeta+ \bar \eta^{(2)} )
 +c_3 a  } \cdot
\\
& \qquad \cdot \bket{ c_2(z+\bar z)
 +  (c_0Q, \la Q^2 ( \eta^{(2)} + \bar \eta^{(2)}) + a
+ (c_0 Q, \Re F^{(2)})  } + \wt P^{(4)}_2
\\
&= \bke{(\up, \zeta)+ (\um, \bar \zeta)} \cdot c_2(z+\bar z)
\\
& \quad +
\bkt{ (\up, \eta^{(2)} ) + (\um, \bar \eta^{(2)} )
 +c_3 a  }
\cdot c_2(z+\bar z)
\\
& \qquad + \bke{(\up, \zeta)+ (\um, \bar \zeta)} \cdot
\bkt{ a + (c_0Q, \la Q^2 (\eta^{(2)} + \bar \eta^{(2)}) )
+ (c_0 Q, \Re F^{(2)})  }
\\
& \qquad + P^{(4)}_2 ~.
\end{align*}
Here$\wt P^{(4)}_2$ and  $P^{(4)}_2$ denote the remaining error terms.
We first observe that line 4, lines 5--6, and $P^{(4)}_2$ are of orders
$O(z^2)$, $O(z^3)$ and $O(z^4)$, respectively.
Hence line 4 contributes to $c_\al z^\al$, line 5--6 to $d_\beta z^\beta$
and $d_1 bz + d_2 b \bar z$, and $P^{(4)}_2$ to $P^{(4)}$.
We also observe that the coefficients in line 4 are all real.
Since there is a $-i \mu$ factor in front, their contribution to
$c_\al$ are purely imaginary.
Next we observe that the coefficients in lines 5--6 are real except
those involving $\eta^{(2)}$. The terms with lowest $\la$-order are:
\[
(\up, \eta^{(2)} ) \cdot c_2 (z+ \bar z)
\; + \; (\up, \zeta) \cdot
(c_0Q, \la Q^2 (\eta^{(2)} + \bar \eta^{(2)})) ~.
\]
 The first part is of order $\la ^3$ since $(\up, \eta^{(2)} )=
(P_\ka^A \up + O(\la), \PcA \eta^{(2)} + O(\la^2))$, with $\PcA
\eta^{(2)}= O(\la)$.  The second part has order $\la ^2$ terms,
but not on $z^2 \bar z$. We conclude that the contribution to $\Re
d_{21}$ from the second part of $p$-equation is of order $\la
^3$. Finally we observe that line 5--6 give a term $d_{1,2} b z +
d_{2,2}b \bar z$, with real $d_{1,2}=O(1)$ and $d_{2,2}=O(\la)$.
Since there is a $-i \mu$ factor in front, their contribution to
$d_1$ and $d_2$ are purely imaginary.

Now we look at the first part of the $p$ equation. The
contribution to $c_\al z^\al$ is from $F^{(2)}$:
\[
 -i \bket{(\tilde \up, z^\al \phi_\al) +
(\tilde \um, \bar z^\al \phi_\al) } ~.
\]
Clearly all coefficients of $z^\al$ are purely imaginary. Together
with the analysis of second part of $p$-equation, we know $c_\al$
are purely imaginary.

The contribution to $d_1 bz +d_2 b\bar z$ is from $\wt F^{(3)}$:
\[
 - i \bket{ ( \tilde \up, 2 \la b QR (2\zeta+\bar \zeta)
+ ( \tilde \um, 2 \la b QR (\zeta+2 \bar \zeta) }
\]
 with $\zeta =z \up + \bar z \um$. Hence all coefficients of $bz$ and
$b\bar z$ are purely imaginary. Together with the analysis of
second part of $p$ equation, we know $d_1$ and $d_2$ are purely
imaginary.

The contribution to $P^{(4)}$ is from $F^{(4)}$:
\[
P^{(4)}_1 = -i \bkt{ (\tilde \up, F^{(4)}) + (\tilde \um, \bar F^{(4)})} ~,
\]
and we have $P^{(4)} = P^{(4)}_1 -i P^{(4)}_2$.

The contribution to $d_\beta z^\beta$ is from $F^{(3)}$:
\[
    -i [(\tilde \up, F^{(3)}) + (\tilde \um, \bar F^{(3)}] ~.
\]
We only consider $d_{21}$.
A coefficient in $F^{(3)}$ has to have imaginary part
in order to have a real part contribution to $d_{21}$.
Hence only the first group of terms in $F^{(3)}$,
 $2\la Q[(\zeta +\bar \zeta) \eta^{(2)}$, has contribution to $\Re d_{21}$.
Also, when we decompose $\eta^{(2)} = z^\al \eta_{\al}$,
we can disregard $\eta_{(11)}$ since it is real.
Now, $\eta_{(20)} = O(\la)$, $\eta_{(02)}= O(\la^2)$,
$\up=O(1)$ and $\um=O(\la)$, hence the main part of $(\Re d_{21}) z^2 \bar z$
 is in
\[
-i  (\tilde \up,\, 2\la Q (\bar z \up) z^2 \eta_{(20)} )
\]
Summarizing our efforts, we have
\begin{align*}
   \Re d_{21}
&= \Im (\up, 2\la Q \up \eta_{(20)} )  + O(\la^3)
 = \Im ( 2\la Q \up^2 , \eta_{(20)} )  + O(\la^3)
\\
&= ( 2\la Q \up^2 , \Im \frac {-1}{\sA -0i - 2\ka} \PcA \Pi \phi_{20}
)
+ O(\la^3)
\\
 &= -  ( 2\la Q \up^2 , \Im \frac {1}{\sA -0i - 2\ka}
\PcA \Pi \la Q \up^2) + O(\la^3)  ~.
\end{align*}
\myendproof
      \subsection{Normal form}
From Lemma \ref{th:3-2}, we have $\Gamma >0$.

\begin{lemma}  \thlabel{thnormal}
We can rewrite the equation \eqref{p.eq2} of $p$ into a {\bf normal
form}:
\begin{equation} \eqlabel{NF}
   \dot q =  \d_{21} |q|^2 q + d_1 b q + g ~.
\end{equation}
where $q$ is a  perturbation of $p$ given in the proof. The
coefficient $\d_{21}$ satisfying the relation
\begin{equation} %
\Re \d_{21} = \Re d_{21}
\end{equation}
If we assume  the estimate  \eqref{bb} on $b$ and  $M(T)\le 2$, then
the error term  $g(t)$ given by \eqref{g.def} satisfies the  bound
\begin{equation}\eqlabel{gbound}
|g(t)| \le C_1 \tl{t}^{-3/2-\sigma}
\end{equation}
for some constant  $C_1$. Furthermore, there is a positive
constant $\sigma$ such that  $|q(t)|$ is bounded by
\begin{equation} \label{qb}
        (1-\sigma) \tl{t}^{-1/2} \le |q(t)|
        \le  (1+\sigma) \tl{t}^{-1/2} ~.
\end{equation}
and hence
\[
        (1-2\sigma) \tl{t}^{-1/2} \le |z(t)|
        \le  (1+2\sigma) \tl{t}^{-1/2} ~.
\]
\end{lemma}

\bigskip \noindent
{\bf Proof of Theorem \ref{M}} From Lemma \ref{th:p1}, we have
\[
\tl{t}^{1/2} |z(t)| \le (1+2\sigma)
\]
From Lemma \ref{etaglobal} and Lemma \ref{eta3b} we can bound
\[
\tl{t}^{3/4-\sigma} \norm{\eta(t)}_{L^4} + \tl{t}^{1+ \sigma/4}
\norm{\eta^{(3)}_{2-5}(t)}_{L^2 \loc}
\]
by $C \varepsilon^{\sigma/4}$. Since $\varepsilon$ is small,  we have proved
that $M(T)\le 3/2$ and this concludes  Theorem \ref{M}.
\myendproof

We now prove Lemma \ref{th:p1}.

\myproof We have
\[
\dot p = \mu \bkt{ c_\alpha z^\alpha + d_\beta z^\beta
+d_1 b z +d_2 b \bar z + P^{(4)} }
\]
and we want to obtain the normal form \eqref{NF}.
We will repeatedly use the following formula:
\begin{equation}     \eqlabel{eq.IBP}
\mu^m p^\al = \frac d{dt} \bke{ \frac {\mu^m p^\al}{i\ka m} }
- \frac {\mu^m p^\al} {i\ka m} f_\al(z)
\end{equation}
where, if $\al=(\al_0  \al_1)$, ($|\al|=\al_0+\al_1=2,3,4 \cdots$)
\begin{align} \label{f_al.def}
f_\al (z) & = (\al_0 + \al_1 \conj) (p^{-1} \dot p)
\\
&=(\al_0 + \al_1 \conj) z^{-1} \bkt{ c_\alpha z^\alpha + d_\beta z^\beta
+d_1 b z +d_2 b \bar z + P^{(4)} }   \nonumber
\end{align}
and $\conj$ denotes the conjugation operator. The formula is
equivalent to integration by parts.

We first remove
${c_\alpha z^\alpha}$. Let
\[
   p_1 = p - \frac {c_\al}{i\ka (1+[\al])}  \mu z^\al.
\]
Since $[\al]$ is even, $1+[\al] \not = 0$. By \eqref{eq.IBP}
\[
  \dot p_1 = \dot p - c_\alpha \mu z^\al-
\frac {c_\al}{i\ka (1+[\al])} \mu z^\al f_\al(z)
\]
Decomposing $f_\al(z)$ into two parts, we can write
\begin{align*}
d_\beta^+ z^\beta &=
-\frac {c_\al}{i\ka (1+[\al])} \mu z^\al (\al_0 + \al_1 \conj) z^{-1}
c_{\tilde \alpha} z^{\tilde \alpha}
\\
g_1 &=
-\frac {c_\al}{i\ka (1+[\al])} \mu z^\al (\al_0 + \al_1 \conj) z^{-1}
 \bkt{ d_\beta z^\beta +d_1 b z + d_2 b \bar z + P^{(4)} }
\end{align*}
and we get
\[
\dot p_1 = \d_\beta \mu z^\beta
+  d_1 \mu b z +d_2  \mu b \bar z +\mu P^{(4)} + g_1
\]
with $\d_\beta = d_\beta+d_\beta^+$. Since $d_\beta^+$ are purely
imaginary, due to that $c_\al$ are purely imaginary, we have the
key relation
\begin{equation} \eqlabel{drel}
\Re \d_\beta = \Re d_\beta
\end{equation}

Next we remove ${d_2\mu  b \bar z}$. Let
\[
  p_2 = p_1 - \frac {\mu d_2 b \bar z}{2 i\ka} ~.
\]
We have
\begin{align*}
\dot p_2 &= \dot p_1 - \mu d_2 b \bar z
- \frac {\mu^2 d_2  }{2 i\ka}(\dot b \bar p + b \Dot {\Bar p})
\\
&=\mu \d_\beta z^\beta  +d_1 \mu b z
+ (\mu P^{(4)} + g_1 + g_2) ~,
\end{align*}
where
\[
g_2= - \frac {\mu^2 d_2  }{2 i\ka}(\dot b \bar p + b \Dot {\Bar p}) ~.
\]

Now we deal with ${\d_\beta z^\beta}$ terms.
Let
\[
  p_3 = p_2 - \sum_{\beta \not = (21)} \frac{\d_\beta \mu z^\beta}
{i \ka(1+[\beta])} .
\]
Note $1+[\beta] \not = 0$ for $\beta \not = (21)$.
We have
\[
\dot p_3 = \dot p_2 - \mu\d_\beta z^\beta + g_3
=\d_{21} \mu z^2 \bar z
+ \mu d_1  b z + (\mu P^{(4)}+ g_1+g_2+g_3) ~,
\]
with
\[
g_3 =  - \sum_{\beta \not = (21)}
\frac {\d_\beta \mu^{1+[\beta]}} {i \ka(1+[\beta])}
\frac{d}{dt} (p^\beta).
\]

Finally,
since $\eta^{(3)}_1=  U^{-1} e^{-i \theta} \, e^{-i At} U \eta_0 $
is larger than $\eta^{(3)}_{2-5}$ when time $t$ is of order $1$, we
need to extract terms of order $O(z \eta^{(3)}_1)$ from $\mu P^{(4)}$.
Recall
\begin{equation} \label{eq:eta30}
\wt \eta^{(3)}_1 = e^{-i At}  \wt \eta_0 =  e^{-i At} U \eta_0 , \qquad
\eta^{(3)}_1 = U^{-1} e^{-i \theta} \, \wt \eta^{(3)}_1
= U^{-1} e^{-i \theta} \, e^{-i At} U \eta_0 ~.
\end{equation}
Also recall \eqref{p.eq1} and \eqref{p.eq2}.
In $F$ we have a term
$2 \la Q((\zeta + \bar \zeta) \eta^{(3)}_1  + \zeta \wbar {\eta^{(3)}_1} )$,
in $h$ a term $\eta^{(3)}_1$ itself.
Hence in $\mu P^{(4)}$, terms of order $O(z \eta_0)$ are exactly
\begin{align*}
 P_{z \eta_0} \equiv
& -i\mu(\tilde \up,
2 \la Q((\zeta +\bar \zeta) \eta^{(3)}_1  + \zeta \bar \eta^{(3)}_1 ))
 -i \mu \conj (\tilde \um,
2 \la Q((\zeta +\bar \zeta) \eta^{(3)}_1  + \zeta \bar \eta^{(3)}_1 ))
\\
& \quad
-i \mu
\bket{(\up, \eta^{(3)}_1) + \conj (\um, \eta^{(3)}_1) } c_2(z+\bar z)  ~.
\end{align*}
Clearly these terms can be summed in the form
\begin{equation} \label{eq:z-eta0}
\mu (z \phi + \bar z \phi, \eta^{(3)}_1 )
+ \conj \mu^{-1}(z \phi + \bar z \phi, \eta^{(3)}_1 )
\end{equation}
Here $\phi$ stand for different local smooth functions.
Moreover, if we write  $U^{-1}=U_3 + U_4 \conj$ with
 $U_3, U_4 = \tfrac 12 (BA^{-1/2} \pm B^{-1}A^{1/2})$, both self-adjoint,
by \eqref{eq:eta30} and
\[
 U^{-1}(z f + \bar z g)= z(U_3 f + U_4 \bar g) + \bar z (U_3 g + U_4 \bar f)
\]
(cf. \eqref{eq:Uzf}), the above is equal to
\[
 \mu(z \phi + \bar z \phi,  e^{-i \theta}  e^{-i At} U \eta_0)
+ \conj  \mu^{-1}(z \phi + \bar z \phi, e^{-i \theta}  e^{-i At} U \eta_0) .
\]
The first term can be written as
\[
(\phi, e^{-i (A-0i-2\ka)t} e^{-i \theta} \bar p  U \eta_0)
+(\phi, e^{-i A t} e^{-i \theta}  p  U \eta_0)
=\frac d{dt} (p_{4,1}) + g_{4,1}
\]
with
\begin{align*}
p_{4,1}&=
i \bke{ \frac{\phi}{A -0i-2\ka}\, , \,
e^{-i (A-0i-2\ka)t} e^{-i \theta} \bar p  U \eta_0) }
+ i \bke{ \frac{\phi}{A }\, , \,
e^{-i At} e^{-i \theta}  p  U \eta_0) }
\\
g_{4,1}&=
-i \bke{ \frac{\phi}{A -0i-2\ka}\, , \,
e^{-i (A-0i-2\ka)t} \frac d{dt} (e^{-i \theta} \bar p)  U \eta_0) }
- i \bke{ \frac{\phi}{A}\, , \,
e^{-i At} \frac d{dt} (e^{-i \theta}  p)  U \eta_0) }
\end{align*}
We can write the second term similarly as
\[
\conj (z \phi + \bar z \phi, e^{-i \theta}  e^{-i At} U \eta_0) =
\frac d{dt} (\conj p_{4,2}) + \conj g_{4,2}
\]
Thus we have
\[
 P_{z \eta_0} = \dot p_4 + g_4, \qquad
p_4 := p_{4,1}+\conj p_{4,2}, \qquad
g_4 := (g_{4,1}+ \conj g_{4,2}) ~.
\]
Observe that $\frac{\phi}{A -0i-2\ka}$ is not in $L^2$, hence we expect
slower decay for $p_{4}$ and $g_{4}$. Specifically,
\[
|p_{4}| \le \norm{\eta_0} \tl{t}^{-1/2}  \bkA{t}^{-1-\sigma} , \qquad
|q_{4}| \le \norm{\eta_0} \tl{t}^{-1}  \bkA{t}^{-1-\sigma} ~.
\]

Now we let
\[
 q= p_3 -  p_{4} ~.
\]
We have
\begin{align*}
\dot q
&= \d_{21} \mu z^2 \bar z
+ \mu d_1  b z + (\mu P^{(4)}+ g_1+g_2+g_3)
-  P_{z \eta_0} + g_{4}
\\
&= \d_{21} |p|^2p
+ d_1  b p + ((\mu P^{(4)}-P_{z \eta_0})+ g_1+g_2+g_3+g_4)
\\
&= \d_{21} |q|^2 q  + d_1  b q + g ~,
\end{align*}
where
\begin{equation}         \label{g.def}
g = ((\mu P^{(4)}-P_{z \eta_0}) + g_1+g_2+g_3+g_4) +
\d_{21}  (|p|^2p -|q|^2 q)  + d_1 b (p-q) ~.
\end{equation}
Hence we have arrived at the normal form.
(Note that, although we kept $\dot p$ and $\dot b$ in the definition
of $g_2$, it should be replaced by the corresponding equations \eqref{p.eq2}
and \eqref{b.eq0}.)
Also note we have
\begin{equation}         \label{q.def}
q = p - \frac {c_\al}{i\ka (1+[\al])}  \mu z^\al
- \frac {\mu d_2 b \bar z}{2 i\ka}
- \sum_{\beta \not = (21)} \frac{\d_\beta \mu z^\beta}
{i \ka (1+[\beta])} -p_4~.
\end{equation}

Finally, we can check the size of $g$ satisfying the bound
\eqref{gbound}. To conclude this lemma, it remains to prove the estimate
on $q$. But this follows from the next lemma. \myendproof

\subsection{Decay and continuity estimates}

In this subsection we present some calculus lemmas which deal with
the decay of $q(t)$ and its continuity on the error term $g(t)$.
We will write
\[
q(t)= \rho(t) \, e^{i \om (t)} ~,
\]
where $\rho= |q|$ and $\om$ is the phase of $q$. Recall
\[
 \tl{t} = \e^{-2} + 2 \Gamma t, \quad \tl{t} \sim \max \bket{ \e^{-2}, t }.
\]

Before we proceed with proof, we first explain some simple facts
of an ordinary differential equation.

\textbf{Example}. \quad
 We consider real functions $r(t) > 0$ which solve
\[
   \dot r(t) = - \Gamma r(t) ^3   - \e \,(1+t)^{-3} ~.
\]
We have the following facts.

   a. All solutions $r(t)$ satisfy
\[
      r(t) \le (C + 2 \Gamma t)^{-1/2}
\quad \text{ with } r(0) = C^{-1/2} ~.
\]

   b. There is a number $r_1$ such that if $r(0) > r_1$,
then $ r(t) \sim (C + 2 \Gamma t)^{-1/2}$.

   c. There is a unique global solution $r_0(t)$ such that
\[
      r_0(t) \sim t^{-2} \quad \text{as } t \to \infty .
\]

   d. If $r(0) < r_0(0)$, then $r(t) = 0$ in finite time.

   e. If $r(0) > r_0(0)$, then
\[
      \int _0^\infty r(s)^2 ds = \infty.
\]

\begin{lemma} \thlabel{1.1}
Let $\e_0 >0 $ be small. Suppose $\rho(t)$ satisfy
\[
\dot \rho = -  \rho ^3  + \wt g(t) ~,
\]
where $|\wt g(t)| < \e_0 \bkA{t}^{-3/2-\sigma}$, $\sigma >0$. We can
find $0 < \rho_1 \le \rho_2 $ ~, (depending on $\e_0$), such that

\noindent
(a) If $\rho_1 < \rho(0) < \rho_2$, then $|\rho(t)| \sim C
\bkA{t}^{-1/2}$.

\noindent
(b) If $0 < \rho(0) < \rho_2$, then $|\rho(t)| \le C
\bkA{t}^{-1/2}$.

\end{lemma}

The above example shows that the conclusion of (a) cannot be
expected to hold if $\rho(0)$ is too small. This is the main
reason that there are two type of  asymptotic behavior: the
resonance dominated  solutions given by case (a) and
the radiation dominated ones by (b). The proof of this lemma is elementary
and similar to the next lemma and thus we omit it.

\begin{lemma}  \thlabel{th:p2}
 Let $\Gamma>0$, $\sigma>0$, and $C_1>0$ be given constants,
independent of $\e_0$. Suppose  a positive function $\rho(t)$
satisfies
\begin{equation}  \label{eq.ap.1}
   \dot \rho = - \Gamma \rho ^3 + \wt g(t),  ~,
\end{equation}
(a)  \quad Suppose $\rho(0) = \e$
with $0<\e \le \e_0$ and
\[
   |\wt g(t)| \le C_1 \tl{t}^{-3/2-\sigma}, \quad
   \tl{t} \equiv \e^{-2} + 2\Gamma t ~.
\]
Then there is a constant $m=m(\e_0)>1$ such that
\[
   m^{-1} \tl{t}^{-1/2}\le  \rho(t) \le m\tl{t}^{-1/2} ~.
\]
Moreover, $m(\e_0) \to 1^+$ as $\e_0 \to 0^+$.

\noindent
(b) \quad Suppose that $\rho(0) \le \e_0$ and
\[
   |\wt g(t)| \le C_1 \tl{t}_0^{-3/2-\sigma}, \quad
   \tl{t}_0 \equiv \e_0^{-2} + 2\Gamma t ~.
\]
Then we have for some constant $C$
\[
  \rho(t) \le C \tl{t}_0^{-1/2} ~.
\]

\end{lemma}

\myproof
We first prove part (a).
Let $\rho_+ =m \tl{t}^{-1/2}$ and $\rho_-=m^{-1} \tl{t}^{-1/2}$,
with $m>1$ to be determined. We have $\rho_+(0) > \rho(0) > \rho_-(0)$.
Moreover,
\[
   \dot \rho_+ = - \Gamma m^{-2} \rho_+^3
   =  - \Gamma \rho_+^3 +  \Gamma (1-m^{-2}) \rho_+^3
  \ge - \Gamma \rho_+^3 + \wt g ~,
\]
if
\[
 \Gamma (1-m^{-2}) m^3 \tl{t}^{-3/2} \ge C_1 \tl{t}^{-3/2-\sigma}.
\]
Also
\[
   \dot \rho_- = - \Gamma m^{2} \rho_-^3
   =  - \Gamma \rho_-^3 -  \Gamma (m^{2}-1) \rho_-^3
  \le - \Gamma \rho_-^3 + \wt g ~,
\]
if
\[
  \Gamma (m^{2}-1) m^{-3} \tl{t}^{-3/2} \ge C_1 \tl{t}^{-3/2-\sigma}.
\]
Since $\tl{t}^{-\sigma} \le \e^{2\sigma}$,
both inequalities hold if
\[
   \Gamma (1-m^{-2}) m^3 ,\, \Gamma (m^{2}-1) m^{-3} \ge C_1 \, \e_0^{2\sigma} ~.
\]
This is true for
$m>1$ arbitrarily close to $1$ by choosing $\e_0$ sufficiently small.
By comparison, we have $\rho_-(t) < \rho(t) < \rho_+(t)$ for all $t$.

To prove part (b), note we can still define $\rho_+$ as a comparison function.
Since $\rho_+(0) \ge \rho(0)$, the above argument still holds.
\myendproof

The following lemma estimates the continuity of $\rho$ in the error
term $\wt g(t)$. Here we consider the case $\rho(0)=\e>0$ only.
\begin{lemma} \label{th:p3}
 Let $\Gamma>0$, $\sigma>0$, and $C_1>0$ be given constants,
independent of $\e_0$. Suppose  two positive function $\rho_1(t)$ and
$\rho_2(t)$
satisfy
\begin{equation*}
   \dot \rho_i = - \Gamma \rho_i ^3 + \wt g_i(t) \quad (i = 1,2) ~, \qquad
  \rho_1(0) = \rho_2(0)=\e  ~,
\end{equation*}
with $0<\e \le \e_0$ and
\[
   |\wt g_i(t)| \le C_1 \tl{t}^{-3/2-\sigma}, \quad
|\diff \wt g(t)| \le \diff_0 \tl{t}^{-3/2-\sigma} ,
\]
where $\diff \wt g = \wt g_2 - \wt g_1$ and $\diff_0 \ge 0$ is a small number.
Then we have
\[
|\rho_2(t) - \rho_1 (t)|
\le  \diff_0 \e^\sigma (\Gamma \sigma)^{-1} \tl{t}^{-(1+\sigma)/2} ~.
\]
Note that the decay rate is improved.
\end{lemma}

\myproof
Let $r=\diff \rho = \rho _2 - \rho_1$.
Then $r$ satisfies
\[
   \dot r = - G r + \diff \wt g, \quad
   r(0)=0, \quad
   G= \Gamma (3 \rho^2 + 3\rho r + r^2) ~.
\]
We have
\begin{equation} \label{eq.ap.2}
r(t) = \int_0^t e^{-\int_s^t G(\tau) d\tau} \diff \wt g(s) \, ds~.
\end{equation}
Note $|r(t)| < \rho_+(t) - \rho_-(t)$, hence
\[
   G(t) \ge \Gamma [ 3\rho_-^2 - 3\rho_+ (\rho_+ -\rho_-)]
   = 3 \Gamma [m^{-2}-m(m-m^{-1})] \tl{t}^{-1}
   \ge 2\Gamma \tl{t}^{-1}
\]
if $m$ is sufficiently close to $1$.
Hence
\[
\int_s^t G(\tau) d\tau
\ge \int_s^t 2\Gamma \tl{\tau}^{-1} d\tau
= 2 \Gamma \int_s^t (2\Gamma)^{-1}\bkt{\ln \tl{\tau} }_{\tau=s}^t
\ge \frac {1+\sigma}2 \ \bke{\ln  \tl{t} -\ln \tl{s} }
\]
and
\[
e^{ - \int_s^t G(\tau) d\tau } \le
 \tl{t} ^{-(1+\sigma)/2} \cdot  \tl{s} ^{((1+\sigma)/2} ~.
\]
By \eqref{eq.ap.2} we have
\begin{align*}
|r(t)| &\le \int_0^t
 \tl{t} ^{-(1+\sigma)/2}   \tl{s} ^{(1+\sigma)/2}
\diff_0 \tl{s}^{-3/2-\sigma} \, ds
\\
&=\diff_0{ \tl{t}  }^{-(1+\sigma)/2}
\int_0^t  \tl{s}^{-1-\sigma/2} \, ds
\le  \diff_0 \e^\sigma (\Gamma \sigma)^{-1} \tl{t}^{-(1+\sigma)/2} ~.
\end{align*}
\myendproof

\section{Change of mass}

For given $Q_E$ and $h_0$, we want to find an $a(t)=a(E,h_0;t)$ which
satisfies
\[
  a(\infty) = 0, \quad  \dot a = (c_1Q, \Im F(k)), \quad
  |a(t)| \le C  \la \tl{t}^{-1}   ~.
\]

\subsection{The main oscillatory part of $a(t)$}

We use the following equivalent integral equation for $a(t)$:
\[
a(t) =  \int_\infty^ t (c_1Q, \Im F(k)) \, ds ~.
\]
Note $ k =aR+h= aR+\zeta + \eta$.
We want to perform several integrations by parts so that we get the
form
\[
a(t) = O(t^{-1}) + \int_\infty^t \, O( t^{-2}) \, ds = O(t^{-1}) ~.
\]

As in Section 4, we decompose $\eta$ and $a$ as
\[
\eta=\eta^{(2)} + \eta^{(3)} , \quad
a = a_{20}(z^2+\bar z^2) +b.
\]

Recall $k=aR+\zeta + \eta$ and
\[
  \Im F(k)=\Im \bket{ \la Q h^2 + 2 \la QR ah + \la (aR+h)(aR+\bar h)h } ~.
\]
We also denote
\[
\mu^{-1} \dot p =
c_\al z^\al + d_\beta z^\beta + d_1 b z + d_2 b \bar z + P^{(4)}
= c_\al z^\al + P^{(3,4)}
\]

Using the decomposition of $F(k)$, we have
\[
(c_1Q, \Im F) =A^{(2)} +  A^{(za)}+ A^{(z\eta)} +A^{(z^3)} + A^{(4)}+ A^{(5)}
~,
\]
where
\begin{align*}
A^{(2)} &= \bke{c_1 Q, \, \la Q \Im \zeta^2 }
\\
A^{(za)}   &= \bke{c_1Q, \, \Im 2\la  Q R a \zeta}
\\
A^{(z\eta)}&= \bke{c_1 Q, \, \Im 2\la Q \zeta \eta}
\\
A^{(z^3)}  &= \bke{c_1 Q, \,  \Im \la |\zeta|^2 \zeta }
\\
A^{(4)}+ A^{(5)} &= \bke{c_1 Q, \,  \Im \bket{
\la Q \eta^2 + 2 \la Q R a \eta + \la \bkt{ |k|^2k-|\zeta|^2\zeta} } } ~,
\\
A^{(4)} &=\bke{c_1 Q, \, \Im  \bket{\la Q  (\eta^{(2)})^2+ 2 \la  QR
a \eta^{(2)}
+ \la \bkt{  2|\zeta|^2 (aR+\eta^{(2)})+\zeta^2 (aR+\wbar {\eta^{(2)}}) } }}
\\
A^{(5)}&=\bke{c_1 Q, \, \Im  \bket{\la Q  [2\eta^{(2)}\eta^{(3)}+(\eta^{(3)})^2]
+ 2 \la  QR a \eta^{(3)} }}
\\
&\quad +\bke{c_1 Q, \Im \bket{\la \bkt{ 2|\zeta|^2 \eta^{(3)}
+\zeta^2 \wbar {\eta^{(3)}}+ \ell^2 \bar \zeta + 2 |\ell|^2 \zeta
+ \ell^2 \bar \ell}}} ~, \quad
\ell = aR+ \eta ~.
\end{align*}
Since $\zeta = z \up + \bar z \um$, hence
\begin{equation*}
\Im \zeta = \Im z \, ( \up - \um), \quad
\Im \zeta^2= (\Im z^2)\, (\up^2 - \um^2) ~.
\end{equation*}
Therefore
\begin{align*}
A^{(2)} &= C_1 \,\Im z^2, \qquad
C_1 = \bke{ c_1Q, \, \la Q (\up^2 - \um^2)} = O(\la^2) ~,
\\
 A^{(za)} &= C_2\, a \Im z ,\qquad
C_2 = (c_1Q, 2\la QR (\up - \um)) = O(\la) ~.
\end{align*}

First we integrate $A^{(2)}$:
\begin{align*}
\int_\infty^t A^{(2)} \, ds
&=  C_1 \Im \int_\infty^t  \,  z^2 \, ds
= C_1 \Im \int_\infty^t  \,  \mu^{-2} p^2 \, ds
\\
&=  C_1 \Im \frac 1{-2i\ka}
\bket{ { \mu^{-2} p^2 }
- \int_\infty^t  \,  \mu^{-2} 2p \dot p \, ds }
\\
&=  \frac {C_1}{4\ka} \, 2 \Re \bket{  z^2  -  2 \int_\infty^t
\, z \mu^{-1} \dot p \, ds }
\\
&=  a_{20} (z^2 + \bar z^2)
- 4 a_{20} \Re \int_\infty^t  \,   z
\bkt{c_\al z^\al + d_\beta z^\beta
+ d_1 b z + d_2 b \bar z +
P^{(4)} } \, ds
\\
&=  a_{20} (z^2 + \bar z^2)
+\int_\infty^t  \, A_{2,3}  +  A_{2,4} +  A_{2,5} \, ds ~,
\end{align*}
where
\begin{align*}
a_{20} &= \frac {C_1}{4\ka}
= \frac \la {4 \ka } \bke{ c_1Q, \, Q (\up^2 - \um^2)} = O(\la^2) ~,
\\
A_{2,3} &=- 4 a_{20}\Re  z  c_\al z^\al ~,
\\
A_{2,4} &=
 -4 a_{20} \Re  \bkt{ d_\beta zz^\beta + d_1 b z^2 }
\\
A_{2,5} &= -4 a_{20} \Re z  P^{(4)}  ~,
\end{align*}
Note here we have used $\Re z d_2 b \bar z =0$.

As mentioned in Section 4, the term $a_{20} (z^2 + \bar z^2)$
is the main oscillatory part of $a$. We denote the remaining part as $b$
and hence
\[
a = a_{20} (z^2 + \bar z^2) + b ~.
\]
Although $b$ is of the same order $t^{-1}$, its oscillation is slower
since the right side of its equation
\[
\dot b =  (c_1 Q, \Im (F-F^{(2)})) - 4 a_{20} \Re z \mu^{-1} \dot p
\]
which we obtained from the above computation,
is of order $O(z^3)$, (cf.~\eqref{b.eq0}).

Another reason for using $b(t)$ is that it is better suited than $a(t)$
for the iteration scheme in Section 8. We have
\[
  \diff a(t) \sim \diff_0 \tl{t} ^{-1} \log \tl{t}, \quad
  \diff b(t) \sim \diff_0 \tl{t} ^{-1} ~.
\]
This difference arizes because it is harder to estimate the
continuity of the phase.

\subsection{Integration of $O(z^3)$ terms}

we next integrate $ A^{(za)}$:
\begin{align*}
\int_\infty^t  A^{(za)} & =
\int_\infty^t C_2 a \Im z
\\
&= C_2\Im \frac 1{-i \ka}
\bket{ az - \int_\infty^t \mu^{-1} \frac d{dt}(a p) }
\\
&=  \frac {C_2}{ 2\ka} \, 2\Re
\bket{ az - \int_\infty^t z (c_1Q, \Im F)
 + a(c_\al z^\al + d_\beta z^\beta + d_1 b z + d_2 b \bar z + P^{(4)}) }
\\
& = c_{za} a(z+\bar z) + \int_\infty^t A_{za,3} + A_{za,4} +A_{za,5}\, ds ~,
\end{align*}
where
\begin{align*}
c_{za} &=C_2 \, /(2\ka) = (c_1Q, 2\la QR (\up - \um)) \, /(2\ka) = O(\la)
\\
A_{za,3} &=- c_{za} (z+\bar z)\, (c_1Q, \Im F^{(2)})
\\
A_{za,4} &= -c_{za} (z+\bar z)\, (c_1Q, \Im (
2 \la QRa \zeta + 2 \la Q \zeta \eta^{(2)} + \la |\zeta|^2 \zeta ) )
 -2 c_{za} a \Re c_ \al z^\al
\\
A_{za,5} & =
-c_{za}  (z+\bar z)\,(c_1Q, \Im  F^{(4)} )
 -2 c_{za} a \Re ( d_\beta z^\beta + d_1 b z + d_2 b \bar z + P^{(4)})
\end{align*}

We now integrate $A^{(z\eta)}$.
Recall $U=U_+ + U_- \conj$, see \eqref{U.dec}.
We will also write $U^{-1}=U_3+U_4 \conj$,
with $U_3, U_4 = \tfrac 12 (BA^{-1/2} \pm B^{-1}A^{1/2})$,
both self-adjoint.
We will also use the following formulas:
\begin{equation} \label{eq:fCg}
\Re \int dx f (\conj g) = \Re \int dx  (\conj f)g , \quad
\Im \int dx f (\conj g) = -\Im \int dx  (\conj f)g ,
\end{equation}
and (using  $U=U_+ + U_- \conj$)
\begin{equation} \label{eq:Uzf}
 U(z f + \bar z g)= z(U_+ f + U_- \bar g) + \bar z (U_+ g + U_- \bar f)
\end{equation}

Recall $\eta = U^{-1} \eta ^\diamond=U^{-1}e^{-i\theta} \tilde \eta$.
Thus, by \eqref{eq:fCg},
\begin{align*}
 A^{(z\eta)}
&= (c_1 Q, \Im 2 \la Q \zeta \eta )
=  \Im \int dx \,  2 c_1 \la Q^2  \zeta \eta
\\
&=  \Im \int dx \,  2 c_1 \la Q^2  (z \up+\bar z \um ) \,
(U_3 + U_4 \conj)  \eta ^\diamond
\\
&= \Im \int dx \, [(U_3 - U_4 \conj) (z \phi_1 + \bar z \phi_2)] \,
\eta ^\diamond
\end{align*}
where
\[
\phi_1= 2 c_1 \la Q^2 \up , \quad
\phi_1= 2 c_1 \la Q^2 \um .
\]
Hence, by \eqref{eq:Uzf}, (think $U_+=U_3$ and $U_-=- U_4$)
\[
A^{(z\eta)}= \Im \int dx \,  (z \phi_3 + \bar z \phi_4) \eta^\diamond ~,
\]
with
\[
\phi_3 = U_3 \phi_1 - U_4 \phi_2 ,\quad
\phi_4 = U_3 \phi_2 - U_4 \phi_1 .
\]

We rewrite
\[
 \eta^\diamond (\tau)=  e^{-i \theta} \tilde \eta =
 e^{- i\theta} e^{- i\sA \tau} f(\tau)
\]
where $f = e^{i\sA \tau } \tilde \eta = \tilde \eta_0 + \int_0^\tau
e^{is \sA} \cdots ds$.
The reason we work with $f$, instead of $\tilde \eta$, is that
those terms in $\partial_{\tau} f$ of same order ($z^2$) are explicit
and do not involve differentiation, compared to those in
$\partial_{\tau} \eta$.
Now we have
\begin{align*}
\int _\infty^t A^{(z\eta)}
&= \Im \int _\infty^t (\phi_3, z \eta^\diamond)
+ (\phi_4, \bar z \eta^\diamond) \, d \tau
\\
&=
\Im  \int _\infty^t \bke{  \phi_3 ,  e^{-i \tau(\sA + \ka)}
\bke{ p e^{-i \theta} f }  }\, d \tau
+
\Im  \int _\infty^t \bke{  \phi_4 ,  e^{-i \tau(\sA - \ka)}
\bke{ \bar p e^{-i \theta} f }  }\, d \tau
\end{align*}
We deal with the first integral, which is equal to
(note $f \in \Hc(A)$)
\begin{align*}
&= \Im
\bke{  \phi_3 ,  \frac 1{ -i(\sA + \ka)} e^{-i \tau(\sA + \ka)}
\bke{ p e^{-i \theta} f }  }
\\
&\quad - \Im  \int _\infty^t \bke{  \phi_3 ,  \frac 1{ -i(\sA + \ka)}
e^{-i \tau(\sA + \ka)}
\frac d {d \tau} \bke{ p e^{-i \theta} f }  }\, d \tau
\\
&=\Re \bke{   \frac 1{ \sA + \ka} \PcA \Pi \phi_3 ,  z e^{-i \theta}
\tilde \eta  }
\\
& \quad -
\Re  \int _\infty^t \bke{\frac 1{ \sA + \ka}  \PcA \Pi \phi_3 ,
 \bket{ (\dot p/p - i \dot \theta )z e^{-i \theta}
\tilde \eta + z e^{-i \theta}
\PcA \bkt{ e^{i \theta} U \Pi F^{\sharp}
- e^{i \theta} [U,i] \dot \theta \eta } } } \, d \tau
\\
&=\Re \bke{ \phi_5 ,  z  \eta ^\diamond }
 -
\Re  \int _\infty^t \bke{  \phi_5 ,
 \bket{ (\dot p/p - i \dot \theta )z \eta ^\diamond
+ z \PcA \bkt{  U \Pi F^{\sharp} - [U,i] \dot \theta \eta } } } \, d \tau
\end{align*}
where
\[
   \phi_5 = \frac 1{ \sA + \ka}  \PcA \Pi \phi_3 ~.
\]
We are careful in adding $\PcA \Pi$ so that $\phi_5$ makes sense.
We can do so since $f \in \Hc(A)$.

Similarly, the second integral is equal to
\[
=\Re \bke{ \phi_6 ,  \bar z  \eta ^\diamond }
-\Re  \int _\infty^t \bke{  \phi_6 ,
 \bket{ (\wbar{\dot p/p} - i \dot \theta )\bar z \eta ^\diamond
+ \bar z \PcA \bkt{  U \Pi F^{\sharp} - [U,i] \dot \theta \eta } } } \, d
\tau
\]
with
\[
  \phi_6 = \frac 1{ \sA - \ka}  \PcA \Pi \phi_4 ~.
\]

We can rewrite their leading terms in the form
\begin{align*}
\Re ( \phi_5, z \eta^\diamond) + \Re ( \phi_6, \bar z \eta^\diamond)
&=\Re \int dx \, (z\phi_5 + \bar z \phi_6) \, (U_+ + U_- \conj) \eta
\\
&= \Re \int dx \, [(U_+ + U_- \conj) \, (z\phi_5 + \bar z \phi_6)] \,
\eta
\\
&=\Re \int dx \, (z\phi_8+ \bar z \phi_7) \eta.
\\
&= \Re \,(z\phi_7+ \bar z \phi_8, \eta).
\end{align*}
where we have used \eqref{eq:fCg} and \eqref{eq:Uzf}, with
\[
\phi_8=U_+ \phi_5 + U_- \phi_6 ,\quad
\phi_7=U_+ \phi_6 + U_- \phi_5 ~.
\]

The remaining integral has the integrand
\begin{align*}
&=-\Re \bke{  \phi_5 ,
 \bket{ ({\dot p/p} - i \dot \theta ) z U \eta
+  z \PcA \bkt{  U \Pi F^{\sharp} - [U,i] \dot \theta \eta } } }
\\
& \quad -\Re \bke{  \phi_6 ,
 \bket{ (\wbar{\dot p/p} - i \dot \theta )\bar z U \eta
+ \bar z \PcA \bkt{  U \Pi F^{\sharp} - [U,i] \dot \theta \eta } } }
\\
&= - \Re \int dx \,
\bke{ \mu^{-1}\dot p \phi_5 + \mu \wbar{\dot p} \phi_6
- i \dot \theta (z \phi_5+\bar z \phi_6) } U \eta
\\
&\qquad \qquad
+ (z \phi_5 + \bar z \phi_6)\,
 \bkt{  U \Pi F^{\sharp } - [U,i] \dot \theta \eta }
\\
&=- \Re \int dx \,
\bke{ \mu^{-1}\dot p \phi_5 + \mu \wbar{\dot p} \phi_6 } U \eta
+ (z \phi_5 + \bar z \phi_6)\,
 \bkt{  U \Pi F^{\sharp } -  \dot \theta U i\eta }
\\
&=A_{z\eta, 3} + A_{z\eta, 4} + A_{z\eta, 5}
\end{align*}
where $A_{z\eta, 3}$ is from
 $F^\sharp=i z^\al \phi_\al^{\sharp} + F^{\sharp \sharp}$, defined in
\eqref{eq:Fsharp},
\[
F^\sharp = - i \dot \theta \zeta - i F(k) - [(c_1Q,\Im F) + ia
\dot \theta ] R_\Pi :=  i \, z^\al \phi ^\sharp_\al + F^{\sharp
\sharp(3)} + F^{\sharp \sharp (4)}.
\]
\begin{align*}
 A_{z\eta, 3} &=
- \Re \int dx \, (z \phi_5 + \bar z \phi_6)\,
   U \Pi i z^\al \phi_\al^{\sharp}
\\
A_{z\eta, 4} &=
 - \Re \int dx \,
\bke{ \mu^{-1}c_\al z^\al \phi_5 - \mu c_\al \bar z^\al \phi_6
} U \eta^{(2)}
\\
& \quad \qquad
+ (z \phi_5 + \bar z \phi_6)\,
 \bkt{  U \Pi F^{\sharp \sharp(3)} -  c_2(z+\bar z)U i\eta^{(2)}  }
\\
A_{z\eta, 5} &=- \Re \int dx \,
\bke{ \mu^{-1}c_\al z^\al \phi_5 - \mu c_\al \bar z^\al \phi_6 } U \eta^{(3)}
+ \bke{ \mu^{-1}P^{(3,4)} \phi_5 + \mu \wbar{P^{(3,4)}} \phi_6 } U \eta
\\
& \qquad \qquad
+ (z \phi_5 + \bar z \phi_6)\,
\bkt{  U \Pi F^{\sharp \sharp(4)} -
\bke{c_2(z+\bar z)U i\eta^{(3)} + F_\theta U i \eta }  }
\end{align*}
We first observe that, although $e^{i \theta}$ appears in the computation, it
does not show up in the final results.
Also, $A_{z\eta, 3}$ is a sum of monomials $c z^\beta$ with $|\beta|=3$.
There is no $a$ or $\eta$ in $A_{z\eta, 3}$.

Summarizing, we have
\begin{align*}
a& = a_{20}  \bke{z^2 + \bar z^2} +  c_{za} a(z+\bar z)
+ \Re (z\phi_7+ \bar z \phi_8, \eta)
\\
& \quad
+ \int_\infty^t
\bkt{ A^{(z^3)} + A_{2,3}  + A_{(za,3)} + A_{(z\eta,3)}  }
+ \bkt{ A^{(4)} + A_{2,4} + A_{(za,4)} + A_{(z\eta,4)} }
\\
& \qquad \quad
+  \bkt{ A^{(5)} + A_{2,5} + A_{(za,5)} + A_{(z\eta,5)}  } \, ds
\end{align*}

We now write
\begin{align*}
A_{\beta} z^\beta &=
A^{(z^3)} + A_{2,3}  + A_{(za,3)} + A_{(z\eta,3)}
\\
&=
\bke{c_1 Q, \,  \Im \la |\zeta|^2 \zeta }
- 4 a_{20} \Re z c_\al z ^\al
- c_{za} (z+\bar z)\, (c_1Q, \Im F^{(2)})
\\
& \quad
- \Re \int dx \, (z \phi_5 + \bar z \phi_6)\,
   U \Pi i z^\al \phi_\al^{\sharp}
\end{align*}
  From integration by parts,  we have
\begin{align*}
 \int_\infty^t A_{\beta} z^\beta  \, ds
&= a_\beta z^\beta
- \int_\infty^t a_\beta z^\beta f_\beta(z) \, ds , \qquad
a_\beta = \frac {A_{\beta}}{ i[\beta]\ka} ~,
\\
&= a_\beta z^\beta + \int_\infty^t A_{3,4}+A_{3,5} \, ds
\end{align*}
where $f_\beta$ is defined in \eqref{f_al.def},
and, ($\beta=(\beta_0, \beta_1)$)
\begin{align*}
A_{3,4} &=- a_\beta z^\beta (\beta_0 + \beta_1 \conj)
z^{-1}  c_\al z^\al
\\
A_{3,5}&=- a_\beta z^\beta (\beta_0 + \beta_1 \conj)
z^{-1} P^{(3,4)}
\end{align*}
Note $[\beta]\not= 0$. Also note $\wbar{a_\beta} = a_{\bar \beta}$.

Let $a^{(4)}+a^{(5)}$ denote the total of the remaining integrals.
We have
\begin{align}
a(t) &=  a_{20} (z^2+ \bar z^2) + a_\beta z^\beta + c_{za} a(z+\bar z)
+ \Re (z\phi_7+ \bar z \phi_8, \eta) + a^{(4)}+a^{(5)}
\eqlabel{at} \\
a^{(4)} &= \int_\infty ^t
  A^{(4)} + A_{2,4} + A_{(za,4)} + A_{(z\eta,4)} + A_{3,4}
\, ds ~, \eqlabel{a4t}
\\
a^{(5)} &= \int_\infty ^t
  A^{(5)} + A_{2,5} + A_{(za,5)} + A_{(z\eta,5)}+ A_{3,5}
\, ds ~. \nonumber
\end{align}

Now we consider the integrand of $a^{(4)}$. They are
terms of order $z^4$.
\begin{align*}
A^{(4)} &=\bke{c_1 Q, \, \Im  \bket{\la Q  (\eta^{(2)})^2+ 2 \la  QR
a \eta^{(2)}
+ \la \bkt{  2|\zeta|^2 (aR+\eta^{(2)})+\zeta^2 (aR+\wbar {\eta^{(2)}}) } }}
\\
A_{2,4} &=
 -4 a_{20} \Re  \bkt{ d_\beta zz^\beta + d_1 b z^2 }
\\
A_{za,4} &= -c_{za} (z+\bar z)\, (c_1Q, \Im (
2 \la QRa \zeta + 2 \la Q \zeta \eta^{(2)} + \la |\zeta|^2 \zeta ) )
 -2 c_{za} a \Re c_ \al z^\al
\\
A_{z\eta, 4} &=
 - \Re \int dx \,
\bke{ \mu^{-1}c_\al z^\al \phi_5 - \mu c_\al \bar z^\al \phi_6
} U \eta^{(2)}
\\
& \quad \qquad
+ (z \phi_5 + \bar z \phi_6)\,
 \bkt{  U \Pi F^{\sharp \sharp(3)} -  c_2(z+\bar z)U i\eta^{(2)}  }
\\
A_{3,4} &=- a_\beta z^\beta (\beta_0 + \beta_1 \conj)
z^{-1}  c_\al z^\al
\end{align*}

We substitute $a=a_{20}(z^2 + \bar z^2) + b$ and $\eta^{(2)}=z^\al \eta_\al$
in the above integrands.
Note that although terms of order $z^4$ have the following forms
\[
 z^4,~  a^2,~ \eta^2,~ z^2 a,~ z^2 \eta,~ a \eta ,~ z \eta^{(3)} ~.
\]
some of them do not occur in the above integrands.
We have the following Lemma.

\begin{lemma}  \label{th:7-1}
After the substitution $a=a_{20}(z^2 + \bar z^2) + b$ and
$\eta^{(2)}=z^\al \eta_\al$, the integrand of $a^{(4)}$, \eqref{a4t},
can be summed into the form
\[
B_{22} |z|^4 + \Re \bket{A_{40} z^4 + A_{31} z^3 \bar z + A_{b2} b z^2 } ~.
\]
There are no terms of the form $b^2$, $b|z|^2$, or $z\eta^{(3)}$.
Moreover, we have
\begin{equation} \label{eq:b22est}
B_{22} = \frac {c_1}2 \Gamma + O(\la^4), \quad
A_{40}, A_{31} = O(\la^3) , \quad
A_{b2} = O(\la) ~.
\end{equation}
\end{lemma}

\myproof
The first part is obtained by direct inspection. Note we deal with
$A_{z\eta,4}$ in the same way we deal with
$A^{(z\eta)}$. The orders of the coefficients are also obtained by
direct check, with the following table in mind:
\begin{gather*}
c_\al = \la, \quad
d_{30}, d_{12}, d_{03}=\la^2 , \quad
d_{21}=1 ,\quad d_1 = 1
\\
c_1 = \la, \quad R=\la^{-1} ,
\\
a_{20}=\la^2 , \quad a_\beta = \la^2 , \quad
c_{za}=\la , \quad
\phi_7, \phi_8=\la^2
\\
\zeta \sim z + \la \bar z , \quad
aR \sim \rho^2 , \quad
\eta \sim \la \rho^2
\end{gather*}
We note that most $C|z|^4$ terms with $C= O(\la^2)$ are killed
by the $\Im$ operator. The only term that survives, due to resonance,
and becomes the main term in $B_{22}$, is from the last term of
$A^{(4)}$:
\[
\bke{c_1 Q, \, \Im  \la \zeta^2 \wbar {\eta^{(2)}}  }
= \bke{ c_1 Q, \, \Im  \la z^2 \up^2  \wbar{z^2 \eta_{20}} }
+ O(\la^4) \, |z|^4+ \sum_{|\gamma|=4, \, \gamma \not = (22)}
O(\la^3) \, z^\gamma ~,
\]
where the first term is equal to $\frac {c_1}2 \Gamma |z|^4$.
We also note that the most dangerous term,
$ |\zeta|^2 \zeta= z^2 \bar z \up^3+O(\la \rho^3)$,
in all integrands, only contributes $O(\la^4)$ to $B_{22}$.
\myendproof

\subsection{Integration of $O(z^4)$ terms}

Next we proceed to integrate out those oscillatory terms in $a^{(4)}$.
\begin{align*}
\int_\infty ^t A_{40} z^4 + A_{31} z^3 \bar z +& A_{b2} b z^2 \, ds
=\frac {A_{40} z^4}{-4i\ka} + \frac {A_{31} z^3 \bar z}{-2i\ka}
+ \frac {A_{b2}bz^2}{-2i\ka}
\\
& \quad
- \int_\infty^t
\frac {A_{40} z^4}{-4i\ka}f_{40}(z)
+ \frac {A_{31} z^3 \bar z}{-2i\ka} f_{31}(z)
+  \frac {A_{b2}\mu^{-2}}{-2i\ka} \frac d{ds} ( bp^2) \, ds
\end{align*}
where
$\frac d{ds} ( bp^2) = \dot b p^2 + 2 bp \dot p = O(z^5)$.
Let
\[
a_{40}=  \frac {A_{40}} {-4i\ka}=O(\la^3) , \quad
a_{31}=  \frac {A_{31}} {-2i\ka}=O(\la^3) , \quad
a_{b2}=  \frac {A_{b2}} {-2i\ka}=O(\la) ~.
\]
Then
\begin{align*}
a^{(4)} &= \Re \bket{ a_{40} z^4 + a_{31}z^3 \bar z + a_{b2} b z^2 }
+ \int_\infty^t  A_{4,5} \, ds
\\
A_{4,5} &= -\Re \bket{a_{40} z^4 f_{40}(z) + a_{31}z^3 \bar z f_{31}(z)
+ a_{b2}\frac d{ds} ( bp^2) }
\end{align*}

Since $\eta^{(3)}_1=  U^{-1} e^{-i \theta} \, e^{-i At} U \eta_0 $
 is larger than $\eta^{(3)}_{2-5}$ when time $t$ is of
order $1$, we also need to integrate out terms of order $z^2 \eta^{(3)}_1$,
 as we
did for \eqref{eq:z-eta0} in the equation for $p$. Specifically, we have
terms of the form
\[
A_{z^2\eta_0} \equiv \sum_{|\al|=2}
\Re (\phi, z^\al \eta^{(3)}_1) =
\frac d{dt} (a_{z^2\eta_0}) + A_{z^2\eta_0,5}
\]
(for different $\phi$), where
$a_{z^2\eta_0}$ and $A_{z^2\eta_0,5}$ are similar to
$p_4$ and $g_4$, respectively,
\begin{align*}
a_{z^2\eta_0} &
\sim \sum_{|\al|=2} \Re \, ( \phi,
z^\al e^{-i \theta} \, \frac 1{A-0i+[\al]\ka} \, e^{-i At} U \eta_0 ) ,
\\
A_{z^2\eta_0,5} &
\sim \sum_{|\al|=2} \Re \, ( \phi,
\mu^{[\al]}\frac d{ds} \bke{p^\al e^{-i \theta}} \,
\frac 1{A-0i+[\al]\ka} \, e^{-i At} U \eta_0 ) ,
\end{align*}
(for different $\phi$).
Here $\frac 1{A-0i+[\al]\ka}$ indicates that we have some
resonance effect, (as in $p_4$ and $g_4$),
and hence these terms has slower dacay in $t$:
\[
|a_{z^2\eta_0}| \le \norm{\eta_0} \tl{t}^{-1}  \bkA{t}^{-1-\sigma} , \qquad
|A_{z^2\eta_0,5}| \le \norm{\eta_0} \tl{t}^{-3/2}  \bkA{t}^{-1-\sigma} ~.
\]

Now we define
\[
B_5=B_5(z, b, \eta^{(2)}, \eta^{(3)})
=A^{(5)} + A_{2,5} + A_{(za,5)} + A_{(z\eta,5)} + A_{4,5} - A_{z^2\eta_0}
+ A_{z^2\eta_0,5}~.
\]
$B_5$ is of order $O(z^5)$. It is a complicated polynomial in its arguments,
which including $U$, $U^{-1}$ and $((\sA-\ka)^{-1} \phi, \cdot )$,
but it does not contain $e^{i \theta}$.

We conclude
\begin{align*}
a &= a_{20}  \bke{z^2 + \bar z^2} +  b
\\
b &= c_{za} a(z+\bar z) + \Re (z\phi_7+ \bar z \phi_8, \eta)
+ \Re \bket{ a_{40} z^4 + a_{31}z^3 \bar z + a_{b2} b z^2} + a_{z^2\eta_0}
\\
& \qquad
+ \int_\infty^t  B_{22}|z|^4  + B_5 \, ds ~.
\end{align*}

We have obtained the main estimate Theorem
\ref{M} assuming the estimate \eqref{bb} on $b$. We now need to
prove the existence of the solution and check the assumption on
$b$.
Define the mapping $S(b)(t) = S_T(b)(t)$:
\begin{align}
S(b)(t)
 &=\bkt{ \rule{0pt}{4mm}
 c_{za} a(z+\bar z) + \Re (z\phi_7+ \bar z \phi_8, \eta)
+ \Re \bket{ a_{40} z^4 + a_{31}z^3 \bar z + a_{b2} b z^2 }
+ a_{z^2\eta_0}  }_T^t
\nonumber
\\
& \qquad
+ \int_T^t  B_{22}|z|^4  + B_5 \, ds ~. \label{Sb.eq}
\end{align}

Recall the class ${\cal B}_T$ of $b$
\begin{equation} \eqlabel{bb2}
 {\cal B}_T = \bket{ b(t): |b(t)| \le D \tl{t}^{-1}, \, 0
\le t \le T } ~,
\end{equation}
where $D= 2 B_{22}/\Gamma=O(\la)$. %
Our goal is to show that $S(b)$ maps ${\cal B}_T$ into itself.
More precisely, we have the following Lemma.

\begin{lemma}\thlabel{sbb}
Suppose that $M(t) \le 2$ and
$|b(t)| \le D \tl{t}^{-1}, \, 0
\le t \le T$. Recall $D= 2 B_{22}/\Gamma=O(\la)$.
Then we have
\[
|S(b)(t)| \le C_1 (D)  \tl{t}^{-3/2} + \frac {B_{22}}{2\Gamma}
\tl{t}^{-1}
\le \frac D2 \tl{t}^{-1} ~.
\]
Hence $S$ map ${\cal B}_T$ to itself. There is a similar statement
for $a(t)$.
\end{lemma}

Assuming the previous bound on $b$, we can also estimate $\theta$.
Since $\theta(t)$ is given by (cf. \ref{eq:theta})
\[
\theta(t)
= \frac {2 c_2}{\ka}\Im z +
\int_0^t -\frac {2 c_2}{\ka} \Im (\mu^{-1}\dot p)
+ F_\theta \, ds
\]
we have
\[
|\theta(t)| \le C \log \tl{t} ~.
\]
This estimate will
not be used in the rigorous proof, but it provides an idea about its size.
\section{Contraction map}

We review what we have so far.  Up to now, we have not shown the
existence of the solution to the equations setup in section 4 with
the boundary condition \eqref{ab}. Notice we cannot conclude this
from the existence to the Schr\"odinger equation as the boundary
condition of $a$ is set at the time $t=\infty$ \eqref{ab}. We have obtained
the main estimate Theorem \ref{M} assuming the estimate
\eqref{bb2} on $b$. We also proved a bound on the map $S$ in Lemma \ref{sbb},
again, assuming an estimate on $b$.
We now need to prove the existence of the
solution and prove this bound on $b$. We shall achieve both goals simultaneously
by proving the map $S$ is a
contraction map  on the space ${\cal B}= {\cal B}_{T = \infty}$. Once we have proved this,
we have constructed rigorously the solution needed in part (1) of Theorem \ref{mainthm}
and established all the upper bounds in part (1). To conclude part (1) of
Theorem \ref{mainthm}, it remains to prove the lower bound in $\zeta$. The size of
$\zeta$ is given by $z$ and thus by $\rho$. From  part (s) of Lemma \ref{th:p2},
we conclude the lower bound of $\zeta$ provided that we can check the bound on
$g$. Since $g$ is given explicitly in (6.12) and we can estimate all terms in
$g$ from the upper bounds in part (1) of Theorem \ref{mainthm}.
 This concludes part (1) of Theorem \ref{mainthm}
assuming that $S$ is a contraction on the space ${\cal B}$.
The rest of this section is a proof that
$S$ is a contraction.

We first recall the setting.
For each $b \in \cal B$, we can solve our system to get
\[
z, p, q=S_1(b), \quad
\eta=S_2(b), \quad
\theta=S_3(b) ~.
\]
More specifically, for fixed $E$ and $h_0 \perp Q$, our system is
\begin{align*}
\dot q &= \d_{21} |q|^2 q + i c_4 b q + g(b,\eta, \theta;t) ~,
\\
& \qquad \quad
q = p + O(z^2), \text{ given by \eqref{q.def}, }  z=\mu^{-1} p ~,
\\
& \qquad \quad q(0) \text{ given by } p(0)=z(0)=(v,\Re h_0) + i(u, \Im h_0)~,
\\
\tilde \eta(t)
&= e^{-i\sA t} \tilde \eta_0
+ \int_0^t e^{-i\sA (t-s)}
\PcA \bket{e^{i \theta}  U \Pi  F^\sharp  -e^{i \theta}
[U,i] \dot \theta  \eta } \, ds ~,
\\
\eta &= U^{-1} e^{-i \theta} \tilde \eta , \quad
\eta(0) = \PcL h_0~,
\\
\theta(t) &= \frac {2 c_2}{\ka}\Im z +
\int_0^t  \wt F_\theta (b,\eta, \theta;s) \, ds  , \qquad
\wt F_\theta =  -\frac {2 c_2}{\ka} \Im (\mu^{-1}\dot p) + F_\theta ~.
\end{align*}
Recall $F^\sharp$ is given by \eqref{Fsharp.def}, and that
$q = p + c\mu z^\al + c\mu  b \bar z + c\mu  z^\beta-p_4$,
for some constants $c$.
In this system all $a$ are already replaced by $a_{20}(z^2+\bar z^2)+b$.
Hence we do not have $a$. We will also write
\[
 \rho=|q|, \quad q = \rho \, e^{i \om} ~.
\]
From the equation of $q$ we can write $\om$ explicitly as
\begin{equation}\eqlabel{om8}
\om(t) = \int_0^t \left [ \Im \d_{21} \rho ^2 + c_4 b + \Im (g/q) \right ] \, ds
\end{equation}
Assuming the estimate on $b$ in Lemma \ref{sbb},
we can bound $\om$ by
\[
|\om(t)| \le C \log \tl{t} ~.
\]
This provides an idea on the size of $\om$, but we shall not need it
in the following  rigorous proof for $b$.

Return to the proof that $S(b)$ is a contraction. Given $b, b^\prime \in \cal
B$, we can define $z, \eta, \theta$ and $z^\prime$, $\eta^\prime$,
$\theta^\prime$ correspondingly. We use $\diff$ to denote the difference
of quantities in these two sets. For example,
 \[
 \diff b = b - b^\prime , \quad
  \diff \eta = \eta - \eta^\prime , \quad
  \diff |z| =|z| - |z^\prime| ~.
\]
If
\[
|\diff b(t)| \le \diff_0 \tl{t}^{-1} ~,
\]
we want to show
\begin{equation}    \label{Sb.ineq}
|\diff S(b)(t)| \le \tfrac 12 \diff_0 \tl{t}^{-1} ~.
\end{equation}

Let us define
\begin{align*}
 N(T) &= \sup_{0 \le t \le T} \Big \{
\tl{t}^{1/2+\sigma}  \length{\diff |z|(t) }
+ \tl{t}^{1/2} (\log \tl{t})^{-1}  \length{\diff z(t) }
\\
& \qquad
+  \tl{t}^{3/4-\sigma}  \norm{\diff \eta(t) }_{L^4}
+  \tl{t}^{1+\sigma/4}  \norm{\diff \eta^{(3)}_{2-5}(t) }_{L^2 \loc}
\\
& \qquad \qquad
+ (\log \tl{t})^{-1}  \length{\diff \theta(t) }
\Big \}
\end{align*}
The key here is that $\diff |z|$ has a faster decay than $|z|$.
We first want to show that
   $N(T) \le C \diff_0$
uniformly for all $T$. Our strategy is to show that
\begin{equation} \label {N.ineq}
  N(T) \le C \e^\sigma N(T)  + C \diff_0
\end{equation}
for all $T$. After this is proved, by choosing $\e_0$ sufficiently
small we have  $N(T) \le C \diff_0$. We also have
\begin{align*}
\length{ \diff S(b)(t) }
&\le \tl{t}^{-1} N \tl{t}^{-1/2} \log \tl{t}
+ \tl{t}^{-1/2} \diff _0 \tl{t}^{-1}
\\
& \qquad
+ \int_\infty^t \tl{s}^{-3/2} \ N\tl{s}^{-1/2-\sigma}
+ \tl{s}^{-2-\sigma}(N+\diff_0) \ ds
\\
&\le C \diff_0 \tl{t}^{-1-\sigma}
\le \tfrac 12 \diff_0 \tl{t}^{-1}
\end{align*}
which shows \eqref{Sb.ineq}. Now we focus on proving \eqref{N.ineq}.

\bigskip

Let $t \le T$. We will write $N=N(T)$. Also, when we say things like
$|\diff(\rho^2)| \le C |\rho \diff\rho|$, what we really means is
$|\diff(\rho^2)| \le C |\rho \diff\rho|+C |\rho^\prime \diff\rho|$.

1. We first estimate $\diff g$. By \eqref{g.def} and the definitions
of $P^{(4)}$, $g_1$,  $g_2$,  and $g_3$, we have
\[
\length{ \diff g(t)} \le C (N+\diff_0) \tl{t}^{-3/2-\sigma}
\]
Note $C(N+\diff_0) \le C_1$ if $\diff_0$ is sufficiently small.
Hence
\[
  m^{-1} \tl{t}^{-1/2} \le \rho(t), \rho^\prime(t)
  \le m \tl{t}^{-1/2} ~.
\]
By Lemma \ref{th:p3},
\[
\length{\diff \rho(t)} \le C (N+\diff_0) \e^{\sigma} \tl{t}^{-(1+\sigma)/2}
\]

2. From the equation for  $\om$ \eqref{om8},  we can bound the variation of $\om$ by
\begin{align*}
|\diff \om(t)| &\le
\int_0^t C |\diff (\rho ^2)| + C |\diff b| + C | \diff g| \tl{s}^{1/2}
+ C |g/q^2| |\diff q| \, ds
\\
& \le \int_0^t C \tl{s}^{-1 -\sigma} \e^\sigma N + C \tl{s}^{-1}
\diff_0 + C (N+\diff_0)  \tl{s}^{-3/2-\sigma+1/2}
\\
& \qquad \qquad
+  C (N+\diff_0)  \tl{s}^{-3/2-\sigma+1} N \tl{s}^{-1/2} \log \tl{s}
 \, ds
\\
&\le \left [ C \e^\sigma N + C \diff_0 \right ] \log \tl{t}
\end{align*}
Hence the variation of $z$ is bounded by
\[
|\diff z(t)| \le |\rho \diff \om| + |\diff \rho|
\le (C\e^\sigma N + C \diff_0) \tl{t}^{-1/2} \log\tl{t} ~.
\]

3. Since $\theta(t) = c_2[(z+ \bar z)]_0^t + \int_0^t \wt F_\theta \, ds $,
we have
\[
  |\diff \theta(t)|
\le C|\diff z| + \int _0^t \diff \wt F_\theta \, ds
\le (C\e^\sigma N + C \diff_0) \log\tl{t} ~.
\]

4. To estimate $\diff \eta $,
\[
\diff \eta =  \int_0^t  e^{-i\sA(t-s)} \PcA
\diff \bket{ e^{i \theta} U\Pi \bkt{ F^{\sharp} +i\eta^2\bar \eta}
        - e^{i \theta} [U,i] \dot \theta \eta }\, ds
\]
 we note
\[
\norm{\diff \bket{e^{i \theta} U\Pi F^\sharp
  - e^{i \theta} [U,i] \dot \theta \eta }(s) }_{L^{4/3}}
\le
C |\diff \theta| \tl{s}^{-1} + (C \e^\sigma N + C \diff_0) \tl{s}^{-1}
\]
hence
\[
\norm{\diff \eta(t)}_{L^4}
\le \int_0^t \frac C{|t-s|^{3/4}} \,
(C \e^\sigma N + C \diff_0) \tl{s}^{-1} \log \tl{s} \, ds
\le C \e^\sigma (N + \diff_0) \tl{s}^{-3/4+\sigma}
\]

5. To estimate $\diff \eta^{(3)}_{2-5} $,
\begin{align*}
\diff \tilde \eta^{(3)}_{2-5}(t) &=
-\int_0^t e^{-i\sA(t-s)}
\bket{ \mu^{[\al]} \diff \bket{\frac d{ds} ( e^{i \theta} p^\al)}
\wt \eta_\al} \, ds
\\
&\quad
+ \int_0^t  e^{-i\sA(t-s)} \PcA
\diff \bket{ e^{i \theta} U\Pi \bkt{ F^{\sharp \sharp} +i\eta^2\bar \eta}
        - e^{i \theta} [U,i] \dot \theta \eta }\, ds
\\
&\quad
+ \int_0^t  e^{-i\sA(t-s)} \PcA
 \diff \bket{e^{i \theta} U\Pi (-i\eta^2\bar \eta) }  \, ds
\\
&= \diff \tilde \eta^{(3)}_3
  + \diff \tilde \eta^{(3)}_4 + \diff \tilde \eta^{(3)}_5 ~.
\end{align*}
Hence
\[
\norm{\diff \tilde \eta^{(3)}_{2-5}(t)}_{L^2 \loc}
\le (C \e^\sigma N + C \diff_0) \tl{s}^{-1-3\sigma/4}
\]
Note that we estimated $\norm{\eta}_{L^2}$ in section 5
in order to estimate $\norm{|\eta|^2 \eta}_{L^{8/7}}$, which is used
in estimating $\norm{\tilde \eta^{(3)}_5}_{L^2 \loc}$.
However, we do not need $\norm{\diff \eta}_{L^2}$ here since
\[
\norm{\diff (|\eta|^2 \eta)}_{L^{8/7}}
\le C \norm{\eta}_{L^2}^{1/2} \, \norm{\eta}_{L^4}^{3/2} \,
\norm{\diff \eta}_{L^4} ~.
\]
Compare with Subsection 5.3. Other estimates are similar to those in that
subsection.

Since $\eta^{(3)}_{2-5} = U^{-1} e^{-i\theta} \tilde \eta^{(3)}_{2-5} (t)$,
\[
\norm{\diff \eta^{(3)}_{2-5}(t)}_{L^2 \loc}
\le |\diff \theta| \norm{\tilde \eta^{(3)}_{2-5}(t)}_{L^2 \loc} +
\norm{\diff \tilde \eta^{(3)}_{2-5}(t)}_{L^2 \loc}
\le C \e^{\sigma/4} (N +  \diff_0) \tl{s}^{-1-\sigma/2}
\]

5. Summarizing, we have shown \eqref{N.ineq}. Choosing $\e_0$ sufficiently
small we have $N \le C \diff_0$.

6. The above shows that the map $S(b)$ is a contraction mapping.
hence there is a fixed point $b=S(b)$, which gives us a solution.

\section{Dynamical renormalization of mass}

As explained in the introduction, we expect that the solution to be of the form
\[
   \psi(t) = [ Q_{E(t)} + h(t) ] e ^{i \Theta(t)}
\]
with a changing $E(t)$, and we use $[Q_E + a R_E + h] e ^{i \Theta(t)}$
as an approximation.
In Theorem 5.1 we conclude that $M(T)\le 2$ assuming $|b(t)|\le D
\tl{t}^{-1}$ for $t \in [0, T]$. We cannot extend this result beyond
$T$. The reason is that, since $E(t)$ is changing and we expect
that $E(t)$ will converge to some $E_\infty$ which
is likely to be different from $E (0)$, after certain time
$[Q_E + a R_E + h] e ^{i \Theta(t)}$ is no longer a good approximation and
 $|a(t)|$ will no longer decay.
To overcome this difficulty, for each time step $T= k \dt$, we propose to
re-choose $E=E_k$ so that the new $a(t)$ satisfies $a(T)=0$.
Then we prove that the new $a(t)$ has the same estimate as the old $a(t)$
if our
time increment $\dt$ is sufficiently small. Then we estimate the change of
$E(t)$, especially taking into account of its oscillation, by studying
$a(t)$, and prove that $E(t)$ does converge.
The rest of the proof essentially follows that of Theorem \ref{mainthm}.

\subsection{Renormalization lemma}

\begin{lemma}  \thlabel{th:QaRhdcmp}
Suppose $\norm{\psi - Q_E} _2 \ll 1$, then we can rewrite
\[
\psi = ((1+a)Q_E + k) e^{i \theta} = (Q_E + a'R_E + h) e^{i \theta}
\]
uniquely with $h,k \perp Q$. Moreover, $a$, $a'$, $\theta$, $k$ and $h$
are small.
\end{lemma}

\myproof
First we can write $\psi = (1+ \al)Q_E + k_1$, with $k_1 \perp Q_E$,
and $\al$ a complex number.
Since $\psi - Q_E$ is small, $\al$ and $k_1$ are both small. Hence we can find
small real $a$ and $\theta$ such that $(1+ \al)=(1+ a) e^{i \theta}$. Let
$k= k_1 e^{-i \theta}$. Then we have $\psi = ((1+a)Q_E + k) e^{i \theta}$.
We have $R_E = c_1^{-1}c_0 Q_E + \Pi_E R_E$, and hence
$a Q_E =  c_1 c_0^{-1}a (R_E - \Pi_E R_E)$. Let $a'=  c_1 c_0^{-1}a$
and $h=k-  a' \, \Pi_E R_E$. We have $h \perp Q_E$ and
$a Q_E + k = a' R_E + h$.
\myendproof

\begin{lemma} \thlabel{th:renormal}
Let $\tau_0$ and $\d_0$ be sufficiently small and
suppose  $E\in {\cal I}_\la$ with dist$(E, \pd {\cal I}_\la)> C\tau_0 \la$.
Suppose that $\psi = [Q_E + a R_E + h] \, e^{i \Theta}$,
with $|\la^{-1}a| = \tau < \tau_0$, $\norm{h}_{L^2} = \d< \d_0$.
Then we can find $\wt{E}$ with
$|\wt{E}-(E+a)| \le  \tau \la/2$ such that, if we write uniquely
\begin{equation}   \label{renormal}
\psi = [Q_{\wt{E}} + \wt{a} R_{\wt{E}} + \wt{h}] \, e^{i \wt{\Theta}} , \quad
\wt{h} \perp Q_{\wt{E}}
\end{equation}
using Lemma \ref{th:QaRhdcmp}, then $\wt{a}=0$ and we have
$\norm{h-\wt{h}}_{L^2}  < \tau /2$, and
$|\Theta- \wt{\Theta}| <  \tau /2$.
In fact, $(\wt{E}, \wt{\Theta})$ is the unique solution of
\begin{equation}   \label{Eprime.eq}
 \bke{\psi - Q_{E'} \, e ^{i \Theta'}\, , \, Q_{E'} } = 0 \, , \qquad
 ( |E- E'| \le 2 \la \tau_0),
\end{equation}
for $\norm{\psi-Q_E}_{L^2} $ sufficiently small.
\end{lemma}

\myproof
We will define a sequence $\bket{E_k}_{k=1,2,3 \cdots}$ which converges to $\wt{E}$.
Let $E_1=E+a$. We can write uniquely
\[
\psi = [Q_{E_1} + a_1 R_{E_1} + h_1] \, e^{i \Theta_1} , \quad
h_1 \perp Q_{E_1}
\]
using Lemma \ref{th:QaRhdcmp}.
Denote $\d Q = Q_E + a R_E - Q_{E_1}$ and $\d \Theta = \Theta-\Theta_1$.
We have $\norm{Q_E  - Q_{E_1}} \le O( a R) \le C |a| \la^{-1} \le C \tau$, and
$\norm{\d Q} \le C a^2 (\pd_E^2 Q) \le C a^2 \la^{-2} \le C \tau ^2$.
Note we have
\[
 a_1 R_{E_1} + h_1
= Q_{E_1}(e^{i\d \Theta} -1) + [ \d Q + h] \,e^{i\d \Theta} ~.
\]
Hence
$0 = \Im (Q_{E_1},  a_1 R_{E_1} + h_1  ) = O(\d \Theta) + O((Q_{E_1},\d Q + h))$,
and hence
\[
|\d \Theta| \le O(\d Q) + C|(Q_{E}, h)|+C|(Q_{E_1}-Q_E, h)| \le C \tau^2 + 0 + C \tau \d
~.
\]
Also,
\[
O(\la^{-1}) \, a_1 = (Q_{E_1},  a_1 R_{E_1} + h_1  )
= c_0^{-1} (e^{i\d \Theta}-1)
+ O((Q_{E_1},\d Q + h)) ~,
\]
hence
\[
\la^{-1}|a_1| \le C|e^{i\d \Theta}-1| + C \tau^2 + C\tau \d
\le C \tau^2 + C\tau \d ~.
\]
Finally,
\begin{align*}
\norm{h_1-h}_{L^2}  &\le
\norm{h_1-h \, e^{i\d\Theta} }_{L^2}
+ \norm{h \, e^{i\d\Theta}-h}_{L^2}
\\
& = \norm{ (Q_{E_1}(e^{i \d \Theta}-1)
+ (\d Q) e^{i\d\Theta} - a_1R_{E_1}}_{L^2}
+ \norm{h }_{L^2}\, |e^{i\d\Theta}-1|
\\
&\le C\tau^2 + C\tau \d ~.
\end{align*}
If we choose $C \tau + C\d \le 1/3$, then we have
\[
\la^{-1}|a_1|, \; |\d \Theta|, \; \norm{h_1-h}_{L^2}  \le \tau/3 ~.
\]

Now for $k \ge 2$ we define $E_k = E_{k-1} + a_{k-1}$, and define $a_k$, $h_k$
and $\Theta_k$ correspondingly by Lemma \ref{th:QaRhdcmp}.
We follow the previous estimates
to get
\[
\la^{-1}|a_k|, \; |\Theta_k - \Theta_{k-1}| , \; \norm{h_k-h_{k-1}}_{L^2}  \le  3^{-k} \tau
\]
Note that the size of $h_k$ may increase in the process, but since
\[
C |a_k| + C\norm{h_k}_{L^2}  \le C \frac{|a_{k-1}|}3 + C(\norm{h_{k-1}}_{L^2}  + \frac{|a_{k-1}|}3)
\le C |a_{k-1}| + C\norm{h_{k-1}}_{L^2}  \le \cdots \le 1/3
\]
our condition for estimates is always satisfied.
Hence $E + \sum_k a_k$ converges to a
limit $\wt{E}$ with $|\wt{E}-E| \le 3\la \tau/2$
and $|\wt{E}-E-a| \le\la \tau/2 $.
Writing $\psi$ in the form \eqref{renormal} with respect to $\wt{E}$,
 we have $\wt{a}= \lim_k  a_k=0$ and
$|\wt{\Theta}-\Theta| \le \sum_k |\Theta_{k+1}-\Theta_k| \le \tau /2$.
We conclude  $(\wt{E}, \wt{\Theta})$ is a solution of \eqref{Eprime.eq}.

To show the uniqueness of \eqref{Eprime.eq}, we first note that it is
locally unique by inspecting the equations obtained by taking
derivatives of \eqref{Eprime.eq} with respect to ${E}'$ and ${\Theta}'$.
Now suppose we may write
\[
\psi = [Q_1 + h_1] e^{i \Theta_1} = [Q_2 + h_2] e^{i \Theta_2}
\]
with $h_1 \perp Q_1$ and $h_2 \perp Q_2$, . Then we have
$[Q_1 + h_1] = [Q_2 + h_2] e^{i \d \Theta}$. Since both $h_1$ and $h_2$
are small, taking projection on $Q_1$ we get $\d \Theta$ is small.
The local uniqueness implies the claim.

\myendproof

\subsection{Proof of Theorem \ref{main2}}

First we proceed an induction argument to find the desired renormalization
at each time step and the corresponding estimates.
Denote the space of initial data
\[
Y = H^2 \cap W^{2,1}(\R^3).
\]
Our initial datum is
 $\psi_0=Q_{E \ini}+ a \ini R_{E\ini} + h\ini$,
with $\norm{\psi_0 - Q_{E \ini}}_Y  \le C^{-1}\e$.
Applying Lemma \ref{th:QaRhdcmp} to $\psi_0$, we can find $E_0$,
$h_0$, and $\Theta_0$ such that
$\psi_0 = \bkt{Q_{E_0}  + h_0} \, e^{i \Theta_0}$.
Moreover, we have $|E_0 - E\ini| \le C ^{-1} |a \ini| \le C^{-1} \la \e$,
and hence $\norm{Q \ini - Q_0 e^{i \Theta_0(0)} }_Y  \le \tfrac 12 \e$.
 This is the beginning of our induction argument.
We will choose $\dt>0$ sufficiently small. Then
we define $T_k=k \dt$ and will define $E_k$, $h_k$, etc. at time $T_k$.
Also, we abbreviate $Q_k = Q_{E_k}$, $R_k=R_{E_k}$, etc..

Note, Theorem 5.1 remains valid if we replace the assumption
$|b(t)| \le D \tl{t}^{-1}$ by $|a(t)| \le D \tl{t}^{-1}$.
Also recall that $D=O(\la)$ by Lemma \ref{th:7-1}.

Assume for $T_k=k \dt$ we have found $E_k$, $a_k(t)$, $h_k(t)$,
$\Theta_k(t)$ so that
\begin{equation}  %
\left \{
\begin{array}{l}
\psi(t) = [Q_{k}+ a_k(t) R_k + h_k] \, e^{i \Theta_k}, \quad
E_k \in {\cal I}_\la ~,
\\
\\
h_k(t) \perp Q_k  \quad
\text{for } t\in [0, T_k] ~ ; \quad
\norm{\psi_0 - Q_k e^{i \Theta_k(0)}}_Y \le \e ~,
\\
\\
|a_k(t)| \le D \tl{t}^{-1} , \quad
a_k(T_k)=0   ~.
\end{array}
\right .
\label{ind.hyp}
\end{equation}
This is our induction hypothesis.
By Theorem 5.1 we have $M_k(T_k) \le 2$.
Since $a_k(T_k)=0$, we have
\begin{equation}  \label{ak.eq}
a_k(t) = \bkt{ a_{20}(E_k)(z^2_k + \bar z^2_k) +\cdots }_T^t
+ \int_T^t B_{22}(E_k) |z_k|^4 + \cdots
\end{equation}
The direct estimate of Lemma \ref{sbb} gives us
\[
|a_k(t)| \le   \frac 12 D \tl{t}^{-1} \qquad
\text{for } t\in [0, T_k] ~.
\]
Thus $|a_k(0)|\le \frac 12 D \e^2$ and we have
\begin{align*}
\norm{\psi_0 - Q_k e^{i \Theta_k(0)}}_Y
&\le
\norm{\psi_0 - Q_0 e^{i \Theta_0(0)}}_Y
+ \norm{ Q_0 e^{i \Theta_0(0)}-Q_k e^{i \Theta_k(0)}}_Y
\\
& \le \frac 12 \e + \la^{-1} |a_k(0)|
\le \frac 34 \, \e ~.
\end{align*}

By continuity, there is a $\sigma > 0$ such that
\[
|a_k(t)| \le   D \tl{t}^{-1} \qquad
\text{for } t\in [0, T_k +\sigma] ~.
\]
Theorem 5.1 ensures that
\[
M_k(T_k +\sigma) \le 2 ~.
\]
We now return to \eqref{ak.eq}.
Since the derivatives of the terms outside
the integral (say, $z^2_k + \bar z^2_k$) are bounded by $\tl{t}^{-1}$,
for $t \in [T_k, T_k + \sigma]$ we have
\begin{align}
|a_k(t)| &\le   C T^{-1} (t-T) + CT^{-2}(t-T)   \label{eq:9-5}
\\
&\le   C_3 T^{-1} (t-T)                 \nonumber
\le \frac 12 {D} \tl{T}^{-1}
\le \frac 58 { D} \tl{t}^{-1}
\end{align}
if $(t-T) \le D/(2 C_3)$ and $(t-T) \le \tl{T}/4 \le \e^{-2}/4$.
Now we define
\[
\dt = \min \bket{ D/(2 C_3) \, , \, \e^{-2}/4 }  ~.
\]
Recall $T_{k+1} = T_k + \dt$.
We have thus proved that $\sigma \ge \dt$ and
\[
|a_k(t)| \le \ \frac 58  D \tl{t}^{-1} ,\qquad
\text{for } t\in [0, T_{k+1}] ~
\]
and $M_k(T_{k+1}) \le 2$.
By Lemma \ref{th:renormal}, we can find $E_{k+1}$ with
\begin{equation} \label{eq:9-6}
|E_{k+1} - E_k | \le \tfrac 32 |a_k(T_{k+1})| ~,
\end{equation}
such that we can rewrite
\[
\psi(t)= [Q_{k+1}+ a_{k+1}(t) R_{k+1} + h_{k+1}(t)] \,
e^{i \Theta_{k+1}(t)} , \quad
h_{k+1}(t) \perp Q_{k+1}
\]
with $a_{k+1}(T_{k+1}) =0$.
Now we compare these two sets of data with respect to $E_k$ and $E_{k+1}$,
for $t \in [0, T_{k+1}]$.
As in the proof of lemma \ref{th:renormal}, we have
\begin{align}
|\Theta_{k}(t) - \Theta_{k+1}(t) |
&\le C|(Q_{k+1}, h_k(t))|
= C|(Q_k + O(\la^{-1} a_k(T_{k+1})) , h_k(t))|   \nonumber
\\
&\le 0 + C \la \tl{T}^{-1} \tl{t}^{-1/2}
\le C \la \tl{t}^{-3/2}    \label{eq:9-7}
\end{align}
and since
\[
 a_{k+1}(t) R_{k+1} + h_{k+1}(t) =
[Q_k + a_k(t) R_k + h_k(t)] \,e^{i \Theta_k(t)-i\Theta_{k+1}(t)} - Q_{k+1}
\]
Taking inner product with $Q_{k+1}$
we get
\[
\la^{-1} |a_{k+1}(t)-a_k(t)|
\le C | \Theta_k(t) - \Theta_{k+1}| + C|a_k(t)|^2 + C \la \tl{T}^{-1}
\norm {h_k(t)} \le C \la \tl{t}^{-3/2}
\]
Hence
\[
|a_{k+1}(t)| \le |a_{k}(t)| + |a_{k+1}(t)-a_k(t)|
\le \frac 58 D \tl{t}^{-1} + \frac 18 D \tl{t}^{-1}
=\frac 34 D \tl{t}^{-1} ~,
\]
if $\e$ is sufficiently small.
By \eqref{eq:9-6} and \eqref{eq:9-7}, we have
\[
\norm{\psi_0 - Q_{k+1} e^{i \Theta_{k+1}(0)}}_Y
\le \norm{\psi_0 - Q_{k} e^{i \Theta_{k}(0)}}_Y
+ C \la \tl{t}^{-1} \le  \e ~.
\]
Also note
\[
|E_0 - E_{k+1}| \le \frac 32 \, |a_{k+1}(0)| \le 2 D \e^2 \le C \la \e^2 ~.
\]
Hence $E_{k+1} \in {\cal I}_\la$.
We have thus proved the induction hypothesis \eqref{ind.hyp} for
$t=T_{k+1}$. The induction argument is completed.

\bigskip

Next we proceed to show that $E_k$ has a limit.
Let $n=2^{k} $ and $T=T_n$.
Since $a_n(T)=0$, we have
\[
a_n(t) =  \bkt{ a_{20}(E_n)(z^2_n + \bar z^2_n) + \cdots }_T^t
+ \int_T^t B_{22}(E_n) |z_n|^4 + \cdots
\]
for $t \in [0,T_n]$. In particular,
\[
|a_n(T/2)| \le C\tl{T}^{-1} + \int^T_{T/2} |B_{22}|\, \tl{s}^{-2} \, ds
\le 2 D \tl{T}^{-1}
\]
(Note that the argument here differs from \eqref{eq:9-5}.)
Hence, by considering $\psi=\psi(T/2)$, $E=E_n$ and $\wt{E} = E_{n/2}$
 in Lemma
\ref{th:renormal}, we have
\[
|E_{n/2}-E_n|
\le 2 |a_n(T/2)| \le  4 D \tl{T}^{-1} ~.
\]
This estimate shows that the sequence $E_{2^k}$ converges to
a limit $E_\infty$. Moreover, choose $k_1$ so that
$2 \Gamma 2^{k_1} \dt > \e^{-2}$
and we have
\[
|E_0-E_\infty| \le |E_0-E_{k_1}| + |E_{k_1}-E_\infty |
\le C \la \e^2 + C D \sum_{k> k_1} \tl{2^k\dt}^{-1}
\le C \la \e^2 ~.
\]
Write $Q_\infty = Q_{E_\infty}$, $R_\infty = R_{E_\infty}$, and
\[
\psi(t) = \bkt{Q_\infty + a_\infty(t) R_\infty + h_\infty(t) } \,
e^{i \Theta_\infty(t)}
\]
and compare this set of data with data at $T_k$, $t \in [0,T_k]$.
As before we get
$|a_{\infty}(t)-a_k(t)| \le C D \tl{t}^{-3/2}$ and hence
\[
|a_\infty(t)| \le \frac 34 D \tl{t}^{-1}  ~.
\]
Since $k$ is arbitrary, this estimate is true for all $t \in [0, \infty)$.
Since $|E_\infty - E_0| \le O(\la \e^2)$, we have
$\norm{\psi_0 - Q_\infty}_Y \le \e$.
Theorem 5.1 shows that
\[
   M_\infty(T) \le 2 \qquad \text{for all } T ~.
\]
The first part of Theorem \ref{main2} is thus proved.

Suppose the assumption of part (2) of  Theorem \ref{main2} holds,
that is,
\[
0< |z\ini|=\e \le \e _0 , \quad
\| \eta\ini\|_Y  \le  C \e ^{3/2} , \quad
\la^{-1} |a \ini| \le C \e ^{2} . %
\]
Let $a_{\infty}=a_{\infty}(0)$, $z_{\infty}=z_{\infty}(0)$, and
$\eta_{\infty}=\eta_{\infty}(0)$.
Apply Lemma \ref{th:renormal} to
$(\psi,E)=(\psi_0, E\ini)$ and $(\psi,E)=(\psi_0, E_\infty)$
respectively. In both cases, $\wt{E}=E_0$. We have
\begin{gather*}
  |a \ini| \le C \la \e^{2}, \quad |a _\infty| \le C \la \e^2
\\
  |E_0-E \ini| \le C |a \ini|, \quad |E_0 - E _\infty| \le C|a_\infty|
\end{gather*}
Therefore $Q\ini - Q_\infty$ is of order
$ \la^{-1}|a\ini-a_\infty|=O( \e^{2})$,
and hence so is $h \ini -h_\infty$.
Denote $P_\infty=P_c(E_\infty)\Pi_\infty$ and
$P \ini=P_c(E \ini)\Pi \ini$, we have
\begin{align*}
\norm{P_\infty h_\infty}_Y
&\le \norm{P\ini h\ini}_Y
+ \norm{P\ini (h_\infty-h\ini)}_Y
+ \norm{(P_\infty-P\ini) h_\infty}_Y
\\
&\le \e^{3/2} + O( \e ^{2}) + |E_\infty-E\ini| \e
\le 2 \e^{3/2} ~.
\end{align*}

By the last statement of Theorem 5.1, we have
$|z_\infty(t)| \ge c \tl{t}^{-1/2}$ for all time $t$.
The last statement of Theorem \ref{main2} is proved.
\myendproof

\section{Radiation dominated solutions}

In this section we prove part (2) of Theorem \ref{mainthm}.
For a given profile $\xi_\infty$,
we want to construction solutions of the form \eqref{psidec}
such that
\[
\chi(x,t) = h(x,t) e^{i \Theta(t)} \longrightarrow e^{i t \Delta} \xi_\infty
\]
as $t$ goes to infinity.
Let $W$ denote the wave operator for $\L $,
\[
W: L^2 \to \Hc(\L), \qquad
W \phi = \lim_{t \to \infty}  e^{-t \L} \, e^{-i(-\Delta - E)t} \phi
~.
\]
We have
\[
e^{t \L} W \chi \longrightarrow e^{i(\Delta + E)t} \, \chi
, \qquad \text{as } t \to \infty,
\]
for general $\chi \in L^2$. See subsection 3.2.
For the profile $\chi_\infty$ given in part (2) of
Theorem \ref{mainthm}, we
let
\[
\xii = U W \chi_\infty ~.
\]

We recall that we set
\[
   \psi = (Q + a R + h) \, e^{i[-Et + \theta(t)]} ~,
\]
and $h$ satisfies
\[
     \partial _t h = -\dot a R -a i Q + \L\original h -i F(aR+h)
- i \dot \theta (Q+aR+h) ~.
\]
To ensure $h(t) \perp Q$ all the time, we set
\begin{align*}
   \dot a &=(c_1 Q, \Im  F(k) ) ~, \qquad c_1 = (Q, R)^{-1} ~,
\\
   \dot \theta &=  - (1 + c_0c_1^{-1}a) ^{-1} \,
\bkt{ a + (c_0Q, \la Q^2(h + \bar h))  +(c_0Q, \Re F(k) )} ~.
\end{align*}
(Recall $c_0=(Q, Q)^{-1}$.)
Then the equation of $h$ is
\begin{align*}
\partial _t h &= \L h + \Pi F_{all} ~,
\\
\L h &= -i \bket{ H h + \Pi \la Q^2  \Pi (h+ \bar h) } ~,
\\
F_{all} &=- i \dot \theta h
  -i F  - \bkt{ (c_1 Q, \Im F) + i a \dot \theta } R_\Pi ~,
\quad R_\Pi := \Pi R ~,
\\
F &= F(aR+h)
\\
&=
\la Q(2|h|^2 + h^2) + 2 \la aQR (2h+ \bar h) + 3\la a^2 QR^2 +
\la (aR+ h)^2(aR + \bar h) ~.
\end{align*}

We first rewrite the equation of $h$.
The equation for $h$ is
\begin{align}
   \partial _t h &= \L h -  i \dot \theta h
+\Pi  F^ \sharp ~,
\nonumber
\\
F ^\sharp &=
- i F(k) - [(c_1Q,\Im F) + ia \dot \theta ] R_\Pi ~.
\nonumber %
\end{align}
Let
\[
   h ^\diamond = U h ~.
\]
Since $\L = U^{-1} (-i \sA) U$, we have
\begin{align*}
\partial _t h^\diamond &=  -i \sA h^\diamond
-  U i \dot \theta U^{-1} h^\diamond
+ U \Pi  F^ \sharp
\\
&=-i \sA h^\diamond -  i \dot \theta h ^\diamond
-   [U,i]\dot \theta U^{-1} h ^\diamond
+  U \Pi  F^ \sharp ~.
\end{align*}

Let $\wt  h = e^{i \theta} h^\diamond$ and use
$U^{-1}  h ^\diamond = h$,
we get
\begin{equation}        %
\partial _t \wt  h  = -i\sA \wt  h
+e^{i \theta}  U \Pi  F^\sharp
- e^{i \theta}  [U,i]\dot \theta  h ~.
\end{equation}
For a specificed datum $\xii$ at infinity, we define a solution by
the equations
\begin{align*}
\wt  h(t) &= \xi(t) + g(t)
\\
\xi(t) &=e^{-i\sA t} \xii
\\
g(t) &= \int_\infty^t  e^{-i\sA (t-s)}
 \bket{e^{i \theta}  U \Pi  F^\sharp  -e^{i \theta}
[U,i] \dot \theta  h } \, ds ~.
\end{align*}
We also let
$h = U^{-1} e^{-i \theta} \wt  h $.

Suppose we have a solution, then the main term in $F^\sharp$ is of order
$t^{-3}$, hence $g \sim t^{-2}$, $a \sim t^{-2}$,
$\dot \theta \sim t^{-3/2}$, $ \theta \sim t^{-1/2}$.

Note that $\xi(t)$ is fixed.
We consider $\xii$ small in the following norm
\[
   \norm{\xii}_{H^2 \cap W^{2,1}} \le \e
\]
with $\e$ sufficiently small. Then
\begin{equation}  \label{xi.est1}
\norm{\xi(t)}_{H^2} \le C_1 \e , \qquad
\norm{\xi(t)}_{W^{2, \infty}} \le C_1 \e |t|^{-3/2}.
\end{equation}
Also, for the leading term $\la |\xi|^2 \xi$ in $F$, we have, for $t>1$,
\begin{align*}
\norm{|\xi|^2 \xi}_{L^2}
&\le C \norm{\xi(t)}_{\infty}^2 \norm{\xi(t)}_{2} \le C \e^3 \bkA{t}^{-3},
\\
\norm{\nabla^2(|\xi|^2 \xi)}_{L^2}
&\le C \norm{\xi(t)}_{\infty}^2 \norm{\nabla^2\xi(t)}_{2}
+ C \norm{\xi(t)}_{\infty} \norm{\nabla\xi(t)}_{4}^2
\\
&\le C (\bkA{t}^{-3/2})^2 \cdot 1
+ C \bkA{t}^{-3/2} \cdot (\bkA{t}^{-3/4})^2
\le C \e^3 \bkA{t}^{-3} .
\end{align*}
It $t \le 1$, we simply have
$\norm{|\xi|^2 \xi}_{H^2}\le C \norm{\xi}_{H^2}^3 \le C \e^3$.
We conclude
\begin{equation} \label{xi.est2}
\norm{|\xi|^2 \xi}_{H^2} \le C \e^3 \bkA{t}^{-3} .
\end{equation}
Similarly, if $\norm{g(t)}_{H^2} \le C \e \bkA{t}^{-2}$,
one can prove, for example,
\[
\norm{|\xi+g|^2 (\xi+g)}_{H^2} \le C \e^3 \bkA{t}^{-3} , \quad
\norm{Q \xi g}_{H^2}\le C \e^2 \bkA{t}^{-7/2} , \quad \text{etc}.
\]

\bigskip

Now we proceed to construct a solution. For convenience, we introduce
a new variable $\om = \dot \theta$. (It should not be confused with
the $\om$ used in Section 8.)
We will define a Cauchy sequence on the space
\begin{align*}
{\cal A}= \bigg \{
& (\om, \theta, a, g): [0,\infty) \to \R \times \R \times \R \times H^2 ,
\\
& \quad
|\om(t)| \le \e^{1/2} \bkA{t}^{-3/2}, \;
|\theta(t)| \le \e^{1/2} \bkA{t}^{-1/2}, \;
|a(t)| \le \e \bkA{t}^{-2}, \;
\norm{g(t)}_{H^2} \le \e \bkA{t}^{-2}  \bigg \}
\end{align*}
Here $H^2 = W^{2,2}(\R^3)$.
We define our map by
\begin{align*}
\om ^\triangle (t) &:=
- (1 + c_0c_1^{-1}a) ^{-1} \,
\bkt{ a + (c_0Q, \la Q^2(h + \bar h))  +(c_0Q, \Re F )}
\\
\theta^\triangle(t) &:=
\int_\infty^t  \om \, ds
\\
a^\triangle(t) &:=
\int_\infty^t   (c_1 Q, \Im  F ) \, ds
\\
g^\triangle(t) &:=
 \int_\infty^t  e^{-i\sA (t-s)}
 \bket{e^{i \theta}  U \Pi
 \bket{- i F - [(c_1Q,\Im F) + ia \om ] R_\Pi }
  -e^{i \theta} [U,i] \om h } \, ds ~,
\end{align*}
where
\[
F=F(aR+h) ,\quad
h(t):= U^{-1} e^{-i \theta} \bke{ e^{-i\sA t} \xii + g(t) } ~.
\]

Our initial data are
\[
\om(t) \equiv 0 , \quad
\theta(t) \equiv 0, \quad
a(t) \equiv 0, \quad
g(t) \equiv 0 ~.
\]

Given $(\om, \theta, a, g) \in {\cal A}$, using this assumption and
\eqref{xi.est1}--\eqref{xi.est2}, we have
\begin{align*}
\norm{|g|^2g} _{H^2} &\le \norm{g}_{H^2}^3 \le C \e^3 \bkA{t}^{-6}
\\
\norm{F}_{H^2} & \le   C \e^2 \bkA{t}^{-3}
\\
|\om ^\triangle(t)| & \le C \e \bkA{t}^{-3/2} \le \e^{1/2} \bkA{t}^{-3/2}
\\
|\theta ^\triangle (t)| & \le \int_\infty^t \e^{1/2} \bkA{s}^{-3/2} \, ds
\le \e^{1/2} \bkA{t}^{-1/2}
\\
|a^\triangle (t)| & \le \int_\infty^t  C \e^2 \bkA{s}^{-3} \, ds
\le \e \bkA{t}^{-2}
\\
\norm{g^\triangle (t)}_{H^2} &
\le \int_\infty^t  C \e^2 \bkA{s}^{-3} + C \e^{3/2} \bkA{s}^{-3}\, ds
\le \e \bkA{t}^{-2}
\end{align*}
We have shown that
 $(\om^\triangle, \theta^\triangle, a^\triangle, g^\triangle) \in {\cal A}$,
that is, our mapping maps ${\cal A}$ into itself.

\bigskip

Next we show that it is a contraction.
Given $(\om_1, \theta_1, a_1, g_1),  (\om_2, \theta_2, a_2, g_2)\in {\cal A}$,
we denote
\[
\diff_0 = \sup_t \bket{ \bkA{t}^3|\diff \om(t)|^2 +\bkA{t}|\diff \theta(t)|^2
+ \bkA{t}^2|\diff a(t)| +\bkA{t}^2 \norm{\diff g(t)}_{H^2}  }
\]
we know $\diff_0 \le 8 \e$.
Notice that $-i|\xi|^2\xi$ is cancelled in $\diff F$.
\begin{align*}
\norm{\diff(|g|^2g)} _{H^2} &\le \norm{g}_{H^2}^2
 \norm{\diff g}_{H^2} \le C \e^2 \diff_0 \bkA{t}^{-6}
\\
\norm{\diff F}_{H^2} & \le   C \e \,\diff_0 \bkA{t}^{-7/2}
\le   C \e \,\diff_0 \bkA{t}^{-3}
\\
|\diff \om ^\triangle(t)| & \le C \diff_0 \bkA{t}^{-3/2}
\le \frac 18 \diff_0^{1/2} \bkA{t}^{-3/2}
\\
|\diff \theta ^\triangle (t)| & \le \int_\infty^t \diff_0^{1/2}
\bkA{s}^{-3/2} \, ds
\le \frac 18 \diff_0^{1/2} \bkA{t}^{-1/2}
\\
|\diff a^\triangle (t)| & \le \int_\infty^t  C \e \diff_0 \bkA{s}^{-3} \, ds
\le \frac 18 \diff_0 \bkA{t}^{-2}
\\
\norm{\diff g^\triangle (t)}_{H^2} &
\le \int_\infty^t  C \e \, \diff_0 \bkA{s}^{-3}
+ C \e^{1/2} \diff_0 \bkA{s}^{-3}\, ds
\le \frac 18 \diff_0 \bkA{t}^{-2}
\end{align*}
Therefore we have
\[
\sup_t \bket{   \bkA{t}^3|\diff \om^\triangle(t)|^2
+\bkA{t}|\diff \theta^\triangle(t)|^2
+ \bkA{t}^2|\diff a^\triangle(t)|
+\bkA{t}^2 \norm{\diff g^\triangle(t)}_{H^2} }
\le \frac 12 \diff_0 ~.
\]
These show that our map is a contraction mapping.
We conclude that we do have solutions  $\wt{h}$ with the main
profile $e^{i \sA t}\xii$.

Now
\[
\chi(x,t) =  e^{i \Theta} h
= e^{i (-Et+\theta)} U^{-1} e^{-i\theta} \wt{h}
= e^{-iEt} \bket{U^{-1} + [U,e^{i\theta}]\, e^{-i\theta} }
(e^{-it\sA} \xii + g)
\]
Since $\norm{g(t)}_{L^2}=O(t^{-2})$ and
$[U,e^{i\theta}]=O(\theta(t))=O(t^{-1/2})$,
we have
\[
\chi(x,t) \to e^{-iEt} U^{-1} e^{-it\sA} \xii, \qquad
\text{as } t \to \infty.
\]
However,
\begin{align*}
e^{-iEt} U^{-1} e^{-it\sA}\xii
&= e^{-iEt}  U^{-1} e^{-it\sA} U W \chi_\infty
= e^{-iEt} e^{t \L} W \chi_\infty
\\
&\longrightarrow  e^{-iEt} e^{i (\Delta +E) t} \chi_\infty
=e^{i \Delta  t} \chi_\infty
\end{align*}
as $t \to \infty$.
Hence part (2) of Theorem \ref{mainthm} is proved.

\bigskip  {\bf Remark on radiation dominated solutions to
Klein-Gordon equations }

We now sketch a construction for radiation dominated solutions to
Klein-Gordon equations.  We follow the notation in the
introduction. For a specified profile $\eta_\pm$, let
\[
   u = \xi + g
\]
where
\[
 \xi(t) = e^{i B t} \eta_+ + e^{-i B t} \eta_-
\]
and $g$ denotes the rest. Then we have
\begin{align*}
(\partial_t^2  + B^2) \xi &= 0,
\\
(\partial_t^2  + B^2) g &= \la ( \xi + g)^3 ~.
\end{align*}
Hence $g(t)$ satisfies
\[
g(t) = \int_\infty^t \bket{ e^{i B (t-s)} - e^{-i B (t-s)} } \frac
1{2i B}  \la (\xi+g)^3  ds ~,
\]
Since the main source term is
\[
 \norm{\xi^3}_2  \le  \norm{\xi}_\infty^2  \norm{\xi}_2
\le C t^{-3}
\]
we get $\norm{g}_2 \le \int_\infty^t C s^{-3} \, ds = C t^{-2}$.
Then we proceed as in section 10 to construct one such solution by
a contraction mapping argument.

\subsection*{Acknowledgments}~
It is a great pleasure to thank M.I. Weinstein for
explaining to us the beautiful  ideas of the work \cite{SW2}.
Part of this work was done when both authors were visiting the
Center for Theoretical Sciences, Taiwan, in the summer 2000.


\begin{thebibliography}{XXXX}

\bibitem {B} J. Bourgain and W. Wang,
Construction of blowup solutions for the nonlinear Schrödinger equation
with critical nonlinearity,
Ann. Scuola Norm. Sup. Pisa Cl. Sci. (4) 25  no. 1-2, 197--215 (1998).



\bibitem{BP} V.S.~Buslaev and G.S.~Perel'man,
On the stability of solitary waves for nonlinear
Schrödinger equations. Nonlinear evolution equations, 75--98,
Amer. Math. Soc. Transl. Ser. 2, 164,
Amer. Math. Soc., Providence, RI, 1995.


\bibitem{C}
C.~Cuccagna, Stabilization of solutions to nonlinear Schr\"odinger
equations, preprint.


\bibitem{FTY} J. Fr\"ohlich, T.-P. Tsai and H.-T. Yau,
On the point-particle (Newtonian) limit of the non-linear Hartree
equation, preprint.


\bibitem{HSS}
W.~Hunziker,  I.M.~Sigal and A.~Soffer, Minimal escape velocities.
Comm. Partial Differential
Equations {\bf 24} (1999), no. 11-12, 2279--2295.


\bibitem{JK} A. Jensen and T.~Kato,
Spectral properties of Schr\"odinger operators and time-decay
of the wave functions,
{\it Duke Math. J.} {\bf 46} no. 3, (1979), 583--611.


\bibitem{JSS}
J.-L.~Journe, A.~Soffer and C.D.~Sogge,
Decay estimates for Schr\"odinger operators.
\textit{Comm. Pure Appl. Math.} {\bf 44} (1991), no. 5, 573--604.

\bibitem{M} M. Murata,
Asymptotic expansions in time for solutions of Schr\"odinger-type
equations, {\it J. Funct. Anal.} {\bf 49}, (1982), 10--56.

\bibitem{RS}
 M.~Reed and B.~Simon, {\it Methods of modern mathematical physics},
%
III, scattering theory,
Academic Press, New York, San Francisco, London, 1975.


\bibitem{Sk} E.~Skibsted,
Propagation estimates for $N$-body Schroedinger operators. Comm.
Math. Phys. {\bf 142} (1991), no. 1, 67--98.


\bibitem{SW1}
A.~Soffer and M.I.~Weinstein,
Multichannel nonlinear scattering theory for nonintegrable equations I, II,
{\it Comm. Math. Phys.} {\bf 133} (1990), 119--146;
{\it J. Diff. Eqns.} {\bf 98}, (1992), 376--390.

\bibitem{SW2}
A.~Soffer and M.I.~Weinstein,
Resonances, radiation damping and instability
in Hamiltonian nonlinear wave equations,
{\it Invent.~math.} {\bf 136}, (1999), 9--74.


\bibitem{W1}
M.I.~Weinstein, Modulational stability of ground states
of nonlinear Schr\"odinger equations,
\textit{SIAM~J. Math. Anal.} \textbf{16} (1985), no. 3, 472--491.

\bibitem{W2}
M.I.~Weinstein, Lyapunov stability of ground states
of nonlinear dispersive evolution equations,
\textit{Comm. Pure Appl. Math.} {\bf 39}, (1986), 51--68.

\bibitem{Y}
K.Yajima, The $W^{k,p}$ continuity of wave operators for
Schr\"odinger operators,
{\it J.Math. Soc. Japan} {\bf 47}, no. 3, (1995), 551--581.




\end{thebibliography}
\end{document}